\documentclass[a4paper,11pt]{article}
\usepackage[a4paper]{geometry}
\usepackage{amssymb,amsmath,graphicx,epsfig,amscd,array,color}
\usepackage{amsthm}
\usepackage[round]{natbib}
\usepackage{enumerate}
\usepackage{mathrsfs}
\usepackage{authblk} 
\usepackage{subfigure}

\newtheorem{main}{Main result}

\newcommand\norm[1]{\lVert#1\rVert}

\DeclareMathOperator{\Cov} {Cov}
\DeclareMathOperator{\diag}{diag}

\title{Multivariate Gaussian Random Fields Using Systems of Stochastic Partial Differential Equations}
\author[1]{Xiangping Hu\footnote{Corresponding author. Email: \texttt{Xiangping.Hu@math.ntnu.no}}}
\author[1]{Daniel Simpson}
\author[2]{Finn Lindgren}
\author[1]{H\aa{}vard Rue}
\affil[1]{Department of Mathematical Sciences, Norwegian University of Science and Technology, N-7491 
Trondheim, Norway}
\affil[2]{Department of Mathematical Sciences, University of Bath, BA2 7AY, United Kingdom}

\date{July 4, 2013}
\begin{document}

\maketitle

\begin{abstract}
In this paper a new approach for constructing \emph{multivariate} Gaussian random fields (GRFs)
using systems of stochastic partial differential equations (SPDEs) has been introduced and applied to simulated data and real data. By solving a system of SPDEs, we can construct multivariate GRFs.
On the theoretical side, the notorious requirement of non-negative definiteness for the covariance matrix of the GRF is satisfied since the constructed covariance matrices with this approach are automatically symmetric positive definite.
Using the approximate stochastic weak solutions to the systems of SPDEs, multivariate GRFs are represented by multivariate Gaussian \emph{Markov} random fields (GMRFs) with sparse precision matrices. 
Therefore, on the computational side, the sparse structures make it possible to use numerical algorithms for sparse matrices to do fast sampling from the random fields and statistical inference. 
Therefore, the \emph{big-n} problem can also be partially resolved for these models. These models out-preform existing multivariate GRF models on a commonly used real dataset.

\noindent \textbf{Keywords}: Multivariate Gaussian random fields; Gaussian Markov random fields; covariance matrix; stochastic partial differential equations; sparse matrices
\end{abstract}

\section{Introduction} \label{sec: multi_introduction}
Gaussian random fields (GRFs) have a dominant role in spatial modelling and there exists a wealth of applications in areas
such as geostatistics, atmospheric and environmental science and many other fields \citep{cressie1993statistics,stein1999interpolation,diggle2006model}.
GRFs are practical since the normalizing constant can be computed explicitly. They also process good properties
since they can be explicitly specified through the mean function $\boldsymbol{\mu}(\cdot)$ and covariance function $\boldsymbol{C}(\cdot, \cdot)$. In $\mathbb{R}^d$,
with $\boldsymbol{s} \in \mathbb{R}^d$, $x(\boldsymbol{s})$ is a continuously indexed GRF if all the finite subsets $\left( x(\boldsymbol{s}_i) \right)_{i = 1}^n$ jointly have Gaussian
distributions. A GRF is said be stationary or homogeneous if the mean function $\mu(\cdot)$  is constant and the covariance function $\boldsymbol{C}(\cdot, \cdot)$
is a function only of the distance between the coordinates \citep{adler2007random}.  
It is common to use covariance functions that are isotropic. The isotropic covariance functions are functions of just the Euclidean distance between two locations.

\subsection{Mat\'{e}rn covariance functions and Multivariate GRFs} \label{sec: MaternCov}
The Mat\'ern family of covariance functions is a class of commonly used isotropic covariance functions introduced by Mat\'{e}rn \citep{matern1986spatial}. This family of covariance functions is usually parametrized by
$\sigma^2 M(\boldsymbol{m}, \boldsymbol{n}|\nu,a)$, where, $M(\boldsymbol{m}, \boldsymbol{n}|\nu,a)$ is the Mat\'{e}rn correlation function between locations
$\boldsymbol{m}$, $\boldsymbol{n}$ $\in \mathbb{R}^d$ and has the form
\begin{equation}
M(\boldsymbol{h} |\nu, a) = \frac{2^{1-\nu}}{\Gamma(\nu)}(a \| \boldsymbol{h} \|)^\nu K_\nu(a \| \boldsymbol{h} \|),
\end{equation}
where $\sigma^2$ is the marginal variance. $\| \boldsymbol{h} \|$ denotes the Euclidean distance between $\boldsymbol{m}$
and $\boldsymbol{n}$. $K_\nu$ is the modified Bessel function of second kind and $a > 0$ is a scaling parameter. The order $\nu $ is
the smoothness parameter. It defines the critical concerns in spatial statistical modelling and simulation \citep{stein1999interpolation},
such as the differentiability of the sample paths and the Hausdorff dimension \citep{handcock1993bayesian, goff1988stochastic}. We follow the common practice to fix the smoothness parameter $\nu$ because
it is poorly identifiable from data \citep{ribeiromodel, lindgren2011explicit}. The range $\rho$ connects the scaling parameter $a$ and the smoothness parameter $\nu$.
Throughout the paper, the simple relationship $\rho = \sqrt{8\nu}/a $ is assumed from the empirically derived definition \citep{lindgren2011explicit}.
We refer to \citet{matern1986spatial}, \citet{diggle1998model}, \citet{guttorp2006studies} and \citet{lindgren2011explicit} for detailed information about Mat\'{e}rn covariance functions.
One can show  that the Mat\'{e}rn covariance function can be reduced to the product of a polynomial and an exponential function when the smoothness parameter
$\nu$ is equal to an integer $m$ plus $\frac{1}{2}$ \citep{gneitingmatern}. The Mat\'ern covariance function nests the popular exponential model since $M(\boldsymbol{h}|\frac{1}{2}, a) = \exp(-a\| \boldsymbol{h} \|) $. 

Analogously to the univariate case, we need to specify a mean vector function $\boldsymbol{\mu}(\boldsymbol{s})$ and a covariance function $\boldsymbol{C}(\|\boldsymbol{h}\|)$ that assigns to 
each distance $\|\boldsymbol{h}\|$ a $p\times p$ symmetric non-negative definite covariance matrix in order to specify an isotropic $p$-dimensional multivariate Gaussian random field $\boldsymbol{x}(\boldsymbol{s})$.  
It is known that it is quite difficult to construct flexible covariance functions  $\boldsymbol{C}(\cdot)$ that satisfy this requirement.

\citet{gneitingmatern} presented an approach for constructing multivariate random fields using matrix-valued covariance functions, where each constituent
component in the matrix-valued covariance function is a Mat\'{e}rn covariance function. Define $\boldsymbol{x}(\boldsymbol{s}) = (x_1(\boldsymbol{s}), x_2(\boldsymbol{s}), \cdots, x_p(\boldsymbol{s}) )^\text{T}$, 
with $\boldsymbol{s} \in \mathbb{R}^d $, so that
each location consists $p$ components. $\text{T}$ denotes the transpose of a vector or a matrix.
Assume the process is second-order stationary with mean vector zero and matrix-valued covariance function of the form
\begin{align} \label{eq: CovarianceMatrix_M}
 \ \boldsymbol{C}(\boldsymbol{h})  = \left( \begin{array}{cccc}
  C_{11}(\boldsymbol{h}) & C_{12}(\boldsymbol{h}) & \cdots & C_{1p}(\boldsymbol{h}) \\
  C_{21}(\boldsymbol{h}) & C_{22}(\boldsymbol{h}) & \cdots & C_{2p}(\boldsymbol{h}) \\
  \vdots & \vdots & \ddots &  \vdots \\
  C_{p1}(\boldsymbol{h}) & C_{p2}(\boldsymbol{h}) & \cdots & C_{pp}(\boldsymbol{h}) \\
         \end{array} \right),
\end{align}

\noindent where,
\begin{equation} \label{eq: Cii}
 C_{ii}(\boldsymbol{h}) = \sigma_{ii} M(\boldsymbol{h} |\nu_{ii}, a_{ii}) ,
\end{equation}
is the covariance function $C_{ii}(\boldsymbol{h}) = \mathbb{E}(x_i(\boldsymbol{s}+\boldsymbol{h}) x_i(\boldsymbol{s}))$ within the field $ x_i (\boldsymbol{s})$ and
\begin{equation} \label{eq: Cij}
 C_{ij}(\boldsymbol{h}) = \rho_{ij} \sigma_{ij}  M(\boldsymbol{h} |\nu_{ij}, a_{ij})
\end{equation}
 is the cross-covariance function $C_{ij}(\boldsymbol{h}) = \mathbb{E}(x_i(\boldsymbol{s}+\boldsymbol{h}) x_j(\boldsymbol{s}))$ between the fields $x_i(\boldsymbol{s})$ 
and $x_j(\boldsymbol{s})$. $\rho_{ij}$ is the co-located correlation coefficient, $\sigma_{ii} \geq 0$ is the marginal variance,
and $\sigma_{i}$ and $\sigma_{j}$ are the corresponding standard deviations, $1 \leq i \neq j \leq p$. We use the following notations through out the paper
\begin{equation}
\begin{split}
\sigma_{ii} & = \sigma_i \sigma_i ,\\
\sigma_{ij} & = \sigma_i \sigma_j .
\end{split}
\end{equation}

\citet{gneitingmatern} presented some conditions to ensure the matrix-valued covariance function in Equation \eqref{eq: CovarianceMatrix_M}
is symmetric and non-negative definite, with focus on the bivariate case. They claimed that the parameters in the parametric family of matrix-valued covariance function given in Equation \eqref{eq: CovarianceMatrix_M} 
for multivariate random fields are interpretable in terms of smoothness, correlation length, variances of the processes and co-located coefficients. It was shown that a parsimonious bivariate Mat\'{e}rn model, 
with fewer parameters than the full bivariate Mat\'{e}rn model, is preferred to the traditional linear model of coregionalization (LMC)  \citep{gelfand2004nonstationary}. 
Even though the GRFs have convenient analytical properties, the covariance-based models for constructing GRFs are hindered by computational issues, or the so-called ``\emph{big-$n$} problem'' \citep{banerjee2004hierarchical}.
This is because inference with these models requires the factorization of dense $n \times n $ covariance matrices, which requires $\mathcal{O}(n^2)$ storage 
and $\mathcal{O}(n^3)$ floating point operations. It follows that  this kind of model is not suitable for problems with realistically large amounts of data.

Statistical inference for large datasets within feasible time is still a challenge in modern
spatial statistics. The size of modern datasets typically overwhelm the traditional models in spatial statistics and a great deal of effort has been expended trying to construct good models that scale well computationally.
In this paper, we extend the models proposed by \citet{lindgren2011explicit}, which exploited an explicit link between GRFs and GMRFs through stochastic partial differential equations (SPDEs). They showed that for some univariate Gaussian
fields with Mat\'{e}rn covariance functions, it is possible to do the modelling with GRFs, but do the computations using GMRFs. 

A GMRF is a discretely indexed Gaussian random field $\boldsymbol{x}(\boldsymbol{s})$ with the property that its full conditional distributions $\{ \pi (x_i|{\boldsymbol{x}_{-i}}); i = 1, \dots, n \}$, only depend on the set of neighbors
$\partial i$ to each node $i$. It is obvious that if $i \in \partial j$ then $j \in \partial i$ from consistency. Rather than parameterizing GMRFs through their covariance matrix, 
they are usually parametrized through their precision matrix (the inverse of the covariance matrix). The Markov property implies that the precision matrix $\boldsymbol{Q}$ 
is sparse due to the fact that $Q_{ij} \neq 0  \Longleftrightarrow i \in \partial j \cup j$ \citep{rue2005gaussian}. This allows for the use of numerical methods for 
sparse matrices for fast sampling and also for fast statistical inference \citep{rue2001fast,rue2005gaussian, lindgren2011explicit}.
The general cost of factorizing the precision matrix $\boldsymbol{Q}$ for a spatial Markovian structure is $\mathcal{O}(n)$ in one dimension, $\mathcal{O}(n^{3/2})$ in two dimensions and  $\mathcal{O}(n^2)$
in three dimensions \citep{rue2005gaussian}.

\citet{lindgren2011explicit} pointed out that using the link between the GRFs and GMRFs can open new doors to modelling difficult problems with
simple models. The SPDE approach can be extended to Gaussian random fields on manifolds, non-stationary random fields, random fields with oscillating
covariance functions and non-separable space-time models.

\subsection{Mat{\'e}rn covariance models through SPDEs}
In this paper we use the following characterization of Mat\'{e}rn random fields, originally due to \citet{whittle1954stationary,whittle1963stochastic}, 
that formed the basis for the methods of \citet{lindgren2011explicit}. A GRF $x(\mathbf{s})$ with
the Mat{\'e}rn covariance function can be described as a solution to the linear fractional SPDE \citep{whittle1954stationary, whittle1963stochastic, lindgren2011explicit}

\begin{equation} \label{eq: spde_simple}
 (\kappa^2 - \Delta)^{\alpha/2} x(\boldsymbol{s}) = \mathcal{W}(\boldsymbol{s}),
\end{equation}
where $(\kappa^2 - \Delta)^{\alpha/2}$ is a pseudo-differential operator and $\alpha = \nu + d/2$, $\kappa > 0$, $\nu > 0$. The innovation process $\mathcal{W}(\boldsymbol{s})$ is a spatial standard Gaussian white noise. 
$\Delta = \sum_{i=1}^d \frac{\partial^2}{\partial x_i^2}$ is the Laplacian on $\mathbb{R}^d$.
Applying the Fourier transform to the (fractional) SPDE given in $\mathbb{R}^d$ in \eqref{eq: spde_simple} yields
\begin{equation} \label{eq: spde_simple_fourier}
 \left\{ \mathscr{F}{(\kappa^2 - \Delta)^{\alpha/2}}\phi \right\} (\boldsymbol{k}) = \left(\kappa^2 + \|\boldsymbol{k}\|^2\right) ^{\alpha/2} \left(\mathscr{F} \phi\right) (\boldsymbol{k}),
\end{equation}
where $\mathscr{F}$ denotes the Fourier transform, $\phi$ is a smooth, rapidly decaying function in $\mathbb{R}^d$.
See for example, \citet{lindgren2011explicit} for detailed description. Equation \eqref{eq: spde_simple_fourier} is used in Section \ref{sec: Model_Comparisions} for model comparison.
One might think that the Mat{\'e}rn covariance function seems rather restrictive
in statistical modelling, but it covers the most commonly used models in spatial statistics \citep{lindgren2011explicit}. \citet[Page 14]{stein1999interpolation} 
has a practical suggestion ``use the Mat{\'e}rn model''. For more information about the Mat{\'e}rn family, see, for example \citet[Section $3.4.1$]{diggle1998model} and \citet[Section $2.6$]{stein1999interpolation}.

\subsection{Outline of the paper}
The rest of the paper is organized as follows. In Section \ref{sec: Model_constr} we discuss the construction of  multivariate GRFs using systems of SPDEs,
we call this approach the SPDEs approach. Additionally, a brief introduction of how to construct the multivariate GRFs using the covariance-based models is also included in this section.
Section \ref{sec: Model_Comparisions} contains a detailed model comparisons between the SPDEs approach and the covariance-based models presented by \citet{gneitingmatern}.
Section \ref{sec: Sampling} discusses how to sample from these models, and statistical inference for simulated data and real dataset is presented in Section \ref{sec: applications}.
The paper ends with a general discussion in Section \ref{sec: discussion}.

\section{Model construction} \label{sec: Model_constr}
In this section we discuss how to use the SPDEs approach to construct multivariate GRFs. One of the appealing properties of this approach is that
the SPDE specification \emph{automatically} constructs valid covariance functions. This means that if the solution to a system of SPDEs exists, then it will
construct a matrix-valued covariance function which fulfills the symmetric non-negative definiteness property. The parameters in the parametric model 
from the SPDEs approach are interpretable in terms of co-located correlation coefficients, smoothness, marginal variances and correlations. 
And there is a correspondence to the parameters in the covariance matrix based models.

\subsection{Multivariate GRFs and the SPDEs approach}
 A zero mean $d$--dimensional multivariate (or $d$--dimensional $p$-variate) GRF is a collection
of continuously indexed multivariate normal random vectors
$$
\boldsymbol{x}(\boldsymbol{s}) \sim MVN(\boldsymbol{0},\boldsymbol{\Sigma}(\boldsymbol{s}) ),
$$
where $\boldsymbol{\Sigma}(\boldsymbol{s})$ is a non-negative definite matrix that depends on the point $\boldsymbol{s} \in \mathbb{R}^d$.
Define the system of SPDEs
\begin{align} \label{eq: SPDEs_system}
\begin{pmatrix}
\mathcal{L}_{11} & \mathcal{L}_{12} & \ldots & \mathcal{L}_{1p}\\
\mathcal{L}_{21} & \mathcal{L}_{22} & \ldots & \mathcal{L}_{2p}\\
\vdots & \vdots & \ddots&\vdots\\
\mathcal{L}_{p1} & \mathcal{L}_{p2} &\ldots & \mathcal{L}_{pp}
\end{pmatrix}
\begin{pmatrix}
x_1(\boldsymbol{s})\\ x_2(\boldsymbol{s})\\ \vdots \\ x_p(\boldsymbol{s})
\end{pmatrix}
=
\begin{pmatrix}
f_1(\boldsymbol{s}) \\ {f}_2(\boldsymbol{s}) \\ \vdots \\ f_p(\boldsymbol{s})
\end{pmatrix},
\end{align}
where $\{ \mathcal{L}_{ij} = b_{ij}(\kappa_{ij}^2 - \Delta)^{\alpha_{ij}/2}$; $\alpha_{ij} = 0 \text{ or } 2, 1 \leq i, j \leq p \}$, are differential operators and $\{ f_i(\boldsymbol{s}); 1 \leq i, j \leq p\} $
are independent but not necessarily identically distributed noise processes.
It turns out that the solution to \eqref{eq: SPDEs_system} defines a multivariate GRF $\boldsymbol{x} = \left( x_1(\boldsymbol{s}), x_2(\boldsymbol{s}), \cdots, x_p(\boldsymbol{s}) \right)^\text{T}$.
Define the operator matrix
\begin{equation} \label{eq: spde_operatormatrix}
 \mathscr{L} = \begin{pmatrix}
\mathcal{L}_{11} & \mathcal{L}_{12}& \ldots & \mathcal{L}_{1p}\\
\mathcal{L}_{21} & \mathcal{L}_{22}& \ldots & \mathcal{L}_{2p}\\
\vdots & \vdots & \ddots&\vdots\\
\mathcal{L}_{p1} & \mathcal{L}_{p2} &\ldots & \mathcal{L}_{pp}
\end{pmatrix}
\end{equation}
and
$$
\boldsymbol{f}(\boldsymbol{s}) = (f_1(\boldsymbol{s}), f_2(\boldsymbol{s}), \cdots , f_p(\boldsymbol{s}))^\text{T},
$$
and then the system of SPDEs \eqref{eq: SPDEs_system} can be written in compact matrix form as
\begin{equation} \label{eq: SPDEs_system_matrix}
\mathscr{L}(\boldsymbol{\theta}) \boldsymbol{x}(\boldsymbol{s}) = \boldsymbol{f}(\boldsymbol{s}),
\end{equation}
where $\boldsymbol{\theta} = \{ \boldsymbol{\alpha}, \boldsymbol{\kappa}, \boldsymbol{b}\}$ denotes the collection of parameters. Similarly as 
Equation \eqref{eq: spde_simple_fourier} we apply Fourier transforms to Equation \eqref{eq: SPDEs_system_matrix}, and it yields
\begin{equation} \label{eq: SPDEs_system_matrix_Fourier}
\mathscr{H}(\boldsymbol{\theta}) \boldsymbol{\hat{x}}(\boldsymbol{k}) = \boldsymbol{\hat{f}}(\boldsymbol{k}),
\end{equation}
where $\boldsymbol{k}$ is the frequency, $\boldsymbol{\hat{x}}(\boldsymbol{k}) = \mathscr{F}\left(\boldsymbol{x}(\boldsymbol{s})\right)$ and 
$ \boldsymbol{\hat{f}}(\boldsymbol{k}) = \mathscr{F}\left(\boldsymbol{f}(\boldsymbol{s})\right)$ are the Fourier transforms of
the random fields and the noise processes, respectively, and $\mathscr{H}(\boldsymbol{\boldsymbol{\theta}})$ is the matrix formed from the Fourier transforms of the operator matrix $\mathscr{L}(\boldsymbol{\theta})$. Each element has 
the form $\mathcal{H}_{ij}(\theta_{ij}) = b_{ij}(\kappa_{ij}^2 + \| \boldsymbol{k} \|^2)^{\alpha_{ij}/2}$
with $\theta_{ij} = \{\alpha_{ij}, \kappa_{ij}, b_{ij}\}$, $1 \leq i,j \leq p$.
From Equation \eqref{eq: SPDEs_system_matrix_Fourier} one can find the spectral process as
 $\boldsymbol{\hat{x}}(\boldsymbol{k})= \mathscr{H}^{-1}(\boldsymbol{\theta}) \boldsymbol{\hat{f}}(\boldsymbol{k})$.
The corresponding power spectrum is defined as
$ \boldsymbol{S}_{\boldsymbol{x}}(\boldsymbol{k}) = \mathbb{E}\left( \boldsymbol{\hat{x}\hat{x}}^{\mbox{H}} \right)$, where $\mbox{H}$ denotes the Hermitian transpose of a matrix.
Simple calculations using the above mentioned formulas yield
\begin{equation} \label{eq: PowerSpectrum_x}
 \boldsymbol{S}_{\boldsymbol{x}}(\boldsymbol{k}) = \mathscr{H}^{-1}(\boldsymbol{\theta}) \boldsymbol{S}_{\boldsymbol{f}}(\boldsymbol{k}) \mathscr{H}^{-H}(\boldsymbol{\theta}),
\end{equation}
where $\mathscr{H}^{-H}$ denotes the inverse of the Hermitian transpose of Fourier transform of the operator matrix $\mathscr{L}$.
The Equation \eqref{eq: PowerSpectrum_x} can be written explicitly as
\begin{align} \label{eq: SPDEs_Fourier_M}
 \ \boldsymbol{S}_{\boldsymbol{x}}(\boldsymbol{k})  = \left( \begin{array}{cccc}
  S_{x_{11}}(\boldsymbol{k}) & S_{x_{12}}(\boldsymbol{k}) & \cdots & S_{x_{1p}}(\boldsymbol{k}) \\
  S_{x_{21}}(\boldsymbol{k}) & S_{x_{22}}(\boldsymbol{k}) & \cdots & S_{x_{2p}}(\boldsymbol{k}) \\
  \vdots & \vdots & \ddots &  \vdots \\
  S_{x_{p1}}(\boldsymbol{k}) & S_{x_{p2}}(\boldsymbol{k}) & \cdots & S_{x_{pp}}(\boldsymbol{k}) \\
         \end{array} \right).
\end{align}
Let $\boldsymbol{S}_{\boldsymbol{f}}(\boldsymbol{k})$ denote the power spectrum matrix of the noise processes, 
$\boldsymbol{S}_{\boldsymbol{f}}(\boldsymbol{k}) = \mathbb{E}\left( \boldsymbol{\hat{f}\hat{f}}^{\mbox{H}} \right)$. If the noise processes are mutually independent, $\boldsymbol{S}_{\boldsymbol{f}}$ is a diagonal matrix and we write it as
$$\boldsymbol{S}_{\boldsymbol{f}}(\boldsymbol{k})  = \diag(S_{f_{11}},S_{f_{22}}, \cdots, S_{f_{pp})},$$ 
where $\{ S_{f_{i}} = \mathbb{E}\left( \hat{f_i}\hat{f_i}^{\mbox{H}} \right); i = 1, \cdots, p \}$. This means it is easy to obtain this matrix.
In general, with Equations \eqref{eq: SPDEs_system} to \eqref{eq: PowerSpectrum_x} we can compute all the components in \eqref{eq: SPDEs_Fourier_M}. We show how to do this for bivariate random fields in Section 
\ref{sec: bivariate_GRF}.

\subsection{Covariance-based model for multivariate GRFs} \label{sec: covarmodel}
It is also possible to construct a multivariate GRF using the covariance-based model, but the notorious non-negative definiteness restriction makes it hard. 
\citet{gneitingmatern} discussed one such approach in detail.
The main aim of their approach is to find the proper constrains for $\nu_{ij}, a_{ij}, \rho_{ij}$ which result in valid matrix-valued covariance functions 
for second-order stationary multivariate GRFs with a covariance function of the form in Equation \eqref{eq: CovarianceMatrix_M}, 
that is, a symmetric non-negative definite covariance function.
Some useful theorems for constructing the covariance functions for bivariate case are presented in \citet[Section $2.2$]{gneitingmatern}.

In this covariance-based model, the components $C_{ij}(\boldsymbol{h})$ in the matrix-valued covariance function $\boldsymbol{C}(\boldsymbol{h})$ given in Equation \eqref{eq: CovarianceMatrix_M} 
are modelled directly. \citet{gneitingmatern} try to find conditions on the parameter space
which result in valid multivariate Mat{\'e}rn models. In the bivariate case when $p = 2$, a full characterization of the parameter space is achieved. For $p\geq 3$ \citet{gneitingmatern} suggested that a parsimonious model should be used in practice.
This kind of model has more restrictions on the smoothness parameter and the scale parameters, such that $a_{ij} = a$, where $a>0$ is the common scale parameter and $\nu_{ij} = \frac{1}{2}(\nu_i +\nu_j)$ for $1 \leq i \neq j \leq p$.
We refer to \citet{gneitingmatern} for more information about different kinds of conditions that yield the valid matrix-valued covariance functions.

Assume that the components of the covariance matrix are known. The power spectrum
can also be obtained from in this covariance-based approach. This can in turn be used to compare parameters with the SPDEs approach.
The covariance matrix of a second-order stationary multivariate Mat\'ern GRF was given in Equation \eqref{eq: CovarianceMatrix_M}. The power spectrum in $\mathbb{R}^d$
of the cross-covariance function for $x_i(\boldsymbol{s})$ and $x_j(\boldsymbol{s})$ is defined as $R_{x_{ij}}(\boldsymbol{k}) = \mathscr{F}\left(\Cov(x_i\left(\boldsymbol{s}), x_j(\boldsymbol{s})\right)\right)$, $1 \leq i, j \leq p$,
which is given by

\begin{equation} \label{eq: Rij_iit}
 R_{x_{ij}}(\boldsymbol{k}) = \frac{1}{(2\pi)^d} \int_{\mathbb{R}^d} e^{-i\boldsymbol{kh}} C_{ij}(\|{\boldsymbol{h}}\|) d\boldsymbol{h}.
\end{equation}
Applying the Fourier transform to the covariance matrix in Equation \eqref{eq: CovarianceMatrix_M} yields
\begin{align} \label{eq: CovarianceMatrix_Fourier_M}
 \ \boldsymbol{R}(\boldsymbol{k})  = \left( \begin{array}{cccc}
  R_{x_{11}}(\boldsymbol{k}) & R_{x_{12}}(\boldsymbol{k}) & \cdots & R_{x_{1p}}(\boldsymbol{k}) \\
  R_{x_{21}}(\boldsymbol{k}) & R_{x_{22}}(\boldsymbol{k}) & \cdots & R_{x_{2p}}(\boldsymbol{k}) \\
  \vdots & \vdots & \ddots &  \vdots \\
  R_{x_{p1}}(\boldsymbol{k}) & R_{x_{p2}}(\boldsymbol{k}) & \cdots & R_{x_{pp}}(\boldsymbol{k}) \\
         \end{array} \right).
\end{align}
The comparison between the covariance-based approach and the SPDEs approach in Section \ref{sec: Model_Comparisions} is based on the 
following fact: if $S_{x_{ij}}(\boldsymbol{k}) = R_{x_{ij}}(\boldsymbol{k})$, for each  $1 \leq i, j \leq p$, then the multivariate GRFs 
constructed through the SPDEs approach \eqref{eq: SPDEs_system} and through the covariance-based model \eqref{eq: CovarianceMatrix_M} are equivalent.

\subsection{GMRF approximations to GRFs} \label{sec: gmrfs_approximation}
It is known that spatial Markovian GMRFs have good computational properties since the precision matrix $\boldsymbol{Q}$ is sparse. 
The Markov structure here means that $x_i(\boldsymbol{s})$ and $x_j(\boldsymbol{s})$ are independent conditioned on
$x_{-ij}(\boldsymbol{s})$, when $i$ and $j$ are not neighbors, where $x_{-ij}(\boldsymbol{s})$ means for $x_{-\{i, j\}}(\boldsymbol{s})$.
In other words, $Q_{ij} = 0$ if and only if  $x_i(\boldsymbol{s})$ and $x_j(\boldsymbol{s})$ are independent conditioned $x_{-ij}(\boldsymbol{s})$. So it is possible to read off
whether $x_i$ and $x_j$ are conditionally independent or not from the elements of $\boldsymbol{Q}$. Note that the precision matrix $\boldsymbol{Q}$ should also be non-negative definite. The sparse structure of the
precision matrix $\boldsymbol{Q}$ is crucial with our models in Bayesian inference methods on large datasets. We refer to \citet[Chapter $2$]{rue2005gaussian} for more information on the theory for GMRFs.

\citet{rue2002fitting} demonstrated that most of the GRFs in geostatistics can be approximated with GMRFs. \citet{hartman2008fast} proposed to use GMRFs instead of
GRFs, from a computational point of view, when doing spatial prediction using Kriging.  This approach can also be used for spatio-temporal models \citep{allcroft2003latent}.
In this paper we follow the approach presented in \citet{lindgren2011explicit} for GMRF approximations to GRFs
 in order to partially resolve the ``\emph{big $n$ problem}'' with our models.  
\citet{lindgren2011explicit} mainly considered univariate GRFs with  Mat\'{e}rn covariance functions. 
In this case the GMRF representation was constructed explicitly through the SPDE, and the solution to the SPDE driven by the Gaussian white noise is a GRFs with Mat{\'e}rn covariance function. 
They showed how to build a GRF model with a covariance matrix $\boldsymbol{\Sigma}$ theoretically from SPDE, and then use the GMRF to represent the GRF, which means the precision matrix of the GMRF fulfills the condition
$\boldsymbol{{Q}^{-1}} \simeq \boldsymbol{\Sigma}$ on some predefined norm. 
\citet{fuglstad2011spatial} considered a modification of the SPDE with a diffusion matrices to control the covariance structure of the GRF, and create inhomogeneous GRFs, but the GMRF representation is still
the main ingredient for computations. 

With our SPDEs approach we use two steps to construct the multivariate GRFs. The first step is to construct the precision matrices for the noise processes in Equation \eqref{eq: SPDEs_system}, and the second step
is to solve the system SPDEs \eqref{eq: SPDEs_system} with the constructed noise processes.  
We focus on the noise processes, if they are not white noise processes, generated by the SPDE given as follows
\begin{equation} \label{eq: spde_noise}
 (\kappa_{n_i}^2 - \Delta)^{\alpha_{n_i}/2} f_i(\boldsymbol{s}) = \mathcal{W}_i(\boldsymbol{s}), \hspace{1mm} \alpha_{n_i} = \nu_{n_i} +d/2, \hspace{1mm} i = 1,2, \dots, p,
\end{equation}
with the requirements that $\kappa_{n_i} > 0$ and $\nu_{n_i} > 0$. $\kappa_{n_i}$ and $\nu_{n_i}$ are the scaling parameter and smoothness parameter for noise process $f_i(\boldsymbol{s})$.
$\{ \mathcal{W}_i(\boldsymbol{s}); i = 1, 2, \dots, p \}$ are standard Gaussian white noise processes. The generated noise processes in this way are independent but not necessarily identically distributed.

The first step in constructing the GMRF representation for the noise process $f_i(\boldsymbol{s})$ on the triangulated lattice is to find the stochastic weak formulation of Equation \eqref{eq: spde_noise} \citep{kloeden1992numerical}.
In this paper Delaunay triangulation is chosen, and we refer to \citet{hjelle2006triangulations} for detailed discussion about Delaunay triangulations.
Denote the inner product of functions $h$ and $g$ as
\begin{equation} \label{eq: spde_innerproduct}
 \langle h, g \rangle = \int h(\boldsymbol{s})g(\boldsymbol{s})d(\boldsymbol{s}),
\end{equation}
where the integral is over the region of interest. We can find the stochastic weak solution of SPDE \eqref{eq: spde_noise} by requiring that 

\begin{equation} \label{eq: spde_weakformulation}
 \left\{ \langle \phi_k, (\kappa_{n_i}^2 - \Delta)^{\alpha_{n_i}/2} f_i \rangle \right\}_{i=1}^M  \stackrel{d}{=} \left\{ \langle \phi_k, \mathcal{W}_i \rangle \right\}_{i=1}^M.
\end{equation}

In the second step we need to construct a finite element representation of the solution to the SPDE.
We refer to \citet{zienkiewicz2005finite} and \citet{bathe2008finite} for more information about finite element methods.
The finite element representation of the solution to SPDE \eqref{eq: spde_noise} is 
\begin{equation} \label{eq: representing}
 f_i(\boldsymbol{s}) = \sum_{k=1}^{N}{\psi_{k}(\boldsymbol{s})\omega_k},
\end{equation}
where $\psi_{k}(\boldsymbol{s})$ is some chosen basis function, $\omega_k$ is some Gaussian distributed weight, and $N$ is the number of the vertexes in the triangulation. The basis function $\psi_{k}(\boldsymbol{s})$
is chosen to be piecewise linear with value $1$ at vertex $k$ and $0$ at all other vertexes. This means that a continuously indexed solution is approximated with a piecewise linear function defined through the joint distribution 
of $\{ \omega_k; k = 1,2,\dots, N \}$ \citep{lindgren2011explicit}. The chosen basis functions ensure that the local interpolation on a triangle is piecewise linear.

The third step is to choose the test functions. In this paper we follow the setting used by \citet{lindgren2011explicit}. With $M = N$ the test functions are 
chosen as $\phi_k = (\kappa_{n_i}^2 - \Delta)^{1/2} \psi_k$ for $\alpha_{n_i} = 1$ and $\phi_k = \psi_k$ for $\alpha_{n_i} = 2 $, 
which are denoted as a \emph{least squares} and a \emph{Galerkin} solution, respectively. When $\alpha_{n_i} \geq 3$, the approximation can be obtained by setting $\alpha_{n_i} = 2$ at the left-hand side of Equation \eqref{eq: spde_noise}
and replacing the right hand side of Equation \eqref{eq: spde_noise} with a field generated with $\alpha_{n_i} -2$ and let $\phi_k = \psi_k$. This iteration procedure terminates when $\alpha_{n_i} = 1 \text{ or } 2 $. 
This is the essence of the recursive Galerkin formulation. More detailed description can be found in \citet{lindgren2011explicit}.
$\alpha_{n_i}$ can only be integer-valued currently. When $\alpha_{n_i}$ is not an integer, different approximation methods must be used and this is beyond the scope of our discussion. 
Response to the discussion to \citet{lindgren2011explicit} discussed fractional $\alpha$.
We define the required $N \times N$ matrices $\boldsymbol{C}$, and $\boldsymbol{G}$ with entries
\begin{equation}\label{eq: matrix_CG}
C_{mn} =  \langle \psi_m, \psi_n \rangle,  \hspace{2mm} G_{mn} = \langle \nabla \psi_m, \nabla \psi_n \rangle, \hspace{2mm} m,n = 1,2,\dots, N,
\end{equation}

\noindent and
\begin{equation} \label{eq: matrix_K}
 (\boldsymbol{K}_{\kappa_{n_i}^2}) = \left( \kappa_{n_i}^2 \boldsymbol{C} + \boldsymbol{G} \right). 
\end{equation}

Finally, by using these matrices given in \eqref{eq: matrix_CG} and \eqref{eq: matrix_K} together with the Neumann boundary conditions (zero normal derivatives at the boundaries), the precision matrices $\boldsymbol{Q}_{f_i}$ for 
noise process $f_i(\boldsymbol{s})$ can be obtained,
 \begin{gather}
 \begin{cases}
   \mathbf{Q}_{1,\kappa_{n_i}^2} & = \mathbf{K}_{\kappa_{n_i}^2},  \hspace{39mm} \text{ for }   \alpha_{n_i} = 1, \\
   \mathbf{Q}_{2,\kappa_{n_i}^2} & = \mathbf{K}_{\kappa_{n_i}^2}^\text{T} \mathbf{C}^{-1} \mathbf{K}_{\kappa_{n_i}^2},   \hspace{24mm} \text{ for } \alpha_{n_i} = 2,  \\
   \mathbf{Q}_{\alpha_{n_i}, \kappa_{n_i}^2} & = \mathbf{K}_{\kappa_{n_i}^2}^\text{T} \mathbf{C}^{-1} {\mathbf{Q}}_{\alpha_{n_i}-2, \kappa_{n_i}^2} \mathbf{C}^{-1} \mathbf{K}_{\kappa_{n_i}^2}, \text{ for } \alpha_{n_i} = 3,4, \cdots.
\end{cases}
 \end{gather}

As pointed out by \citet{lindgren2011explicit} the inverse of $\boldsymbol{C}$, i.e., $\boldsymbol{C}^{-1}$, 
is usually a dense matrix. This causes the precision matrix $\boldsymbol{Q}_{f_i}$  to be dense and ruins all the effort we have made.
So we actually used a diagonal matrix $\tilde{\boldsymbol{C}}$,  where $\tilde{\boldsymbol{C}}_{ii} =  \langle \psi_i, 1 \rangle $, instead of $\boldsymbol{C}$. 
This diagonal matrix results in a sparse precision matrix and hence sparse GMRFs models can be obtained. 
Using the diagonal matrix $\tilde{\boldsymbol{C}}_{ii}$ yields a Markov approximation to the FEM solution.
The effects of the Markov approximation
have been studied by  \citet{bolin2009wavelet}. They claimed that the difference between the exact representation by the finite element method and the Markov approximation is negligible.

Move to our multivariate GRFs constructed by systems of SPDEs \eqref{eq: SPDEs_system}. In this case we take the same set of basis functions $\left\{ \psi_k; k = 1,2,\dots, M \right\}$ and construct a basis for the solution space for
$(x_1, x_2, \dots, x_p)^\text{T}$ as
\small
\begin{equation}
p \left\{
 \begin{pmatrix}
  \psi_1  \\
   0      \\
   \vdots  \\
   0
 \end{pmatrix},\dots,
 \begin{pmatrix}
  \psi_M  \\
   0      \\
   \vdots  \\
   0
 \end{pmatrix},
 \begin{pmatrix}
   0      \\ 
 \psi_1  \\
   \vdots  \\
   0
 \end{pmatrix},
 \dots,
 \begin{pmatrix}
   0      \\
  \psi_M  \\
   \vdots  \\
   0
 \end{pmatrix},
  \dots,
 \begin{pmatrix}
   0      \\ 
   0      \\
   \vdots  \\
   \psi_1
 \end{pmatrix},
 \dots,
 \begin{pmatrix}
   0      \\
   0      \\
   \vdots  \\
   \psi_M
 \end{pmatrix}, \right.
\end{equation}
\normalsize
where there are a total of $Mp$ basis functions $\{ \boldsymbol{\psi}_i \}$ which are numbered in the order listed above.
The weak solution of the system of SPDEs \eqref{eq: SPDEs_system} requires 
\begin{equation}
 \left[ \langle \boldsymbol{\psi}_i, \mathscr{L} \boldsymbol{x} \rangle \right]_{i = 1}^{Mp} \stackrel{d}{=} \left[ \langle \boldsymbol{\psi}_i, \boldsymbol{f} \rangle \right]_{i = 1}^{Mp}, 
\end{equation}
with $\mathscr{L}$ defined in Equation \eqref{eq: spde_operatormatrix}. The finite element representation to the solution of the system of SPDEs is
\begin{equation}
 \boldsymbol{x}(\boldsymbol{s}) = \sum_{i = 1}^{Mp} \boldsymbol{\psi}_i(\boldsymbol{s}) \omega_i 
\end{equation}

In order to find the precision matrix for the solution we need to define the following matrices,
\begin{equation}  \label{eq: SPDE_multiCKQ}
  \begin{split}
  \boldsymbol{D} & = \begin{pmatrix}
  	     \tilde{\boldsymbol{C}} &  &  &  \\
  	   & \tilde{\boldsymbol{C}} &  &  \\
  	   &  & \ddots &   \\
  	   &  &  & \tilde{\boldsymbol{C}}
  \end{pmatrix}, \hspace{3mm}
  \boldsymbol{K} = \begin{pmatrix}
                \boldsymbol{K}_{11} & \boldsymbol{K}_{12} & \cdots & \boldsymbol{K}_{1p} \\
  	      \boldsymbol{K}_{21} & \boldsymbol{K}_{22} & \cdots & \boldsymbol{K}_{2p} \\
  	      \vdots & \vdots & \ddots &  \vdots \\  
  	      \boldsymbol{K}_{p1} & \boldsymbol{K}_{p2} & \cdots & \boldsymbol{K}_{pp}
               \end{pmatrix}, \hspace{3mm}  \\
  \boldsymbol{Q}_{\boldsymbol{f}} & = \begin{pmatrix}
  	     \boldsymbol{Q}_{f_1} &  &  &  \\
  	   & \boldsymbol{Q}_{f_2} &  &  \\
  	   &  & \ddots &   \\
  	   &  &  & \boldsymbol{Q}_{f_p}
  \end{pmatrix},
  \end{split}
\end{equation}
where $\boldsymbol{D}$ and $\boldsymbol{Q}_{\boldsymbol{f}}$ are block diagonal matrices with $p$ blocks on the diagonal.
$ \{ \boldsymbol{K}_{ij} = b_{ij}(\kappa_{ij}^2\boldsymbol{C} + \boldsymbol{G}); i, j = 1,2, \dots, p \}$ for $\alpha_{ij} = 2$ and 
$\{ \boldsymbol{K}_{ij} = b_{ij} \boldsymbol{I}_{M \times M}; i, j = 1,2, \dots, p \}$ for $\alpha_{ij} = 0$. 

We summarize the results for multivariate GRFs with our systems of SPDEs approach in Main result \textbf{\ref{th: reslt1}}. 

\begin{main} \label{th: reslt1}
 Let $\boldsymbol{Q}$ be the precision matrix for the Gaussian weights $\omega_i$ in the system of SPDEs \eqref{eq: SPDEs_system}, then the precision matrix of the multivariate GMRF is
\begin{equation}
 \boldsymbol{Q} = \boldsymbol{K} \boldsymbol{D}^{-1} \boldsymbol{Q}_{\boldsymbol{f}} \boldsymbol{D}^{-1} \boldsymbol{K}
\end{equation}
with $\boldsymbol{D}$ and $\boldsymbol{K}$ defined in \eqref{eq: SPDE_multiCKQ}.
\end{main}

The form of the Main result \textbf{\ref{th: reslt1}} is similar to the discussion in \citet[Appendix C.$4$]{lindgren2011explicit}. 
With our approach the precision matrices for the multivariate GRFs are sparse and the smoothness of the fields  is mainly controlled by the noise processes. 
This sparse precision matrix is used for sampling the multivariate GRFs in Section \ref{sec: Sampling} and statistical inference 
with simulated data and real data in Section \ref{sec: applications}.

\section{Model comparison and sampling the GRFs} \label{sec: Model_Comparisions}
 In this section we focus on the construction of the bivariate GRF, i.e., $p = 2$ in $\mathbb{R}^d$
using the SPDEs approach and then compare with the covariance-based approach presented by \citet{gneitingmatern}. 
As discussed in Section \ref{sec: covarmodel}, the SPDEs approach can construct the same multivariate GRFs as the covariance-based models if
$S_{x_{ij}}(\boldsymbol{k}) = R_{x_{ij}}(\boldsymbol{k})$ for all $1 \leq i, j \leq q$.
The comparison for the univariate GRF is trivial and the multivariate GRF when $p > 2$ can be done
in a similar way.

\subsection{Bivariate GRF with SPDEs} \label{sec: bivariate_GRF}
 When $p = 2$ in the system of SPDEs \eqref{eq: SPDEs_system}, we can construct bivariate GRFs when $b_{12} \neq 0 \text{ or } b_{21} \neq 0$. In this case, the system of SPDEs has the following form
\begin{equation}  \label{eq: SPDEs}
  \begin{split}
   b_{11}(\kappa_{11}^2 - \Delta)^{\alpha_{11}/2} x_1(\boldsymbol{s}) +  b_{12}(\kappa_{12}^2 - \Delta)^{\alpha_{12}/2} x_2(\boldsymbol{s}) & = f_1(\boldsymbol{s}),   \\
   b_{22}(\kappa_{22}^2 - \Delta)^{\alpha_{22}/2} x_2(\boldsymbol{s}) +  b_{21}(\kappa_{21}^2 - \Delta)^{\alpha_{21}/2} x_1(\boldsymbol{s}) & = f_2(\boldsymbol{s}),   
  \end{split}
\end{equation}
and the solution $\boldsymbol{x}(\boldsymbol{s}) = (x_1(\boldsymbol{s}), x_2(\boldsymbol{s}))^\text{T}$ to the system of equations \eqref{eq: SPDEs} is a bivariate random field.
Since it is convenient to study the properties of the bivariate GRFs in the spectral domain, the Fourier transform is applied,
\begin{equation} \label{eq: SPDEs_Fourier}
   \begin{aligned}
    b_{11}(\kappa_{11}^2 + \| \boldsymbol{k} \|^2)^{\alpha_{11}/2} \hat{x}_1(\boldsymbol{k}) +  b_{12}(\kappa_{12}^2 + \| \boldsymbol{k} \|^2)^{\alpha_{12}/2} \hat{x}_2(\boldsymbol{k}) & = \hat{f}_1(\boldsymbol{k}),   \\
    b_{22}(\kappa_{22}^2 + \| \boldsymbol{k} \|^2)^{\alpha_{22}/2} \hat{x}_2(\boldsymbol{k}) +  b_{21}(\kappa_{21}^2 + \| \boldsymbol{k} \|^2)^{\alpha_{21}/2} \hat{x}_1(\boldsymbol{k}) & = \hat{f}_2(\boldsymbol{k}).
  \end{aligned}
\end{equation}
The matrix form of the differential operator in the spectral domain can be written as
\begin{align} \label{eq: Hmatrix_2}
\mathscr{H}(\boldsymbol{\theta}) = \begin{pmatrix}
 \mathcal{H}_{11}(\theta_{11}) & \mathcal{H}_{12}(\theta_{12 })\\
 \mathcal{H}_{21}(\theta_{21}) & \mathcal{H}_{22}(\theta_{22})
\end{pmatrix}.
\end{align}
If the noise processes are mutually independent (but not necessarily identically distributed), the power spectrum matrix of the noise processes is a block diagonal matrix with the form
\begin{align} \label{eq: noisespectrum_2}
{S}_{\boldsymbol{f}}(\boldsymbol{k})  = \begin{pmatrix}
 S_{f_1}(\boldsymbol{k}) & 0 \\
 0 & S_{f_2}(\boldsymbol{k}) 
\end{pmatrix},
\end{align}
where $S_{f_1}(\boldsymbol{k})$ and $S_{f_2}(\boldsymbol{k})$ are the  power spectra in $\mathbb{R}^d$ for the noise processes
$f_1(\boldsymbol{s})$ and $f_2(\boldsymbol{s})$, respectively.
If the noise processes are white, then the problems is simplified and the corresponding power spectra have the forms
$S_{\mathcal{W}_1}(\boldsymbol{k}) = (2\pi)^{-d} \sigma_{n_1}^2$ and $ S_{\mathcal{W}_2}(\boldsymbol{k}) = (2\pi)^{-d} \sigma_{n_2}^2$,
where $\sigma_{n_1}$ and $\sigma_{n_2}$ are the standard deviations for the white noise processes $\mathcal{W}_1$ and $\mathcal{W}_2$, respectively. 
However, these two parameters are confounded with $\{ b_{i,j}; i,j = 1,2 \}$ and we fix $\sigma_{n_1} = 1$ and $\sigma_{n_2} = 1$.
The conclusion is also valid with other types of noise processes.  
Notice that we have used the new notation $ \{S_{\mathcal{W}_i}; i = 1,2 \}$, because
it is also possible to use more interesting noise processes which are not only the simple white noise. For instance, we can use the noise processes with Mat\'{e}rn covariance functions. 
These kinds of noise processes can be generated (independently) from the SPDEs
\begin{equation} \label{eq: noises_generate}
\begin{split}
 (\kappa_{n_1}^2 - \Delta)^{\alpha_{n_1}/2} f_1 = \mathcal{W}_1, \\
 (\kappa_{n_2}^2 - \Delta)^{\alpha_{n_2}/2} f_2 = \mathcal{W}_2, 
\end{split}
\end{equation}
where $\mathcal{W}_1$ and $\mathcal{W}_2$ are standard Gaussian white noise processes and $\kappa_{n_1}$ and $\kappa_{n_2}$
are scaling parameters. $\alpha_{n_1}$ and $\alpha_{n_2}$ are related to smoothness parameters $\nu_{n_1}$ and $\nu_{n_2}$ for $f_1$ and $f_2$ and $\alpha_{n_i} = \nu_{n_i} + d/2$. 
Apply the Fourier transform to \eqref{eq: noises_generate} and use a similar procedure as defined in Equation \eqref{eq: SPDEs_system_matrix} - Equation \eqref{eq: SPDEs_Fourier_M}, 
and then the power spectra for the noise processes generated from SPDEs \eqref{eq: noises_generate} can be obtained and they have the forms
\begin{equation} \label{eq: noise_fourier_matern}
 \begin{split}
 S_{f_1}(\boldsymbol{k}) = \frac{1}{(2\pi)^d} \frac{1}{(\kappa_{n_1}^2 + \|\boldsymbol{k}\|^2)^{\alpha_{n_1}} }, \\
 S_{f_2}(\boldsymbol{k}) = \frac{1}{(2\pi)^d} \frac{1}{(\kappa_{n_2}^2 + \|\boldsymbol{k}\|^2)^{\alpha_{n_2}} }. 
\end{split}
\end{equation}
The power spectrum matrix of the GRFs presented in \eqref{eq: SPDEs_Fourier_M} can also be simplified further and has the form
\begin{align} \label{eq: SPDEs_Fourier_2}
 \ \boldsymbol{S}_{\boldsymbol{x}}(\boldsymbol{k})  = \left( \begin{array}{cc}
  S_{x_{11}}(\boldsymbol{k}) & S_{x_{12}}(\boldsymbol{k}) \\
  S_{x_{21}}(\boldsymbol{k}) & S_{x_{22}}(\boldsymbol{k})
  \end{array} \right).
\end{align}
By using \eqref{eq: PowerSpectrum_x} together with the exact formula for the differential operators \eqref{eq: Hmatrix_2} and the spectra for the noise processes as defined in \eqref{eq: noise_fourier_matern}, 
we can get an exact symbolic expressions for all the elements in the power spectrum matrix \eqref{eq: SPDEs_Fourier_2}
\begin{equation} \label{eq: S_x_matrix}
 \begin{split}
 S_{x_{11}} & = \frac{S_{{f}_1} |\mathcal{H}_{22}^2| + S_{{f}_2} |\mathcal{H}_{12}^2|}{|(\mathcal{H}_{11}\mathcal{H}_{22}-\mathcal{H}_{12}\mathcal{H}_{21})^2|}, \\
 S_{x_{12}} & =-\frac{\mathcal{H}_{22} S_{{f}_1} |\mathcal{H}_{21}^2|\mathcal{H}_{11} + \mathcal{H}_{12} S_{{f}_2} |\mathcal{H}_{11}^2|\mathcal{H}_{21}}{|(\mathcal{H}_{11}\mathcal{H}_{22}-\mathcal{H}_{12}\mathcal{H}_{21})^2|\mathcal{H}_{21}\mathcal{H}_{11}}, \\
 S_{x_{21}} & =-\frac{\mathcal{H}_{21} S_{{f}_1} |\mathcal{H}_{22}^2|\mathcal{H}_{12} + \mathcal{H}_{11} S_{{f}_2} |\mathcal{H}_{12}^2|\mathcal{H}_{22}}{|(\mathcal{H}_{11}\mathcal{H}_{22}-\mathcal{H}_{12}\mathcal{H}_{21})^2|\mathcal{H}_{22}H_{12}}, \\
 S_{x_{22}} & = \frac{S_{{f}_1} |\mathcal{H}_{21}^2| + S_{{f}_2} |\mathcal{H}_{11}^2|}{|(\mathcal{H}_{11}\mathcal{H}_{22}-\mathcal{H}_{12}\mathcal{H}_{21})^2|}.
\end{split}
\end{equation}

In order to simply the problem, we make the operator matrix in the spectral domain a lower triangular matrix by setting $H_{12}(\theta_{12}, \boldsymbol{k}) = 0$, or equivalently, by setting $b_{12} = 0$.
The system of SPDEs in the spectral domain becomes
\begin{align}
 \label{eq: SPDE1_Fouirer_tri}
 & b_{11}(\kappa_{11}^2 + \| \boldsymbol{k} \|^2)^{\alpha_{11}/2} \hat{x}_1(\boldsymbol{k}) = \hat{f}_1(\boldsymbol{k}),   \\
 \label{eq: SPDE2_Fourier_tri}
 & b_{22}(\kappa_{22}^2 + \| \boldsymbol{k} \|^2)^{\alpha_{22}/2} \hat{x}_2(\boldsymbol{k}) +  b_{21}(\kappa_{21}^2 + \| \boldsymbol{k} \|^2)^{\alpha_{21}/2} \hat{x}_1(\boldsymbol{k}) = \hat{f}_2(\boldsymbol{k}).
\end{align}

\noindent This means expression \eqref{eq: S_x_matrix} becomes
\begin{equation} \label{eq: S_x_tri}
\begin{aligned}
 S_{x_{11}} & = \frac{S_{{f}_1}}{|\mathcal{H}_{11}^2|},  & S_{x_{21}}  & = - \frac{\bar{\mathcal{H}}_{21} S_{{f}_1} }{\bar{\mathcal{H}}_{22} |\mathcal{H}_{11}|^2}, \\
 S_{x_{12}} & = -\frac{S_{{f}_1} \mathcal{H}_{21}}{\mathcal{H}_{22}|\mathcal{H}_{11}|^2},  & S_{x_{22}} & = \frac{|\mathcal{H}_{21}|^2 S_{{f}_1} + |\mathcal{H}_{11}|^2 S_{{f}_2} }{|\mathcal{H}_{11}|^2|\mathcal{H}_{22}|^2},
\end{aligned}
\end{equation}
where $\bar{\mathcal{H}}_{ij}$ denotes the conjugate of $\mathcal{H}_{ij}$.
This is called the \emph{triangular} version of SPDEs in this paper.
If the operators $\mathcal{H}_{12}$ and $\mathcal{H}_{21}$ are real,  we obtain $S_{x_{12}}(\boldsymbol{k}) = S_{x_{21}}(\boldsymbol{k})$.
In other words, we have an imposed symmetry property on the cross-covariance in this case.
With the triangular version of the SPDEs, under some extra conditions, the properties of the multivariate GRFs is easy to interpret. We will see more about this in Section \ref{sec: applications}.
In this paper we focus on the triangular version of the SPDEs. However, the full version of the SPDEs could be handled analogously to the triangular version of the SPDEs. 
If there is no constraint on the operator matrix, i.e., $b_{ij} \neq 0 \text{ for all } i, j = 1,2$, it is called the \emph{full} version of the SPDEs. 
In general, $S_{x_{12}}(\boldsymbol{k})$ and $ S_{x_{21}}(\boldsymbol{k})$ are not necessarily equal. This can release the constraint shared by the Mat\'ern model proposed by \citet{gneitingmatern} and the Linear 
model of coregionalization (LMC), namely that the cross-covariance has an imposed symmetry property, i.e., $C_{ij} (\boldsymbol{h}) = C_{ji}(\boldsymbol{h})$.
This does in general not hold as discussion by \citet[Chapter $20$]{wackernagel2003multivariate}. We refer to \citet{goulard1992linear}, \citet[Chapter $14$]{wackernagel2003multivariate} and \citet{gelfand2004nonstationary} for more information about LMC.
In this paper we assume that all the operators are real which simplifies calculations and discussion. We reorganize \eqref{eq: S_x_tri} and find
\begin{equation} \label{eq: powerspectrum_ratio}
\begin{aligned}
\frac{S_{x_{12}}}{S_{x_{11}}} & = -\frac{\mathcal{H}_{21}}{\mathcal{H}_{22}}, \\
\frac{S_{x_{22}}}{S_{x_{11}}} & = \frac{|\mathcal{H}_{21}^2|}{|\mathcal{H}_{22}^2|} + \frac{|\mathcal{H}_{11}^2|}{\mathcal{H}_{22}^2}\frac{S_{{f}_2}}{S_{{f}_1}}, \\
\frac{S_{x_{22}}}{S_{x_{12}}} & = -\left(\frac{|\mathcal{H}_{11}^2|}{\mathcal{H}_{21}\mathcal{H}_{22}}\right)\frac{S_{{f}_2}}{S_{{f}_1}} - \frac{\mathcal{H}_{21}}{\mathcal{H}_{22}}.
\end{aligned}
\end{equation}
From the results given in \eqref{eq: S_x_matrix} and \eqref{eq: powerspectrum_ratio}, we can notice that $S_{x_{12}}$ and $S_{x_{11}}$ only depend on the power spectrum of the noise process $f_1$ and
$S_{x_{22}}(\boldsymbol{k})$ depends on the noise process power spectra $S_{f_1}(\boldsymbol{k})$ and $S_{f_2}(\boldsymbol{k})$. The ratio between the power 
spectra $S_{x_{12}}$ and $S_{x_{11}}$ is independent of both the noise processes.

When $\{ f_i(\boldsymbol{s}); i = 1,2 \}$  are mutually independent and generated from \eqref{eq: noises_generate}, the elements in the power spectrum matrix \eqref{eq: SPDEs_Fourier_2} can be written down
explicitly by using \eqref{eq: noise_fourier_matern} and \eqref{eq: S_x_tri},

\begin{equation} \label{eq: S_x_full_result}
\begin{split}
 S_{x_{11}}(\boldsymbol{k}) & = \frac{1}{(2\pi)^d (\kappa_{n_1}^2 + \|\boldsymbol{k}\|^2)^{\alpha_{n_1}}{b_{11}^2\left(\kappa_{11}^2 + \| \boldsymbol{k} \|^2 \right)^{\alpha_{11}}}}, \\
 S_{x_{12}}(\boldsymbol{k}) &  = -\frac{b_{21} (\kappa_{n_1}^2 + \|\boldsymbol{k}\|^2)^{-\alpha_{n_1}} (\kappa_{21}^2 + \| \boldsymbol{k} \|^2)^{\alpha_{21} /2}}{(2\pi)^d 
                                   \left(b_{22}(\kappa_{22}^2 + \| \boldsymbol{k} \|^2)^{\alpha_{22}/2} b_{11}^2(\kappa_{11}^2 + \|\boldsymbol{k}\|^2)^{\alpha_{11}}\right)}, \\
S_{x_{22}}(\boldsymbol{k}) & = \frac{ \frac{{b_{21}^2 (\kappa_{21}^2 + \| \boldsymbol{k} \|^2)^{\alpha_{21}}}}{(2\pi)^d  (\kappa_{n_1}^2+\|\boldsymbol{k}\|^2)^{\alpha_{n_1}}}
                          +\frac{b_{11}^2 (\kappa_{11}^2 + \| \boldsymbol{k} \|^2)^{\alpha_{11}}}{(2\pi)^d  (\kappa_{n_2}^2+\|\boldsymbol{k}\|^2)^{\alpha_{n_2}}} }
                            {\left( b_{11}^2 (\kappa_{11}^2 + \| \boldsymbol{k} \|^2)^{\alpha_{11}} \right) \left( b_{22}^2 (\kappa_{22}^2 + \| \boldsymbol{k} \|^2)^{\alpha_{22}} \right)}. \\
\end{split}
\end{equation}

\begin{figure}[hbtp]
    \centering
    \subfigure[]{\includegraphics[width=0.45\textwidth,height=0.45\textwidth]{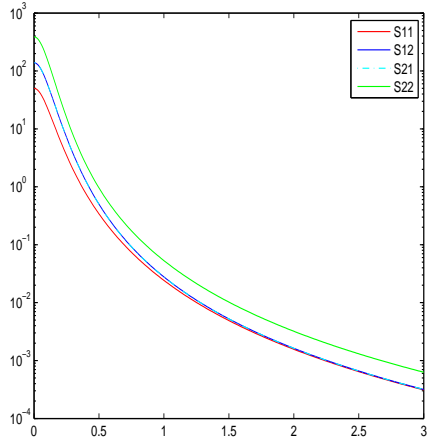}}
    \subfigure[]{\includegraphics[width=0.45\textwidth,height=0.45\textwidth]{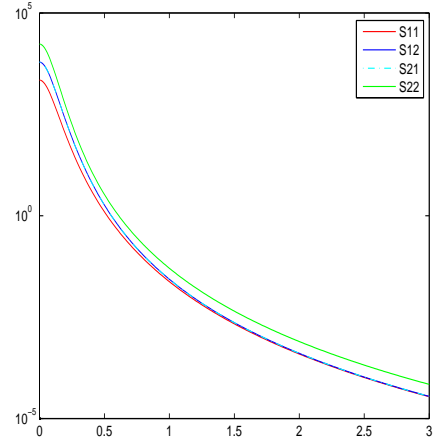}} \\
    \subfigure[]{\includegraphics[width=0.45\textwidth,height=0.45\textwidth]{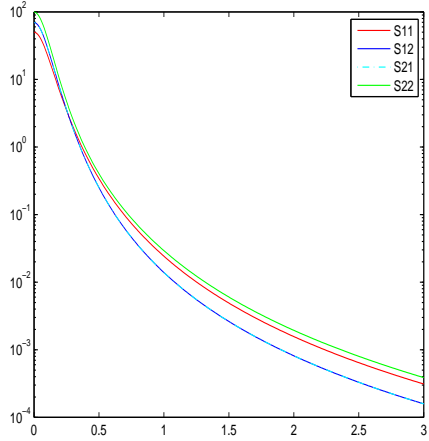}}
    \subfigure[]{\includegraphics[width=0.45\textwidth,height=0.45\textwidth]{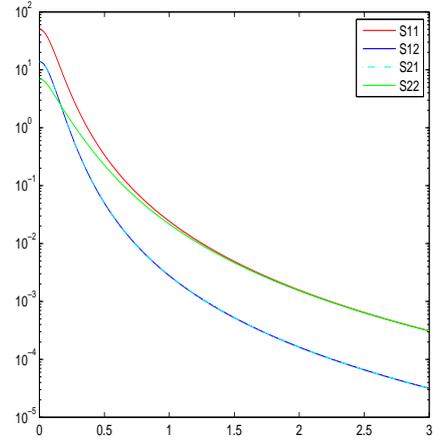}}
    \caption{Asymptotic behaviors of the power spectra corresponding to case $1$ (a), case $2$ (b), case $3$ (c) and case $4$ (d) with parameter values given in Table \ref{tab: para_asym}. }
    \label{fig: Asym1}
 \end{figure}

The asymptotic behavior of the power spectra for the bivariate GRF can be obtained from \eqref{eq: S_x_full_result}. 
For some selected parameter values defined in Table \ref{tab: para_asym}, the power spectra are
shown in Figure \ref{fig: Asym1}. From these figure, we can notice that the parameters $\{ b_{ij}; i, j = 1,2, i >j \}$ controls the correlation between the two fields.
When a smaller absolute value of $b_{21}$ is chosen, the correlation between these two fields decreases rapidly. We can also show that the sign of $b_{21} \cdot b_{22}$
is related the sign of the correlation between the two GRFs. Obviously, $b_{ij}$ are also related to the variance of the GRFs. The parameters 
$\kappa_{ij}$ are related to the range of the two fields. $\alpha_{ij}$ and $\alpha_{n_i}$ are related to the smoothness of the two GRFs.
The asymptotic behavior is curial when dealing with the real-world applications.

\begin{table}
 \centering
\caption{Parameter values for asymptotic behaviors of the power spectra for bivariate GRFs}
\begin{tabular}{l|l|l||l|l|l}
\hline
\hline
\multicolumn{3}{c||}{ case $1$}                             &    \multicolumn{3}{c}{ case $2$}       \\
 \hline
$\boldsymbol{\alpha}$     &  $\boldsymbol{\kappa}$     &   $\boldsymbol{b}$ &  $\boldsymbol{\alpha}$      &  $\boldsymbol{\kappa}$      &   $\boldsymbol{b}$\\
\hline
$\alpha_{11} = 2$     &  $\kappa_{11} = 0.15$  & $b_{11} = 1$   &   $\alpha_{11} = 2$     &  $\kappa_{11} = 0.15$   & $b_{11} = 1$ \\
$\alpha_{12} = 0$     &  $\kappa_{12} = 0$     & $b_{12} = 0$   &   $\alpha_{12} = 0$     &  $\kappa_{12} = 0$      & $b_{12} = 0$ \\
$\alpha_{21} = 2$     &  $\kappa_{21} = 0.5$   & $b_{21} = -1$  &   $\alpha_{21} = 2$     &  $\kappa_{21} = 0.5$    & $b_{21} = -1$ \\
$\alpha_{22} = 2$     &  $\kappa_{22} = 0.3$   & $b_{22} = 1$   &   $\alpha_{22} = 2$     &  $\kappa_{22} = 0.3$    & $b_{22} = 1$  \\
$\alpha_{n_1} = 0$    &  $\kappa_{n_1} = 0.15$ &                &   $\alpha_{n_1} = 1$    &  $\kappa_{n_1} = 0.15$   &               \\
$\alpha_{n_1} = 0$    &  $\kappa_{n_2} = 0.3$  &                &   $\alpha_{n_2} = 1$    &  $\kappa_{n_2} = 0.3$    &               \\
\hline
\multicolumn{3}{c||}{ case $3$}                             &    \multicolumn{3}{c}{ case $4$}       \\
 \hline
$\boldsymbol{\alpha}$     &  $\boldsymbol{\kappa}$     &   $\boldsymbol{b}$ &  $\boldsymbol{\alpha}$      &  $\boldsymbol{\kappa}$      &   $\boldsymbol{b}$\\
\hline
$\alpha_{11} = 2$     &  $\kappa_{11} = 0.15$  & $b_{11} = 1$   &   $\alpha_{11} = 2$     &  $\kappa_{11} = 0.15$   & $b_{11} = 1$    \\
$\alpha_{12} = 0$     &  $\kappa_{12} = 0$     & $b_{12} = 0$   &   $\alpha_{12} = 0$     &  $\kappa_{12} = 0$      & $b_{12} = 0$    \\
$\alpha_{21} = 2$     &  $\kappa_{21} = 0.15$  & $b_{21} = -0.5$&   $\alpha_{21} = 2$     &  $\kappa_{21} = 0.5$    & $b_{21} = -1$ \\
$\alpha_{22} = 2$     &  $\kappa_{22} = 0.3$   & $b_{22} = 1$   &   $\alpha_{22} = 2$     &  $\kappa_{22} = 0.3$    & $b_{22} = 1$     \\
$\alpha_{n_1} = 0$    &  $\kappa_{n_1} = 0.15$ &                &   $\alpha_{n_1} = 0$    &  $\kappa_{n_1} = 0.15$  &                    \\
$\alpha_{n_1} = 0$    &  $\kappa_{n_2} = 0.3$  &                &   $\alpha_{n_2} = 0$    &  $\kappa_{n_2} = 0.3$   &                    \\
\hline
\end{tabular}
  \label{tab: para_asym}
\end{table}

\subsection{Bivariate GRF with the covariance-based model}
As presented by \citet{gneitingmatern}, a multivariate GRFs can be constructed using a covariance-based model. In the bivariate setting, 
the matrix-valued covariance function for the bivariate GRF contains $4$ elements and the corresponding power spectrum matrix becomes
\begin{align} \label{eq: CovarianceMatrix_Fourier_2}
 \ \boldsymbol{R}(\boldsymbol{k})  = \left( \begin{array}{cc}
  R_{x_{11}}(\boldsymbol{k}) & R_{x_{12}}(\boldsymbol{k}) \\
  R_{x_{21}}(\boldsymbol{k}) & R_{x_{22}}(\boldsymbol{k})
 \end{array} \right).
\end{align}

Applying integral given in Equation \eqref{eq: Rij_iit}, together with the expression for the marginal variances in Equation \eqref{eq: Cii} and cross-covariance in Equation \eqref{eq: Cij},
we can find closed form for the elements in the power spectrum matrix,
\begin{equation} \label{eq: R_ii}
  \begin{split}
 R_{x_{ii}}(\boldsymbol{k}) & = \frac{1}{(2\pi)^d} \left[ a_{ii}^{2\nu_{ii}} (\sqrt{4\pi})^d \frac{\sigma_{ii} \Gamma(\nu_{ii} + d/2)}{(a_{ii}^2 + \| \boldsymbol{k} \|^2)^{\nu_{ii} + d/2} \Gamma(\nu_{ii})} \right], \\
 R_{x_{ij}}(\boldsymbol{k}) & = \frac{1}{(2\pi)^d} \left[ a_{ij}^{2\nu_{ij}} (\sqrt{4\pi})^d \frac{\rho_{ij} \sigma_{i}\sigma_{j} \Gamma(\nu_{ij} + d/2)}{(a_{ij}^2 + \| \boldsymbol{k} \|^2)^{\nu_{ij}+d/2} \Gamma(\nu_{ij})} \right].
 \end{split}
\end{equation}
Notice that if $d= \{2, 4, ...\}$, then the expressions in \eqref{eq: R_ii} can be simplified even further by (recursively) using the well-known formula 
\begin{displaymath}
 \Gamma(\nu +1) = \nu \Gamma(\nu).
\end{displaymath}
Since all the components $\{ C_{ij}; i,j = 1,2\}$ in the matrix-valued covariance matrix $\boldsymbol{C}$ are from Mat{\'e}rn family, as defined in Equation \eqref{eq: Cii} and Equation \eqref{eq: Cij} \citep{gneitingmatern}, the 
power spectra of the marginal covariance functions $\{C_{ii}; i = 1,2 \}$ and the cross covariance functions $\{ C_{ij}; i, j = 1,2, i \neq j\}$ have similar forms as indicated in \eqref{eq: R_ii}.

\subsection{Parameter matching} \label{sec: parameter_matching}
As mentioned in Section \ref{sec: covarmodel}, the model comparison in this paper is based on the fact that if $S_{ij}(\boldsymbol{k}) = R_{ij}(\boldsymbol{k})$, for each  $1 \leq i, j \leq p$, and
then the multivariate GRFs constructed from the SPDEs approach \eqref{eq: SPDEs_system} and from the covariance-based model \eqref{eq: CovarianceMatrix_M} will be equivalent.

By comparing each element in \eqref{eq: S_x_full_result} and \eqref{eq: R_ii}, we can get the results given as follows,
\begin{equation}
\label{eq: comparsion_eq_1_1}
 \begin{split}
 0 & = \rho_{12}  - \rho_{21},  \\
 \kappa_{11}  = \kappa_{21} = \kappa_{22} = \kappa_{n_1} = \kappa_{n_2}  & = a_{11} = a_{12} = a_{21} = a_{22}, \\
 \alpha_{11} + \alpha_{n_1} & = \nu_{11} + \frac{d}{2}, \\
 \alpha_{11} + \frac{\alpha_{22}}{2} + \alpha_{n_1} - \frac{\alpha_{21}}{2} & = \nu_{12} + \frac{d}{2}, \\
\end{split}
\end{equation}
and
\begin{equation}
\label{eq: comparsion_eq_2}
\begin{split}
& \frac{1}{ b_{11}^2}  = \frac{(4\pi)^{d/2} a_{11}^{2\nu_{n1}}\sigma_{11}\Gamma(\nu_{11}+d/2)}{\Gamma(\nu_{11})}, \\
& -\frac{b_{21}}{ b_{22} b_{11}^2}  = \frac{(4\pi)^{d/2} a_{12}^{2\nu_{12}} \Gamma(\nu_{12}+d/2) \rho_{12} \sigma_1 \sigma_2}{\Gamma(\nu_{12})},   \\
&  \frac{b_{21}^2  + b_{11}^2 }{ b_{11}^2b_{22}^2}
  = \frac{(4\pi)^{d/2} \sigma_{22} a_{22}^{2\nu_{22}}\Gamma(\nu_{22}+d/2)}{\Gamma(\nu_{22})}.
\end{split}
\end{equation}
More equalities can be obtained based on the relationship between $\alpha_{21} + \alpha_{n_2}$ and $\alpha_{11} + \alpha_{n_1}$. 
When $\alpha_{21} + \alpha_{n_2} \leq \alpha_{11} + \alpha_{n_1}$, we can get
\begin{equation}
\label{eq: comparsion_eq_1_2}
\alpha_{n_2}+\alpha_{22}  = \nu_{22} + \frac{d}{2}.
\end{equation}

\noindent When $\alpha_{21} + \alpha_{n_2} \geq \alpha_{11} + \alpha_{n_1}$, we can get 
\begin{equation}
\label{eq: comparsion_eq_1_3}
 \begin{split}
 \alpha_{11} + \alpha_{22} + \alpha_{n_1} - \alpha_{21} = \nu_{22} + \frac{d}{2}.
\end{split}
\end{equation}

\noindent The Equation \eqref{eq: comparsion_eq_1_2} and Equation \eqref{eq: comparsion_eq_1_3} are critical when discussing about the asymptotic behaviors of bivariate GRF and the corresponding power spectra. 
From \eqref{eq: comparsion_eq_1_1}, we can notice that in order to construct equivalent models to the ones as discussed by \citet{gneitingmatern}, we need to have the constraint that all
the scaling parameters should be equal in addition to some constraints for the smoothness parameters given in the third and the fourth equations in \eqref{eq: comparsion_eq_1_1} and 
one more conditional equality given in \eqref{eq: comparsion_eq_1_2} or \eqref{eq: comparsion_eq_1_3}.
From the first equation in \eqref{eq: comparsion_eq_1_1}, we notice the extra symmetry restriction discussed in \citet[Section $4$]{gneitingmatern} also must be fulfilled.  
Additionally, from the second equation in \eqref{eq: comparsion_eq_2}, we notice that there is one important relationship
\begin{equation} \label{eq: comparsion_ineq}
 \begin{split}
 \rho_{12} b_{22} b_{21} & \leq 0,
 \end{split}
\end{equation}
between the co-located correlation coefficient $\rho_{12}$ and $b_{21}$ and $b_{22}$. It shows that the correlation $\rho_{12}$ must have opposite sign to the 
product of $b_{21}$ and $b_{22}$. This is an important information not only for sampling from the bivariate GRFs but also for interpreting the results from inference.

\noindent From the results given in \eqref{eq: comparsion_eq_2}, we can obtain the following results
\begin{equation} \label{eq: comparsion_eq_3}
\begin{split}
 -\frac{b_{22}}{b_{21}} & = \frac{a_{11}^{\nu_{11}}\sigma_1\Gamma({\nu_{11}+d/2})/\Gamma({\nu_{11}})}{a_{12}^{\nu_{12}}\rho\sigma_2\Gamma({\nu_{12}+d/2})/\Gamma({\nu_{12}})}, \\
\frac{b_{22}^2}{b_{11}^2+b_{21}^2} & = \frac{a_{11}^{\nu_{11}}\sigma_{11}\Gamma({\nu_{11}+d/2})/\Gamma({\nu_{11}})}{a_{22}^{\nu_{22}}\sigma_{22}\Gamma({\nu_{22}+d/2})/\Gamma({\nu_{22}})}.
\end{split}
\end{equation}
Notice that when $a_{11} = a_{21} = a_{22}$ and $\nu_{11} = \nu_{21} = \nu_{22}$, the above results in \eqref{eq: comparsion_eq_3} can be simplified to the following form

\begin{equation} \label{eq: comparsion_eq_4}
\begin{split}
 -\frac{b_{22}}{b_{21}} & = \frac{\sigma_1}{\rho \sigma_2}, \\
\frac{b_{22}^2}{b_{11}^2 + b_{21}^2} & = \frac{\sigma_{11}}{\sigma_{22}}.
\end{split}
\end{equation}
These results show that the parameters $\{ b_{ij}; i, j = 1,2\}$ are not only connected to the correlation between these two fields, but also connected with the marginal variance of the 
GRF. From the results given in \eqref{eq: comparsion_eq_1_1} to \eqref{eq: comparsion_eq_4}, we can notice that the parameters in the system of SPDEs have similar interpretations as the 
parameters from the covariance-based model.

From these results we see that it is possible to construct \emph{multivariate} GRFs through the SPDEs approach as the covariance-based approach
proposed by \citet{gneitingmatern}. Additionally, there are three main advantages for our SPDEs approach.
The first advantage is that our new approach does not explicitly depend on the theory of positive definite matrix. We do not need to worry about the 
notorious requirement of positive definite covariance matrices. 
The second advantage is that we can remove the symmetry property which is shared by the covariance-based approach
and the LMC approach \citep{gelfand2004nonstationary, gneitingmatern}. The third advantage, which has not yet been discussed, is that we can construct multivariate GRFs on manifolds, such as the sphere $\mathbb{S}^2$,
by simply reinterpret the systems of SPDEs to be defined on the manifold. \citet[Section $3.1$]{lindgren2011explicit} discussed the theoretical background in the univariate setting, which is basically the same as 
for our multivariate setting. 

Furthermore, we can actually go even future for the multivariate GRFs, such as multivariate GRFs with oscillating covariance functions and non-stationary multivariate GRFs. 
These kinds of multivariate GRFs are under development but are beyond the scope of this paper.

\subsection{Sampling bivariate GRFs and trivariate GRFs} \label{sec: Sampling}
As presented in Section \ref{sec: gmrfs_approximation} we can construct multivariate GRFs by the SPDEs approach theoretically, but do the computations using the GMRF representation. The precision matrix $\boldsymbol{Q}$
is usually sparse and the sparseness of the precision matrices enables us to apply numerical linear algebra for sparse matrices for sampling from the GRFs and for fast statistical inference.
We now assume that the multivariate GMRF has mean vector $\boldsymbol{\mu}$ and precision matrix $\boldsymbol{Q}$, i.e., $\boldsymbol{x} \sim \mathcal{N} (\boldsymbol{\mu},\boldsymbol{Q}^{-1})$.
When sampling from the GMRF, the forward substitution and backward substitution is applied by using the Cholesky triangle $\boldsymbol{L}$,
where by definition $\boldsymbol{Q} =\boldsymbol{L} \boldsymbol{L}^\text{T} $. The commonly used steps for sampling a GRF are as follows.
\begin{enumerate}
 \item Calculate the Cholesky triangle $\boldsymbol{L}$ from the Cholesky factorization;
 \item Sample $\boldsymbol{z} \sim \mathcal{N}(\boldsymbol{0},\boldsymbol{I})$, where $\boldsymbol{I}$ is the identity matrix with the same dimensions as the precision matrix $\boldsymbol{Q}$;
 \item Solve the linear system of equations $\boldsymbol{L} \boldsymbol{v} = \boldsymbol{z}$ for $\boldsymbol{v}$. Then $\boldsymbol{v}$ has the correct precision matrix $\boldsymbol{Q}$, and
       $\boldsymbol{v} \sim \mathcal{N} (\boldsymbol{0},\boldsymbol{Q}^{-1})$;
 \item Finally correct the mean by computing  $\boldsymbol{x} = \boldsymbol{\mu} + \boldsymbol{v}$.
\end{enumerate}
Then $\boldsymbol{x}$ is a sample of the GMRF with correct mean $\boldsymbol{\mu}$ and precision matrix $\boldsymbol{Q}$. 
If $\boldsymbol{Q}$ is a band matrix, a band-Cholesky factorization can be used with the algorithm given by \citet[Algorithm $2.9$, Page $45$]{rue2005gaussian}. 
Different types of parametrization of GMRFs and their corresponding sampling procedures can be found in \citet{rue2005gaussian}.
If the mean of the field is $\boldsymbol{0}$, the Step $4$ is not needed and $\boldsymbol{v}$ is a sample from the GMRF.

\begin{figure} 
    \centering
    \subfigure[]{\includegraphics[width=0.45\textwidth,height=0.45\textwidth]{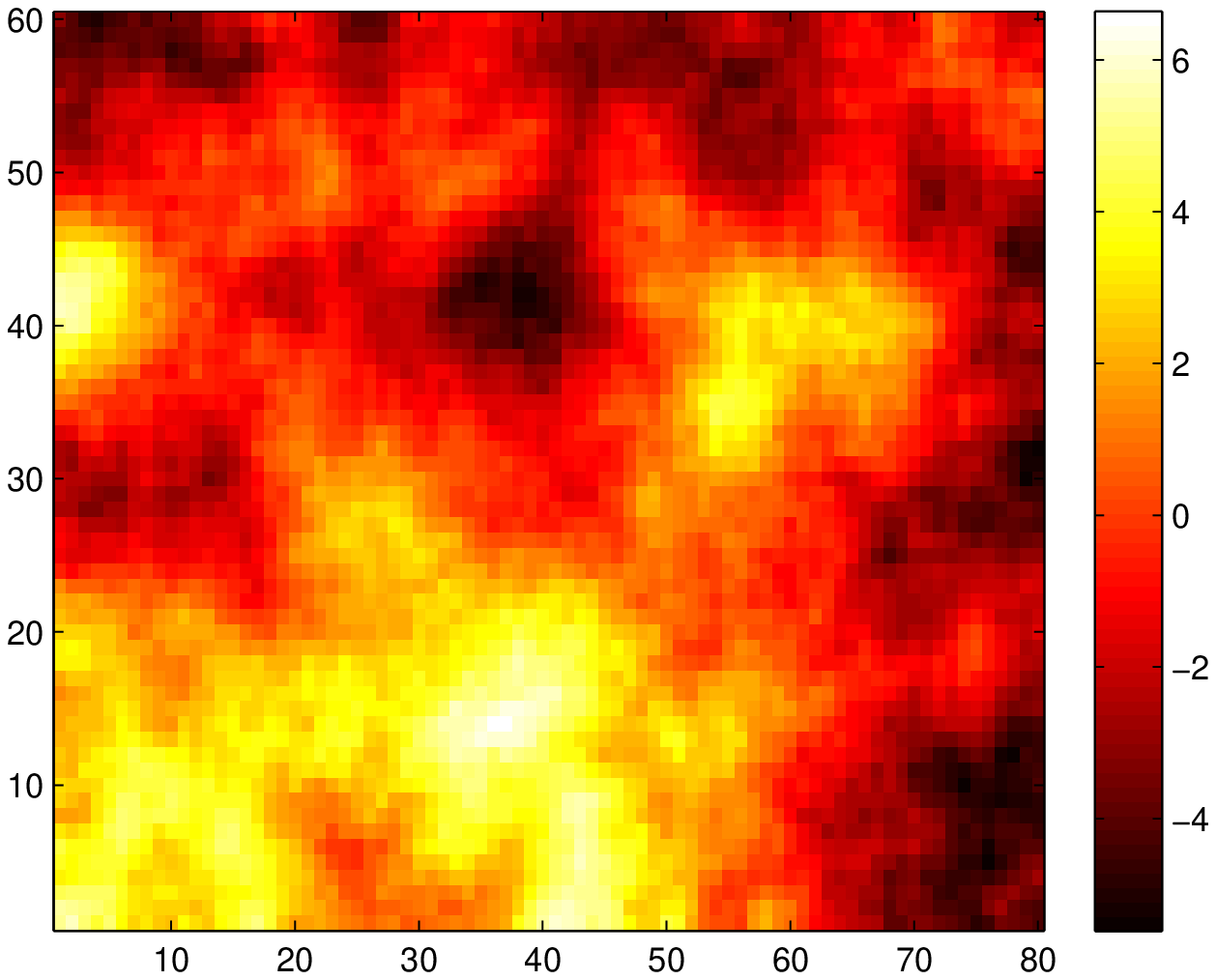}}
    \subfigure[]{\includegraphics[width=0.45\textwidth,height=0.45\textwidth]{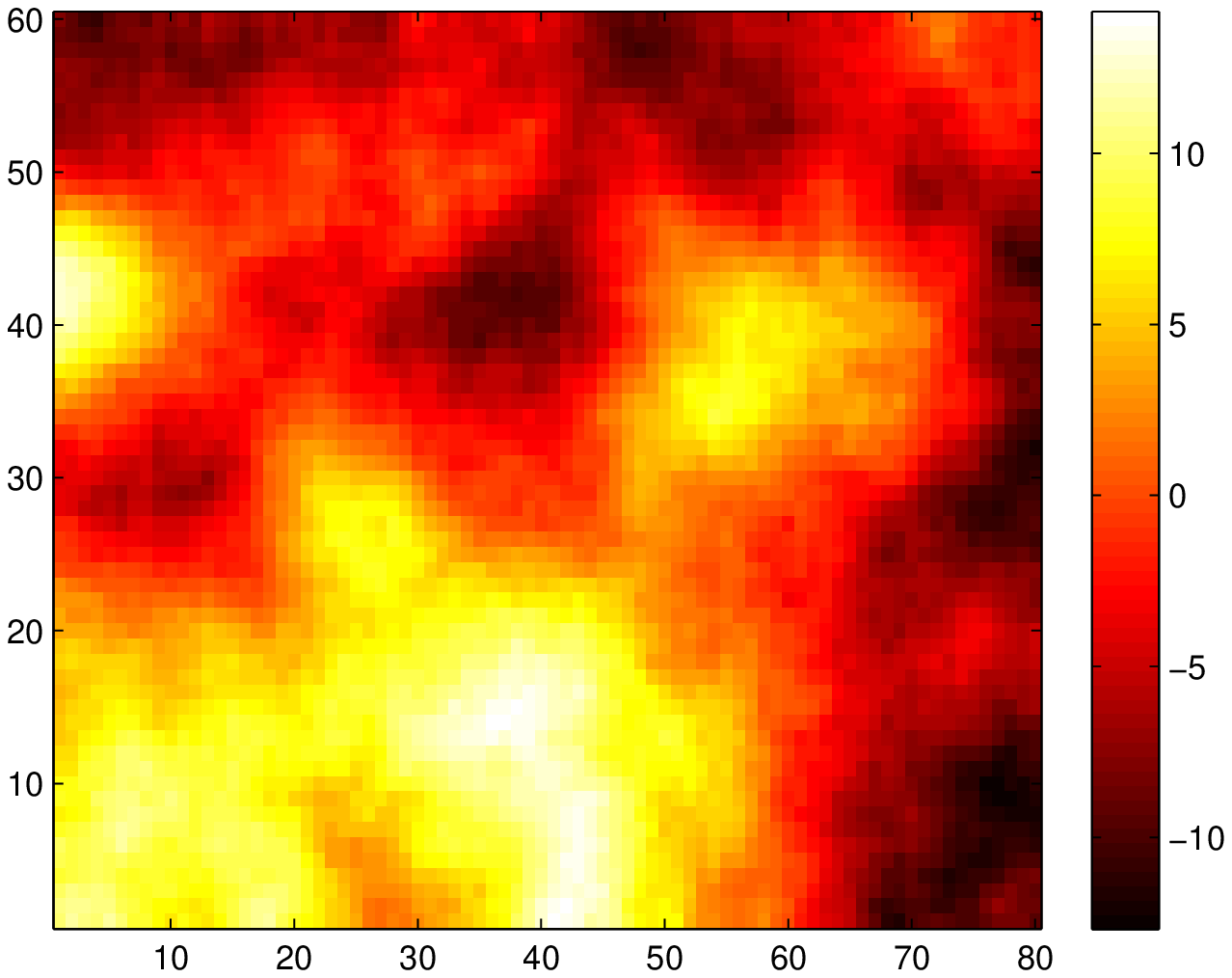}}
    \caption{One simulated realization from the bivariate Gaussian Random Field with positive correlation with parameters $\alpha_{11} = 2, \alpha_{12} = 0, \alpha_{21} = 2, \alpha_{22} = 2, \alpha_{n_1} = 0, \alpha_{n_2} = 0$,
             $\kappa_{11} = 0.15, \kappa_{22} = 0.3, \kappa_{21} = 0.5, \kappa_{n_1} = 0.15, \kappa_{n_2} = 0.3$, $b_{11} = 1, b_{12} = 0, b_{22} = 1$ and $b_{21} = -1$. } 
\label{fig: BiMatern_positive}
 \end{figure}

\begin{figure}
    \centering
    \subfigure[]{\includegraphics[width=0.45\textwidth,height=0.45\textwidth]{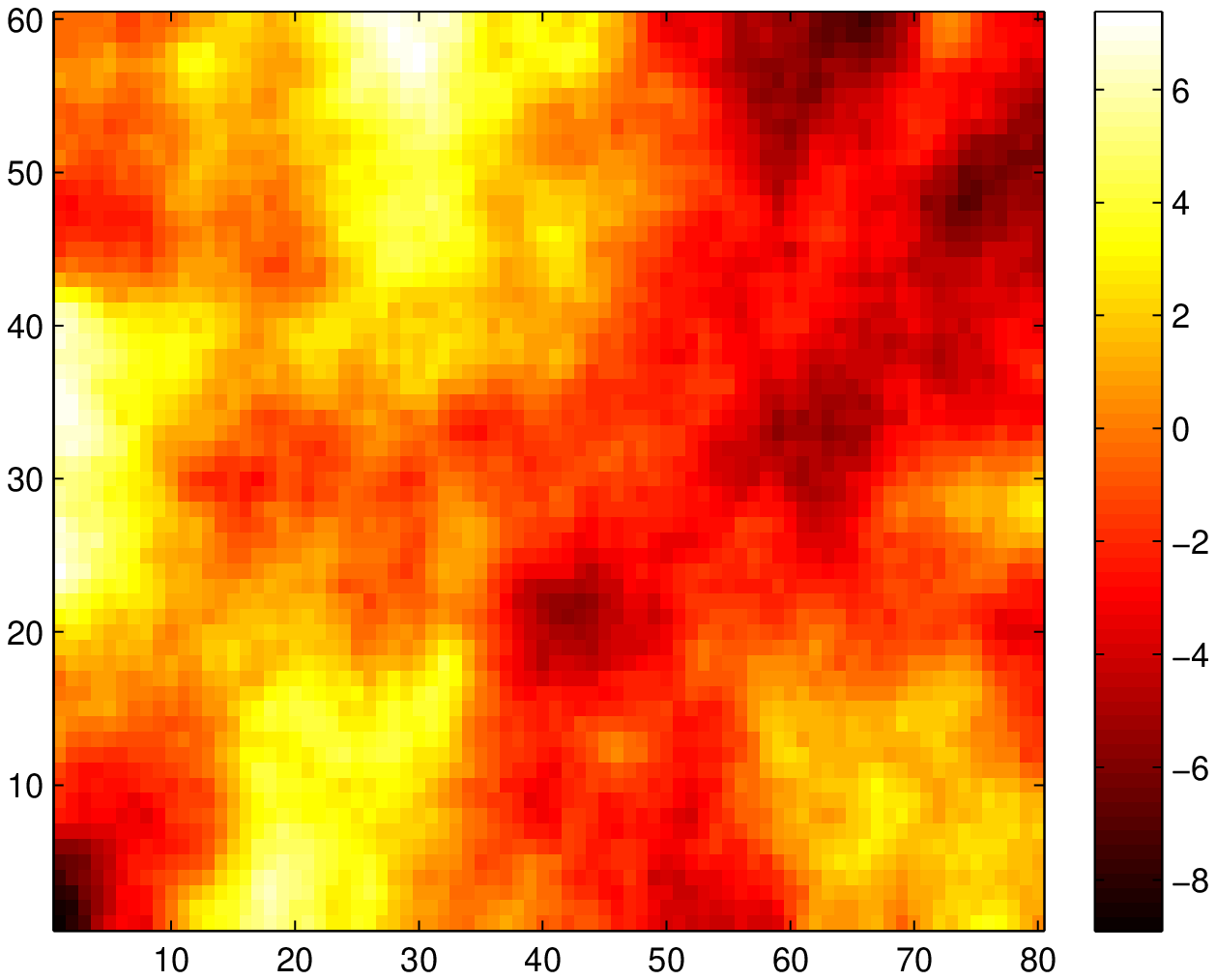}}
    \subfigure[]{\includegraphics[width=0.45\textwidth,height=0.45\textwidth]{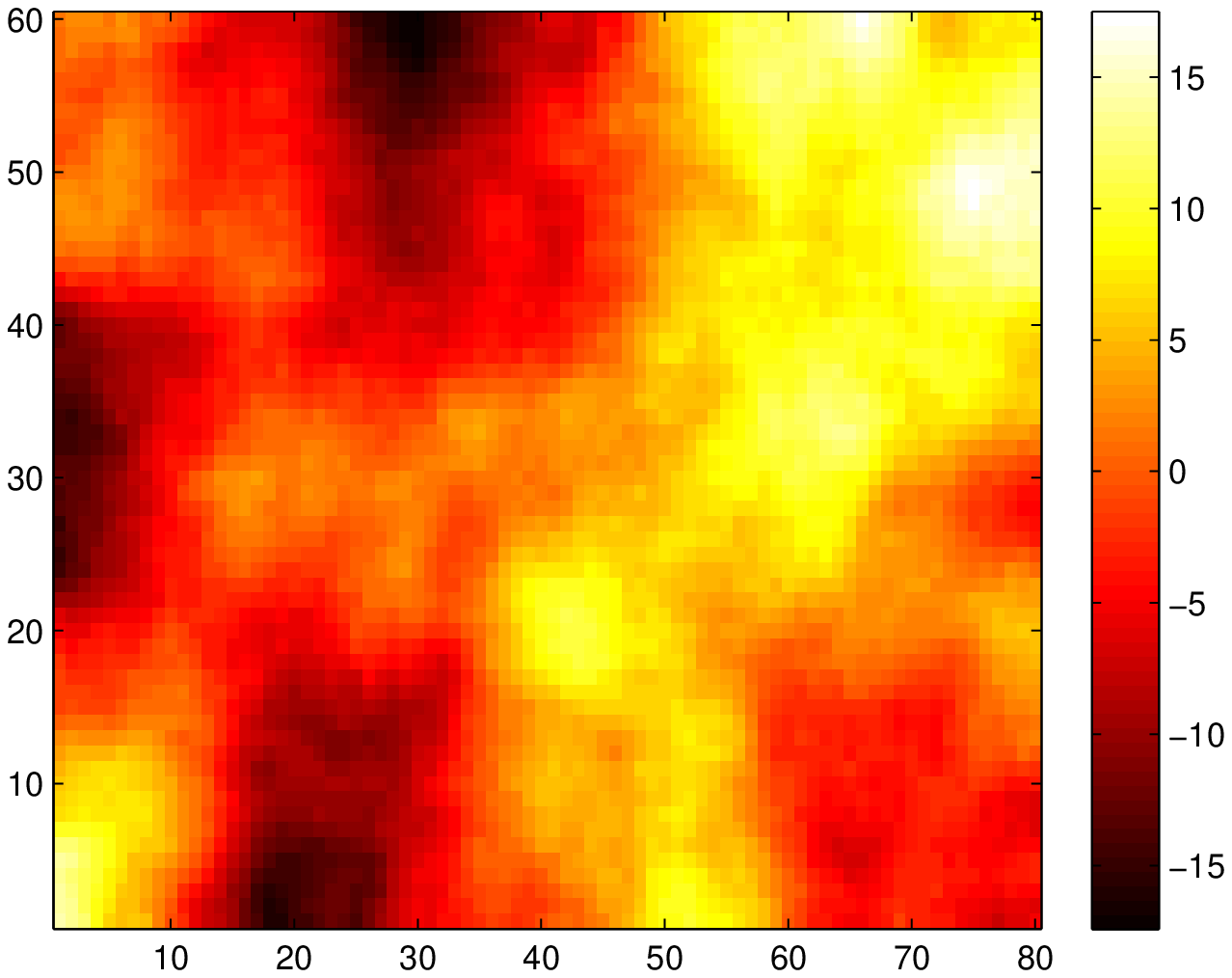}}
    \caption{One simulated realization from the bivariate Gaussian Random Field with negative correlation with parameters $\alpha_{11} = 2, \alpha_{12} = 0, \alpha_{21} = 2, \alpha_{22} = 2, \alpha_{n_1} = 0, \alpha_{n_2} = 0$,
             $\kappa_{11} = 0.15, \kappa_{22} = 0.3, \kappa_{21} = 0.5, \kappa_{n_1} = 0.15, \kappa_{n_2} = 0.3$, $b_{11} = 1, b_{12} = 0, b_{22} = 1$ and $b_{21} = 1$. }
\label{fig: BiMatern_negetive}
 \end{figure}

\begin{figure} 
    \centering
    \includegraphics[width=0.8\textwidth,height=0.6\textwidth]{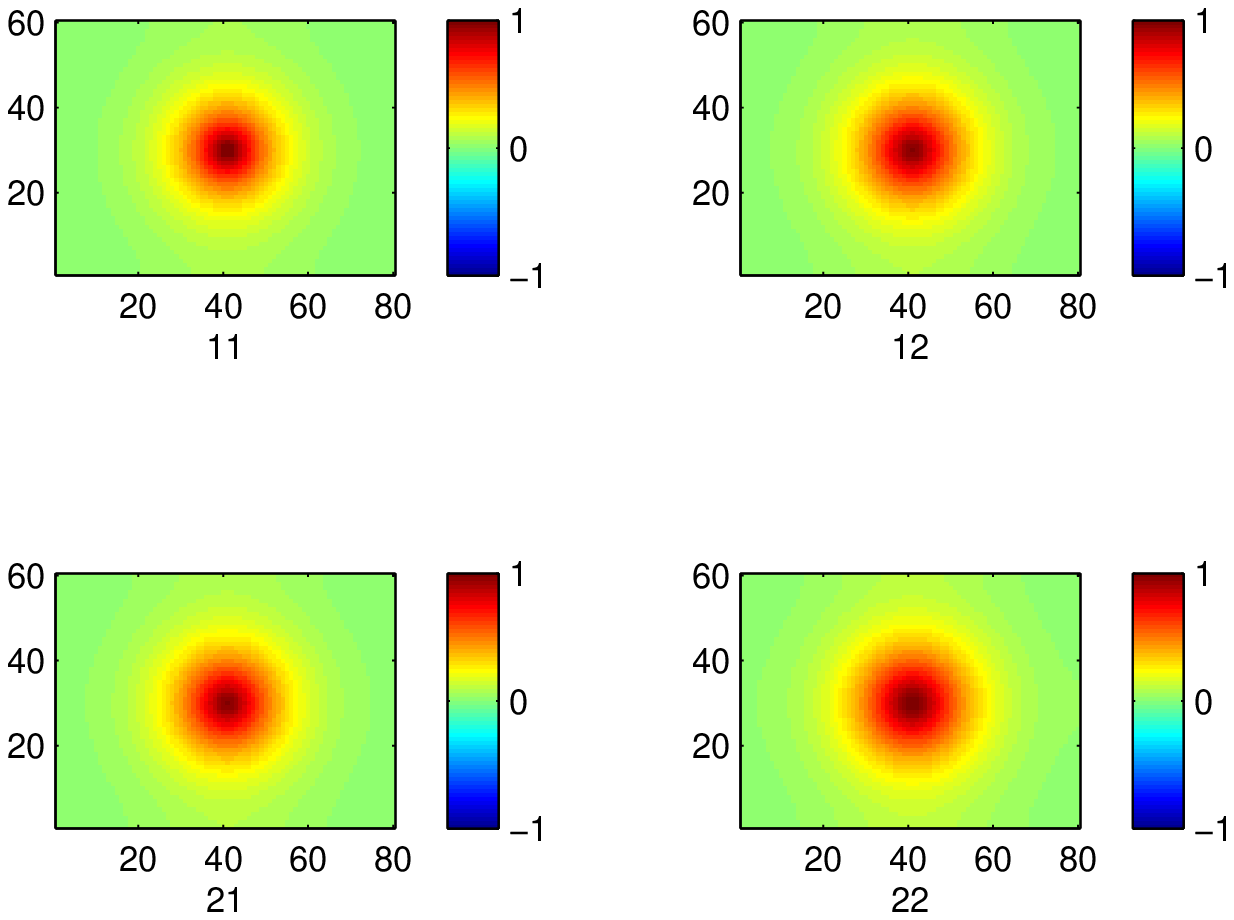}
    \caption{Marginal correlations (diagonal direction) and cross-correlations (anti-diagonal direction) with the reference points in the middle of the two GRFs 
             for the positively correlated random fields.}
   \label{fig: sampling_cov1}
\end{figure}

\begin{figure} 
    \centering
    \includegraphics[width=0.8\textwidth,height=0.6\textwidth]{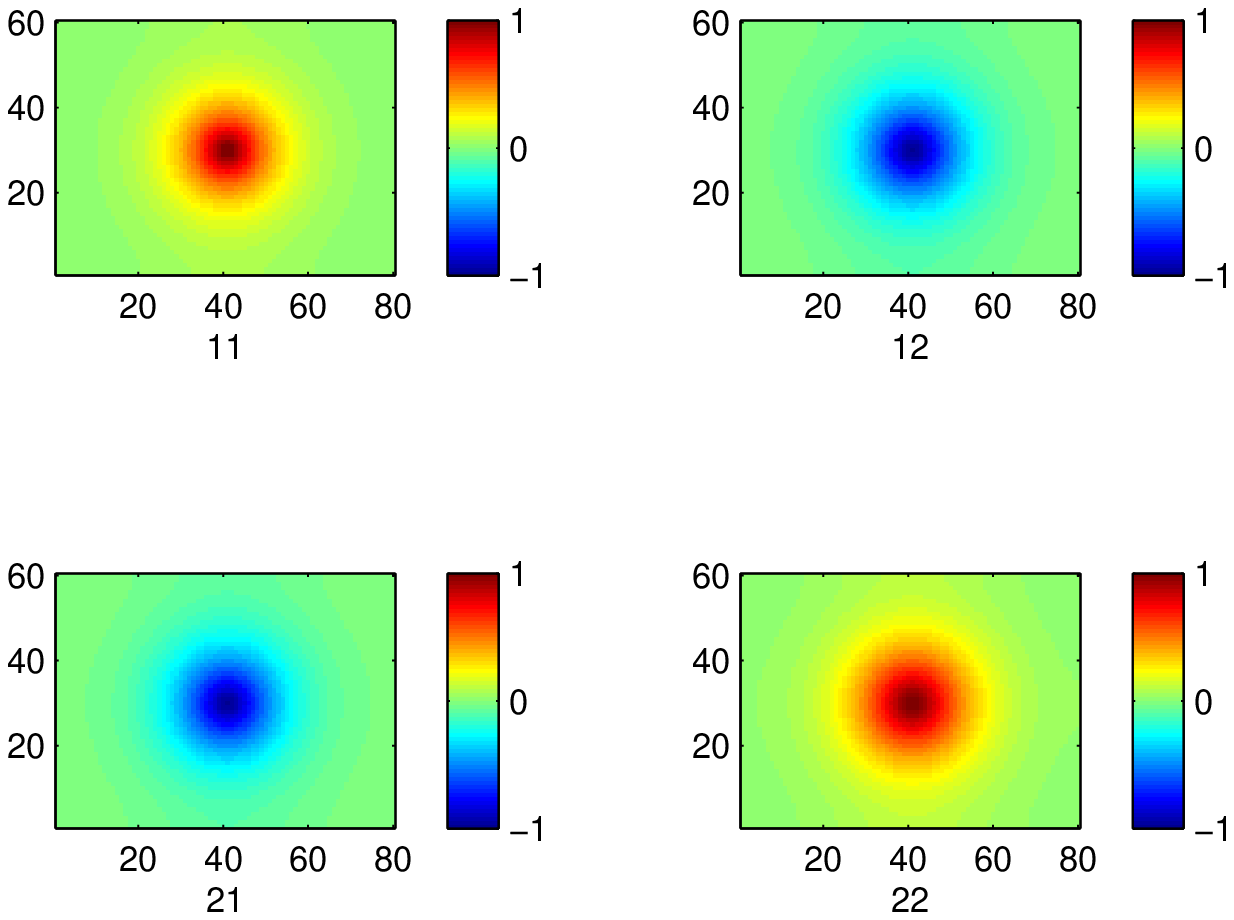}
    \caption{Marginal correlations (diagonal direction) and cross-correlations (anti-diagonal direction) with the reference points in the middle of the two GRFs
              for the negatively correlated random fields.}
   \label{fig: sampling_cov2}
\end{figure}

With white noise at the right hand of SPDEs \eqref{eq: SPDEs}, and the parameters from the
differential operators given in Table \ref{tab: sampling_parameters_bivariate}, we can get samples for positively correlated random fields with $b_{21} < 0 $ and negatively correlated random fields 
with $b_{21} >0$. These two samples are shown in Figure \ref{fig: BiMatern_positive} and Figure \ref{fig: BiMatern_negetive}.
Choosing the reference points in the middle of the two GRFs for these two cases, we can get the corresponding marginal correlations within each of the GRFs 
and the cross-correlations between the two GRFs. The results are shown in Figure \ref{fig: sampling_cov1} and Figure \ref{fig: sampling_cov2}. 
We can see that the random fields are isotropic and have the same correlation range. This is because we have chosen almost the same parameters for our cases, except the signs of $b_{21}$.  

\begin{table}
\centering
\caption{Parameters for sampling bivariate GRFs}
 \begin{tabular}{c|c|c||c|c|c}
\hline
\hline
\multicolumn{3}{c||}{ positively correlated GRFs}          &    \multicolumn{3}{c}{ negatively correlated GRFs}       \\
\hline
$\boldsymbol{\alpha}$     &  $\boldsymbol{\kappa}$        &   $\boldsymbol{b}$ &  $\boldsymbol{\alpha}$      &  $\boldsymbol{\kappa}$        &   $\boldsymbol{b}$\\
\hline
$\alpha_{11} = 2 $    &  $\kappa_{11} = 0.15$     & $b_{11} = 1$   &   $\alpha_{11} = 2$     &  $\kappa_{11} = 0.15$     & $b_{11} = 1$ \\
$\alpha_{12} = 0 $    &  $\kappa_{12} = 0$        & $b_{12} = 0$   &   $\alpha_{12} = 0$     &  $\kappa_{12} = 0$        & $b_{12} = 0$ \\
$\alpha_{21} = 2 $    &  $\kappa_{21} = 0.5$      & $b_{21} = -1$  &   $\alpha_{21} = 2$     &  $\kappa_{21} = 0.5$      & $b_{21} = 1$ \\
$\alpha_{22} = 2 $    &  $\kappa_{22} = 0.3$      & $b_{22} = 1$   &   $\alpha_{22} = 2$     &  $\kappa_{22} = 0.3$      & $b_{22} = 1$  \\
$\alpha_{n_1} = 0$    &  $\kappa_{n_1} = 0.15$    &                &   $\alpha_{n_1} = 0$    &  $\kappa_{n_1} = 0.15$    &               \\
$\alpha_{n_2} = 0$    &  $\kappa_{n_2} = 0.3$     &                &   $\alpha_{n_2} = 0$    &  $\kappa_{n_2} = 0.3$     &                \\
\hline
 \end{tabular}
  \label{tab: sampling_parameters_bivariate}
\end{table}

\begin{figure} 
    \centering
    \includegraphics[width=0.8\textwidth,height=0.6\textwidth]{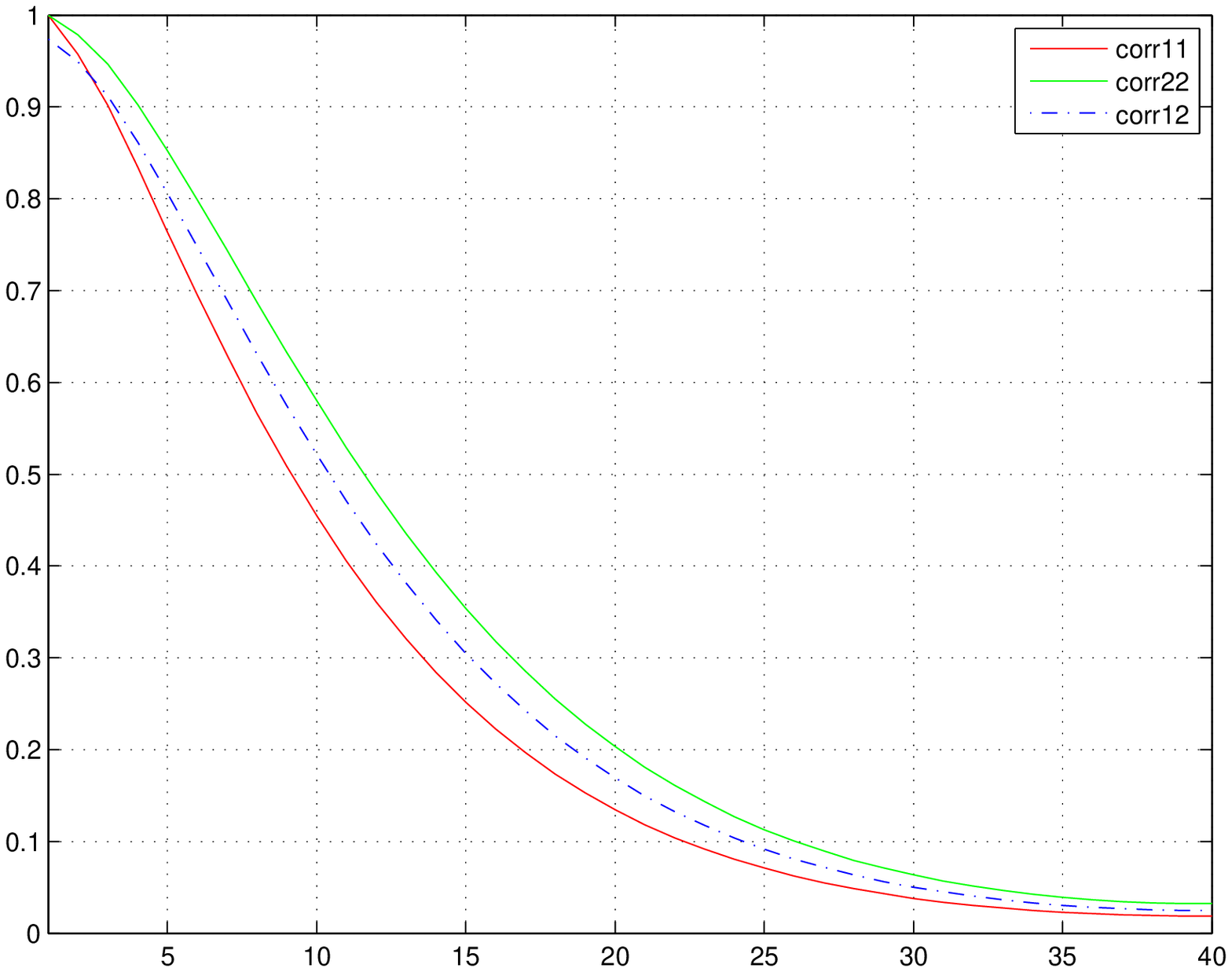}
    \caption{Marginal correlations and cross-correlations with the reference points in the middle of the two GRFs 
             for the positively correlated random fields. 'corr11' means the marginal correlation within random field $1$. 
             'corr22' means the marginal correlation within random field $2$. 'corr12' means the cross-correlation between random fields $1$ and $2$.}
   \label{fig: sampling_correlation1}
\end{figure}

\begin{figure} 
    \centering
    \includegraphics[width=0.8\textwidth,height=0.6\textwidth]{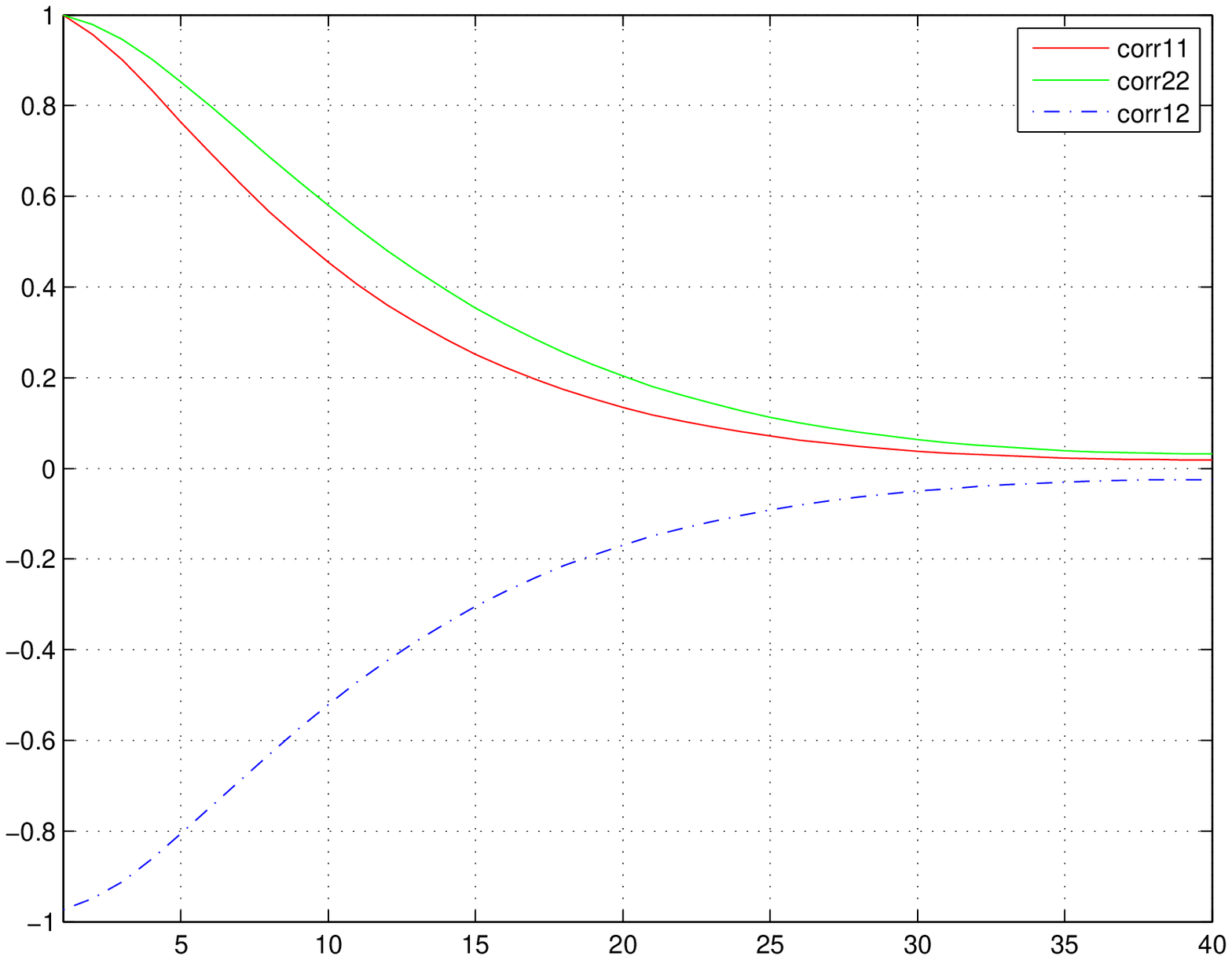}
    \caption{Marginal correlations and cross-correlations with the reference points in the middle of the two GRFs 
             for the negatively correlated random fields. 'corr11' means the marginal correlation within random field $1$. 
             'corr22' means the marginal correlation within random field $2$. 'corr12' means the cross-correlation between random fields $1$ and $2$.}
   \label{fig: sampling_correlation2}
\end{figure}

From Figure \ref{fig: BiMatern_positive} - Figure \ref{fig: sampling_cov2} and many more samples from the bivariate GRFs which we have simulated, 
we can notice that the sign of $b_{21} \cdot b_{22}$ is related to the sign of the correlation between these two GRFs,  
which corresponds to the comparison result in  \eqref{eq: comparsion_ineq}. The smoothness of the GRFs are related to the values $\alpha_{ij}$. These results also verify the conclusion given in 
\eqref{eq: comparsion_eq_1_1}. 

We can also notice that the first GRF is a Mat{\'e}rn random field when we use the white noise process as the driving process 
or under the condition that $\kappa_{n_1} = \kappa_{11}$ with the triangular systems of SPDEs.
The second field, in general, is not a Mat\'ern random field, but it can be relatively close to a Mat\'ern random field. 
With additional conditions, the second random fields could also be a Mat{\'e}rn random field, but this is not focused in this paper.
With the triangular system of SPDEs, together with some other conditions, the correlation range which was mentioned in Section \ref{sec: multi_introduction}
for the first random field can be calculated by the empirically derived formula $\rho = \sqrt{8\nu}/\kappa$ but not for the second random field.
In general we need to find the correlation ranges for the random fields numerically. 

\begin{figure} 
    \centering
    \subfigure[]{\includegraphics[width=0.45\textwidth,height=0.45\textwidth]{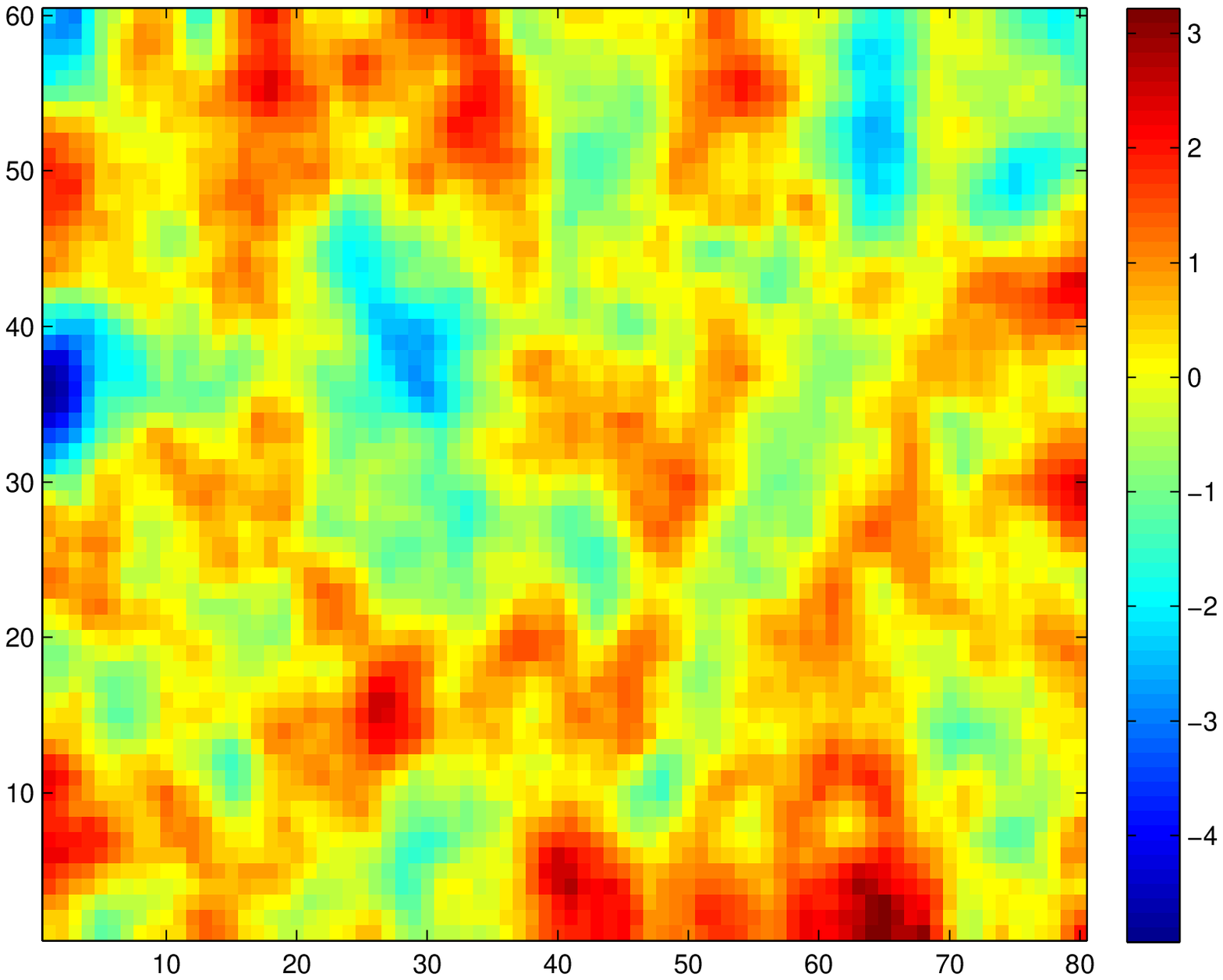}}
    \subfigure[]{\includegraphics[width=0.45\textwidth,height=0.45\textwidth]{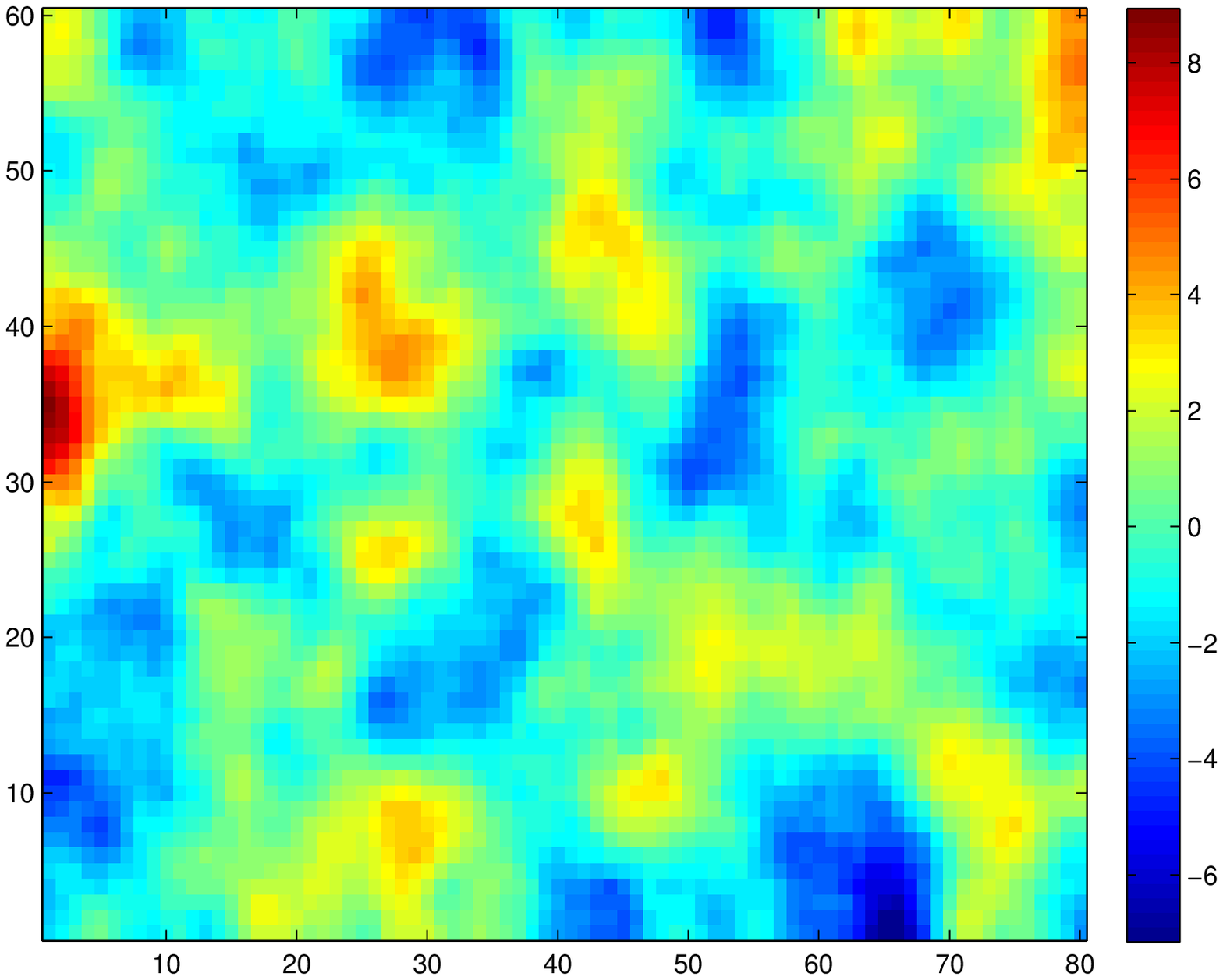}}
    \subfigure[]{\includegraphics[width=0.45\textwidth,height=0.45\textwidth]{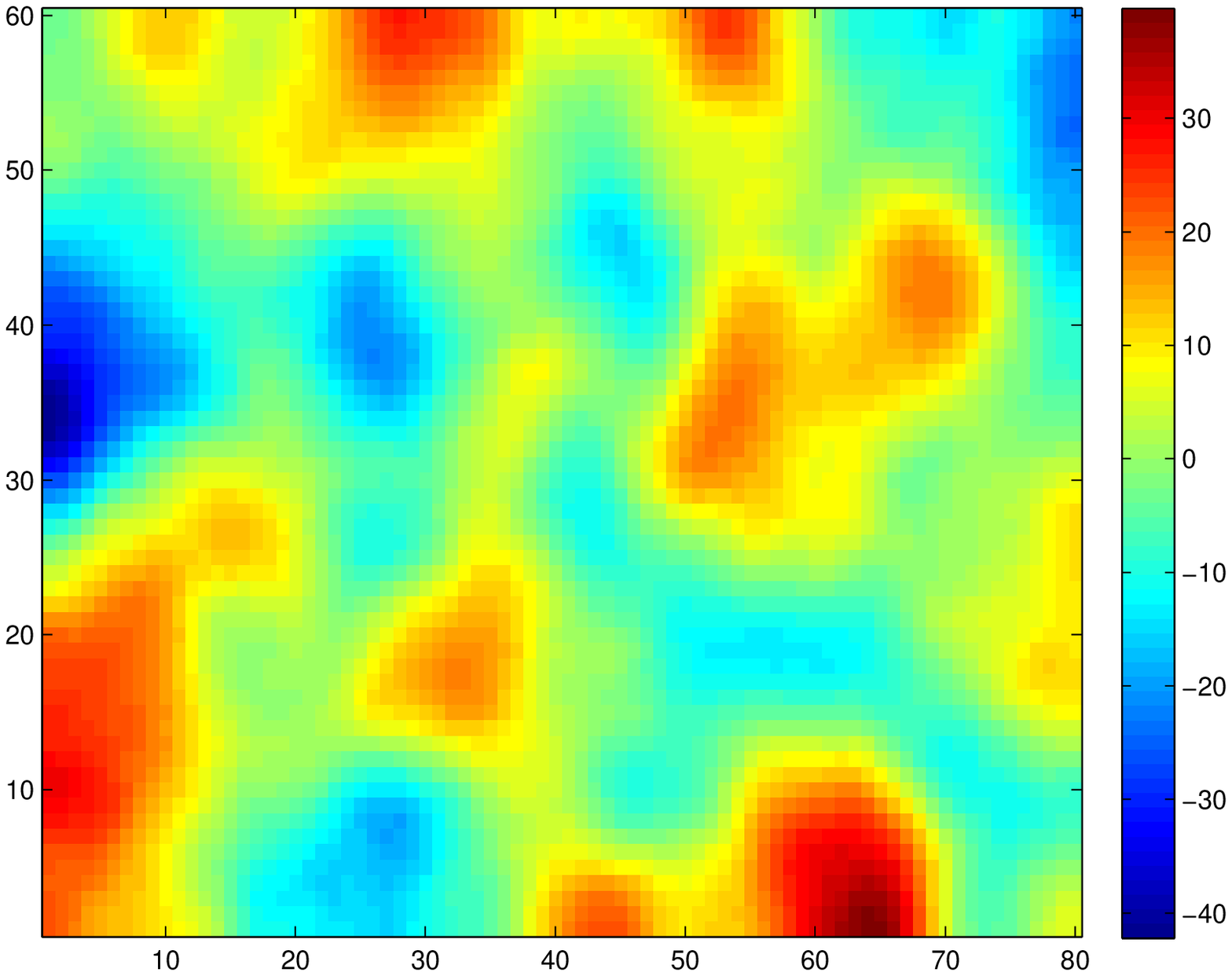}}
    \caption{One simulated realization from the trivariate Gaussian Random Fields with parameters given in Table \ref{tab: sampling_parameters_trivariate}.} 
\label{fig: Tri_Matern}
 \end{figure}

\begin{figure} 
    \centering
    \includegraphics[width=0.8\textwidth,height=0.6\textwidth]{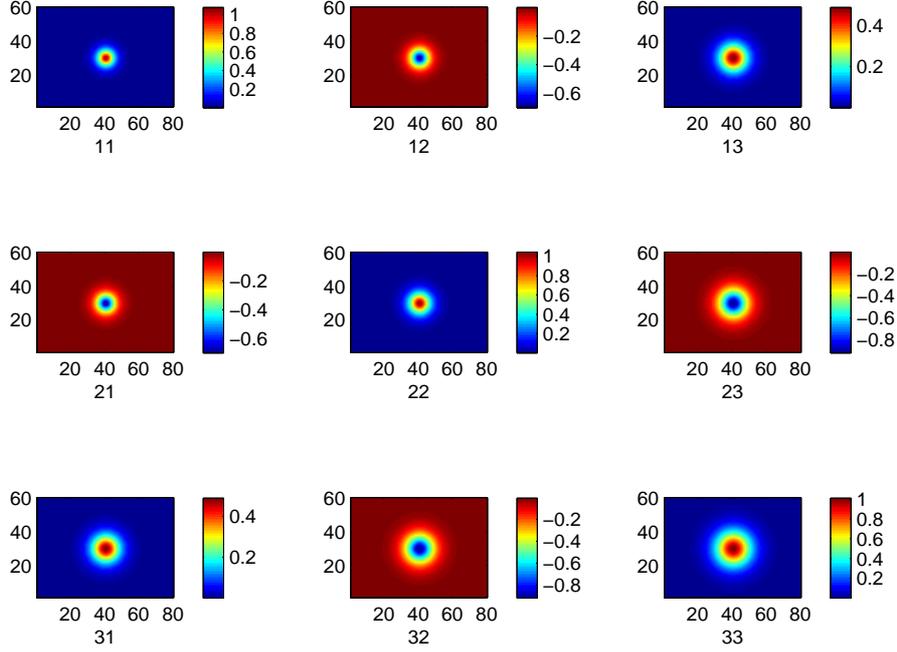}
    \caption{Marginal correlation matrices (diagonal) and cross-correlation matrices (off-diagonal) with each of the reference point in the middle of the three GRFs.}
   \label{fig: sampling_cov3}
\end{figure}

For trivariate GRFs the sampling procedure is exactly the same as for bivariate GRFs. We use the triangular system of the SPDEs. The true values given in Table \ref{tab: simulated_result3} are used.
One sample from the trivariate GRFs is shown in Figure \ref{fig: Tri_Matern} with the corresponding correlation functions given in Figure \ref{fig: sampling_cov3}.
From these figures, we notice that the trivariate random fields have similar interpretation as the bivariate random fields, but are more complicated. 
Since the triangular version of the system of SPDEs has been used,
the sign of $b_{21} \cdot b_{22}$ is related to the sign of the correlation between the first two GRFs. But for the third fields, it is not only related to the sign of  $b_{32} \cdot b_{33}$ but also the influence from
the sign of $b_{21} \cdot b_{22}$. By choosing a different parametrization, we can end up with more interesting models such as the random field $1$ and random field $3$ are positively correlated in some locations and negatively 
correlated in other locations. We are not going to discuss these issues here since this is ongoing research.

\begin{table}
\centering
\caption{Parameters for sampling trivariate GRFs}
 \begin{tabular}{c|c|c}
\hline
\hline
$\boldsymbol{\alpha}$     &  $\boldsymbol{\kappa}$        &   $\boldsymbol{b}$ \\
\hline
$\alpha_{11} = 2 $        &  $\kappa_{11} = 0.5$          & $b_{11} = 1$       \\
$\alpha_{12} = 0 $        &  $\kappa_{12} = 0$            & $b_{12} = 0$       \\
$\alpha_{13} = 0 $        &  $\kappa_{13} = 0$            & $b_{13} = 0$       \\
$\alpha_{21} = 2 $        &  $\kappa_{21} = 0.6$          & $b_{21} = 0.8$      \\
$\alpha_{22} = 2 $    	  &  $\kappa_{22} = 0.4$          & $b_{22} = 1$       \\
$\alpha_{23} = 0 $        &  $\kappa_{23} = 0 $           & $b_{23} = 0$       \\
$\alpha_{31} = 2 $        &  $\kappa_{31} = 0.5$          & $b_{31} = 1$      \\
$\alpha_{32} = 2 $        &  $\kappa_{32} = 1 $           & $b_{32} = 0.9$       \\
$\alpha_{33} = 2 $        &  $\kappa_{33} = 0.3$          & $b_{33} = 1$       \\
$\alpha_{n_1} = 1$        &  $\kappa_{n_1} = 0.5$         &                    \\
$\alpha_{n_2} = 1$        &  $\kappa_{n_2} = 0.4$         &                    \\
$\alpha_{n_3} = 1$        &  $\kappa_{n_3} = 0.3$         &                    \\
\hline
 \end{tabular}
  \label{tab: sampling_parameters_trivariate}
\end{table}

\begin{figure} 
    \centering
    \includegraphics[width=0.8\textwidth,height=0.6\textwidth]{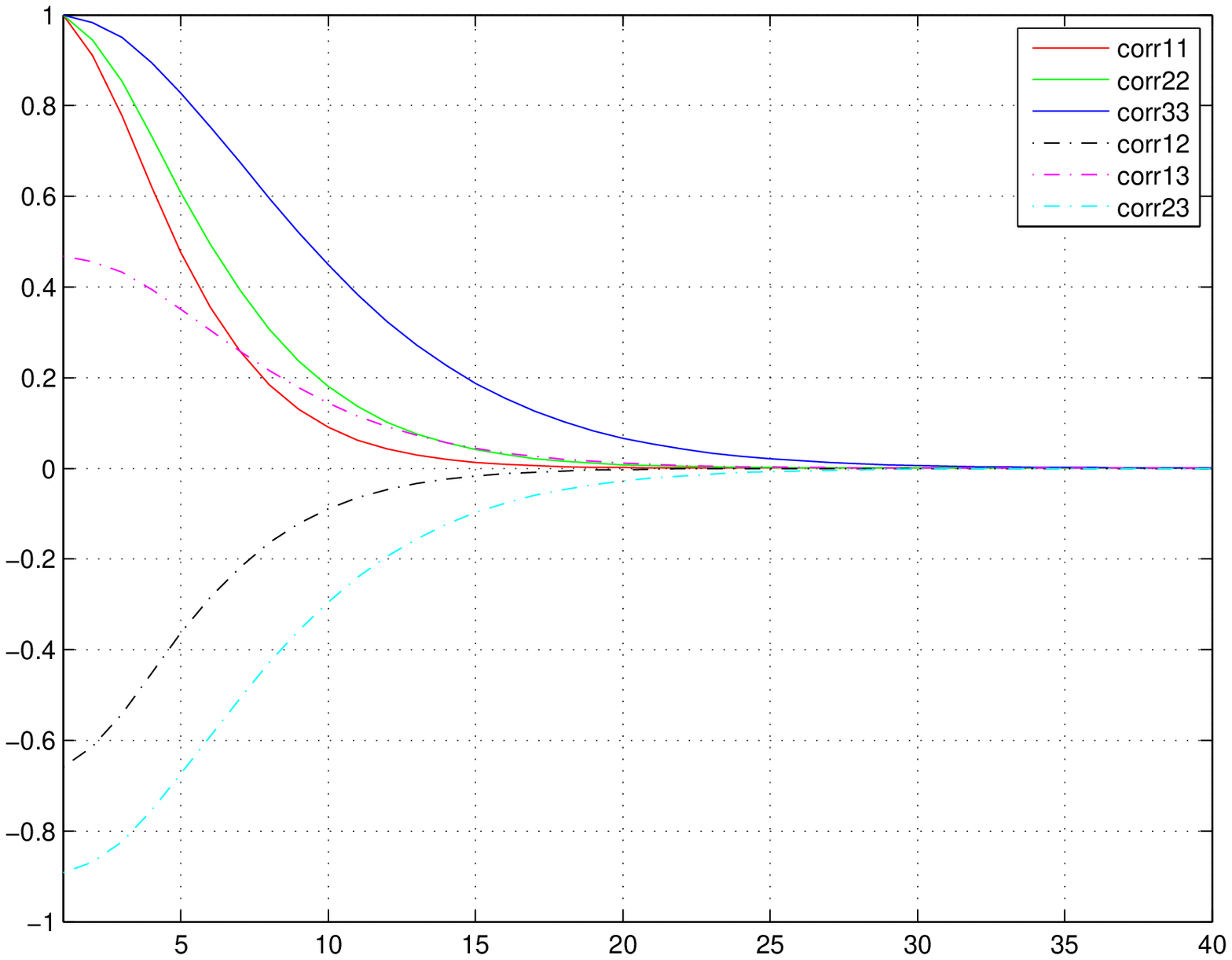}
    \caption{Marginal correlation matrices and cross-correlation matrices with each of the reference point in the middle of the three GRFs.
            'corr11' means the marginal correlation within random field $1$. 'corr22' means the marginal correlation within random field $2$. 
            'corr33' means the marginal correlation within random field $3$. 'corr12' means the cross-correlation between random fields $1$ and $2$.
             'corr13' means the cross-correlation between random fields $1$ and $3$. 'corr23' means the cross-correlation between random fields $2$ and $3$.}
   \label{fig: sampling_correlation3}
\end{figure}

\section{Examples and applications} \label{sec: applications}
In order to illustrate how to use the SPDEs approach for constructing multivariate GRFs and the usefulness of our approach, some examples both with simulated data and real data are chosen. First, 
some basic theory on inference with multivariate GMRFs is given.  We focus on the triangular system of SPDEs in this paper.

We use the bivariate GMRFs as an example. The multivariate GMRFs can be done analogously. 
Assume that we have used $N$ triangles in the discretization for each random field $x_i(\boldsymbol{s}) | \theta_i, i = 1, 2$. 
For the bivariate case $\boldsymbol{x}(\boldsymbol{s}) = \left(x(\boldsymbol{s})_1, x(\boldsymbol{s})_2 \right)^\text{T}$  
has a $2N$-dimensional multivariate Gaussian distribution with probability density function
\begin{equation} \label{eq: bivariate_x|theta}
 \pi(\boldsymbol{x} | \boldsymbol{\theta}) = \left(\frac{1}{2\pi} \right)^{2N} |\boldsymbol{Q}(\boldsymbol\theta)|^{1/2} \exp \left(-\frac{1}{2} \boldsymbol{x}^\text{T} \boldsymbol{Q}(\boldsymbol{\theta}) \boldsymbol{x} \right), 
\end{equation}
where $\boldsymbol{Q}(\boldsymbol\theta)$ is the precision matrix of the bivariate GMRF with the parameters $\boldsymbol{\theta}$. Furthermore, we assume the length of the data $\boldsymbol{y} = (y_1, y_2)^\text{T}$ is $t = k_1 + k_2$ 
where $y_1$ is the observation of $x_1(\boldsymbol{s})$ with length $k_1$ and $y_2$ is the observation of $x_2(\boldsymbol{s})$ with length $k_2$. Then $\boldsymbol{y}$ is a $t$-dimensional random variable with probability density function
\begin{equation} \label{eq: bivariate_y|x,theta}
 \pi(\boldsymbol{y|\boldsymbol{x},\boldsymbol{\theta}}) = \left(\frac{1}{2\pi} \right)^{t}|\boldsymbol{Q}_n|^{1/2}
               \exp \left(-\frac{1}{2} (\boldsymbol{y}-\boldsymbol{Ax})^\text{T} \boldsymbol{Q}_n (\boldsymbol{y}-\boldsymbol{Ax}) \right),
\end{equation}
where $\boldsymbol{Q}_n$ is defined in Section \ref{sec: gmrfs_approximation} with size $t \times t$, and $\boldsymbol{A}$ is a $t \times 2N$ matrix which links the sparse observations to 
our bivariate GMRFs. Notice that the density function $\pi(\boldsymbol{y}|\boldsymbol{x},\boldsymbol{\theta})$ 
is independent of $\boldsymbol{\theta}$. Hence we can write the  probability density function as $\pi(\boldsymbol{y}|\boldsymbol{x})$. 
We first find the probability density function of ${\boldsymbol{x}|\boldsymbol{y}, \boldsymbol{\theta}}$ from \eqref{eq: bivariate_x|theta} - \eqref{eq: bivariate_y|x,theta}
\begin{equation} \label{eq: bivariate_x|y,theta}
\begin{split}
 \pi(\boldsymbol{x}|\boldsymbol{y}, \boldsymbol{\theta}) & \propto \pi({\boldsymbol{x}, \boldsymbol{y} | \boldsymbol{\theta}})   \\
          & = \pi(\boldsymbol{x}|\boldsymbol{\theta}) \pi(\boldsymbol{y}|\boldsymbol{x}, \boldsymbol{\theta}) \\
          & \propto \exp \left( -\frac{1}{2} \left( x^\text{T} (\boldsymbol{Q}(\boldsymbol{\theta}) + \boldsymbol{A}^\text{T} \boldsymbol{Q}_n \boldsymbol{A}) \boldsymbol{x} - 2\boldsymbol{x}^\text{T} \boldsymbol{A}^\text{T}\boldsymbol{Q}_n \boldsymbol{y} \right) \right).
\end{split}
\end{equation}
Let $\boldsymbol{\mu}_c (\boldsymbol{\theta}) =  \boldsymbol{Q}_c^{-1}(\boldsymbol{\theta}) \boldsymbol{A}^\text{T} \boldsymbol{Q}_n \boldsymbol{y} $, 
and $\boldsymbol{Q}_c (\boldsymbol{\theta}) = \boldsymbol{Q}(\boldsymbol{\theta}) + \boldsymbol{A}^\text{T} \boldsymbol{Q}_n \boldsymbol{A}$, and then $\pi(\boldsymbol{x}|\boldsymbol{y}, \boldsymbol{\theta})$ can be denoted as
\begin{displaymath}
 {\boldsymbol{x}|\boldsymbol{y}, \boldsymbol{\theta}} \sim \mathcal{N} \left( \boldsymbol{\mu}_c (\boldsymbol{\theta}), \boldsymbol{Q}_c^{-1} (\boldsymbol{\theta}) \right),
\end{displaymath}
or in the canonical parametrization as
\begin{displaymath}
 {\boldsymbol{x}|\boldsymbol{y}, \boldsymbol{\theta}} \sim \mathcal{N}_c \left(\boldsymbol{b}_c,  \boldsymbol{Q}_c (\boldsymbol{\theta}) \right),
\end{displaymath}
with $\boldsymbol{b}_c = \boldsymbol{A}^\text{T} \boldsymbol{Q}_n \boldsymbol{y}$. Thus ${\boldsymbol{x}|\boldsymbol{y}, \boldsymbol{\theta}}$ is a $2N$-dimensional multivariate Gaussian distribution.
The canonical parametrization for a GMRF is useful with successive conditioning \citep{rue2005gaussian}. 
For more information about the canonical parametrization for the GMRFs, we refer to \citet[Chapter $2.2.3$]{rue2005gaussian}.

The probability density function \eqref{eq: bivariate_x|y,theta} can be used to integrate out $\boldsymbol{x}$ from the joint density of 
$\boldsymbol{x}$, $\boldsymbol{y}$ and $\boldsymbol{\theta}$,
\begin{equation} \label{eq: baysian_y & theta}
\begin{split}
 \pi(\boldsymbol{y}, \boldsymbol{\theta}) & = \frac{\pi(\boldsymbol{\theta}, \boldsymbol{x}, \boldsymbol{y})}{\pi(\boldsymbol{x} | \boldsymbol{\theta}, \boldsymbol{y})} \\
                                      & = \frac{\pi(\boldsymbol{\theta}) \pi(\boldsymbol{x}|\boldsymbol{\theta}) \pi(\boldsymbol{y|\boldsymbol{x},\boldsymbol{\theta}})}{\pi(\boldsymbol{x}|\boldsymbol{y}, \boldsymbol{\theta})},
\end{split}
\end{equation}
where $\pi(\boldsymbol{\theta})$ is the prior distribution of $\boldsymbol{\theta}$.  With \eqref{eq: baysian_y & theta}, the posterior distribution of $\boldsymbol{\theta}$ can be obtained as
\begin{equation} \label{eq: bivariate_theta|y}
 \begin{split}
  \pi(\boldsymbol{\theta} | \boldsymbol{y}) & \propto \pi(\boldsymbol{\theta}) \frac{|\boldsymbol{Q}(\boldsymbol{\theta})|^{1/2} |\boldsymbol{Q_n}|^{1/2}}{|\boldsymbol{Q}_c(\boldsymbol{\theta})|^{1/2}} 
                          \exp \left( -\frac{1}{2} \boldsymbol{x}^\text{T} \boldsymbol{Q}(\boldsymbol{\theta}) \boldsymbol{x} \right) \\
                          & \times \exp \left(-\frac{1}{2} (\boldsymbol{y}-\boldsymbol{Ax})^\text{T} \boldsymbol{Q}_n (\boldsymbol{\theta}) (\boldsymbol{y}-\boldsymbol{Ax}) \right) \\
                          & \times \exp \left( \frac{1}{2} (\boldsymbol{x}-\boldsymbol{\mu}_c (\boldsymbol{\theta}))^\text{T} \boldsymbol{Q}_c (\boldsymbol{\theta}) (\boldsymbol{x}-\boldsymbol{\mu}_c (\boldsymbol{\theta})) \right).
 \end{split}
\end{equation}
The quadratic terms in the exponential functions in \eqref{eq: bivariate_theta|y} can be simplified by using $\boldsymbol{\mu}_c (\boldsymbol{\theta})$ and $\boldsymbol{Q}_c (\boldsymbol{\theta})$. It is also 
convenient to use the logarithm of the posterior distribution of $\boldsymbol{\theta}$. Reorganize \eqref{eq: bivariate_theta|y} to get the formula which will be used for inference 
\begin{equation} \label{eq: bivariate_log(theta|y)}
\begin{split}
 \log(\pi(\boldsymbol{\theta}|\boldsymbol{y})) = & \text{ Const} + \log(\pi(\boldsymbol{\theta})) + \frac{1}{2}\log(|\boldsymbol{Q}(\boldsymbol{\theta})|) \\
              & - \frac{1}{2}\log(|\boldsymbol{Q}_c(\boldsymbol{\theta})|) + \frac{1}{2} \boldsymbol{\mu}_c(\boldsymbol{\theta})^\text{T} \boldsymbol{Q}_c(\boldsymbol{\theta}) \boldsymbol{\mu}_c(\boldsymbol{\theta}).
\end{split}
\end{equation}

\noindent From \eqref{eq: bivariate_log(theta|y)} we can see that it is difficult to handle the posterior distribution of $\boldsymbol{\theta}$ analytically since both the determinants and the quadratic terms are hard to handle.
Thus numerical methods should be applied for the statistical inference in this paper. 

Furthermore, even though it is not the topic for our paper, we point out that it is also possible to obtain the probability density function $\pi(\boldsymbol{x}|\boldsymbol{y})$ by integrate out $\boldsymbol{\theta}$. 
But this is difficult and needs to be obtained numerically using the following expression \eqref{eq: bivariate_x|y}
\begin{equation} \label{eq: bivariate_x|y}
 \pi(\boldsymbol{x}|\boldsymbol{y}) = \int_{\mathbb{R}^m}\pi(\boldsymbol{x}|\boldsymbol{\theta}, \boldsymbol{y}) \pi(\boldsymbol{\theta}|\boldsymbol{y}) d{\boldsymbol{\theta}}
\end{equation}
where $m$ is the dimension of $\boldsymbol{\theta}$. When the dimension of $\boldsymbol{\theta}$ is large, then this integration might be infeasible in practice.

\subsection{Statistical inference with simulated data} \label{sec: simulated_data}
First, we illustrate how to do the statistical inference for simulated data with the SPDEs approach with known true parameter values. 
These datasets both contain one realization with $2000$ observations at different locations. The parameters used  for generating the simulated data are presented 
in Table \ref{tab: simulated_parameters}. To make it simpler, the nugget effects are assumed to be known, $\tau_1 = \tau_2 = 0.001$. As discussed in Section \ref{sec: multi_introduction}
it is hard to estimate the smoothness parameters, so we fix $\{ \alpha_{ij}; i,j = 1,2 \}$ in the system of SPDEs to the known values. The smoothness parameters ${\alpha_n}_1$ and $\alpha_{n_2}$ for the noise processes are also fixed to the 
known values due to the same reason. The scaling parameters $\kappa_{n_1}$ and $\kappa_{n_2}$ for the noise processes are restricted with $\kappa_{n_1} = \kappa_{11}$ and $\kappa_{n_2} = \kappa_{22}$ for the simulated data. 
Thus in the simulated data examples, only $\boldsymbol{\theta} = \{\kappa_{11}, \kappa_{21}, \kappa_{22}, b_{11}, b_{21}, b_{22}\}$ needed to be estimated. 
Since $\kappa_{11}, \kappa_{21}$  and $\kappa_{22}$ have to take positive values, we a priori assign \emph{log-normal} distributions with mean zero and large variances for each of the parameters. 
$b_{11}, b_{21} \text{ and } b_{22}$ are given \emph{normal} priors with mean zero and large variances. 

\begin{figure}
 \centering
\subfigure[]{\includegraphics[width=0.45\textwidth,height=0.45\textwidth]{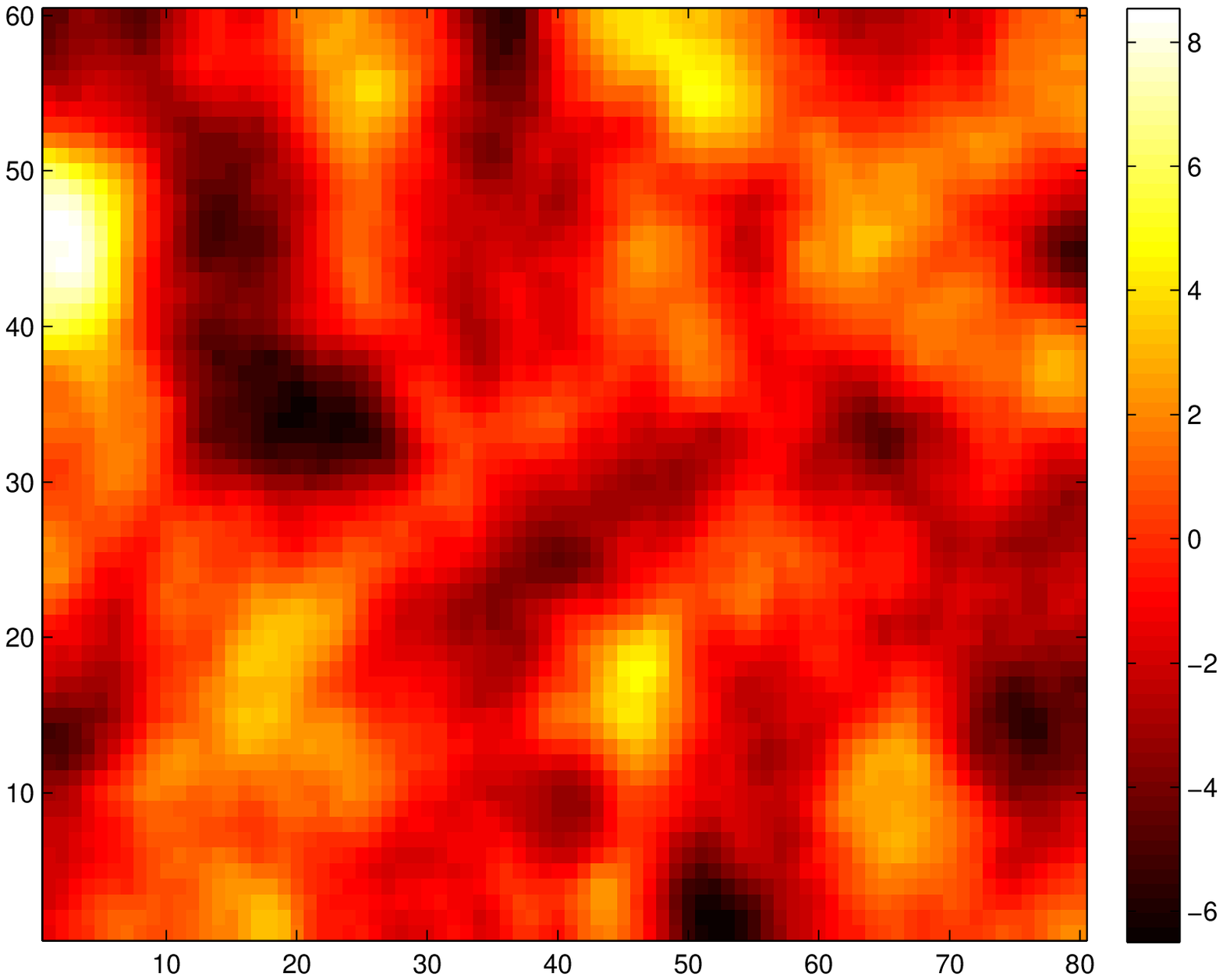} \label{fig: bivariate1_data_simulated_1}}
\subfigure[]{\includegraphics[width=0.45\textwidth,height=0.45\textwidth]{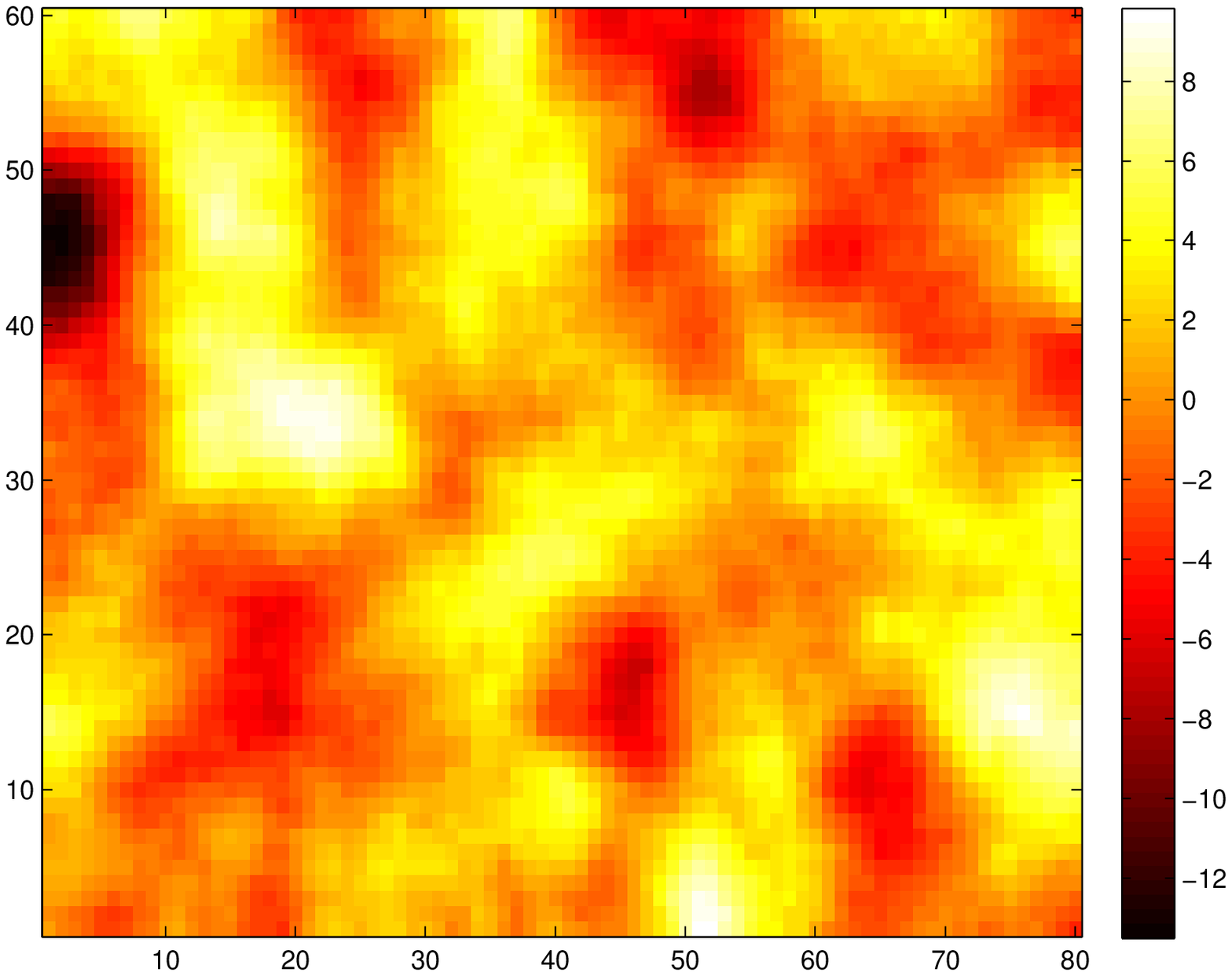} \label{fig: bivariate1_data_simulated_2}} \\
\subfigure[]{\includegraphics[width=0.45\textwidth,height=0.45\textwidth]{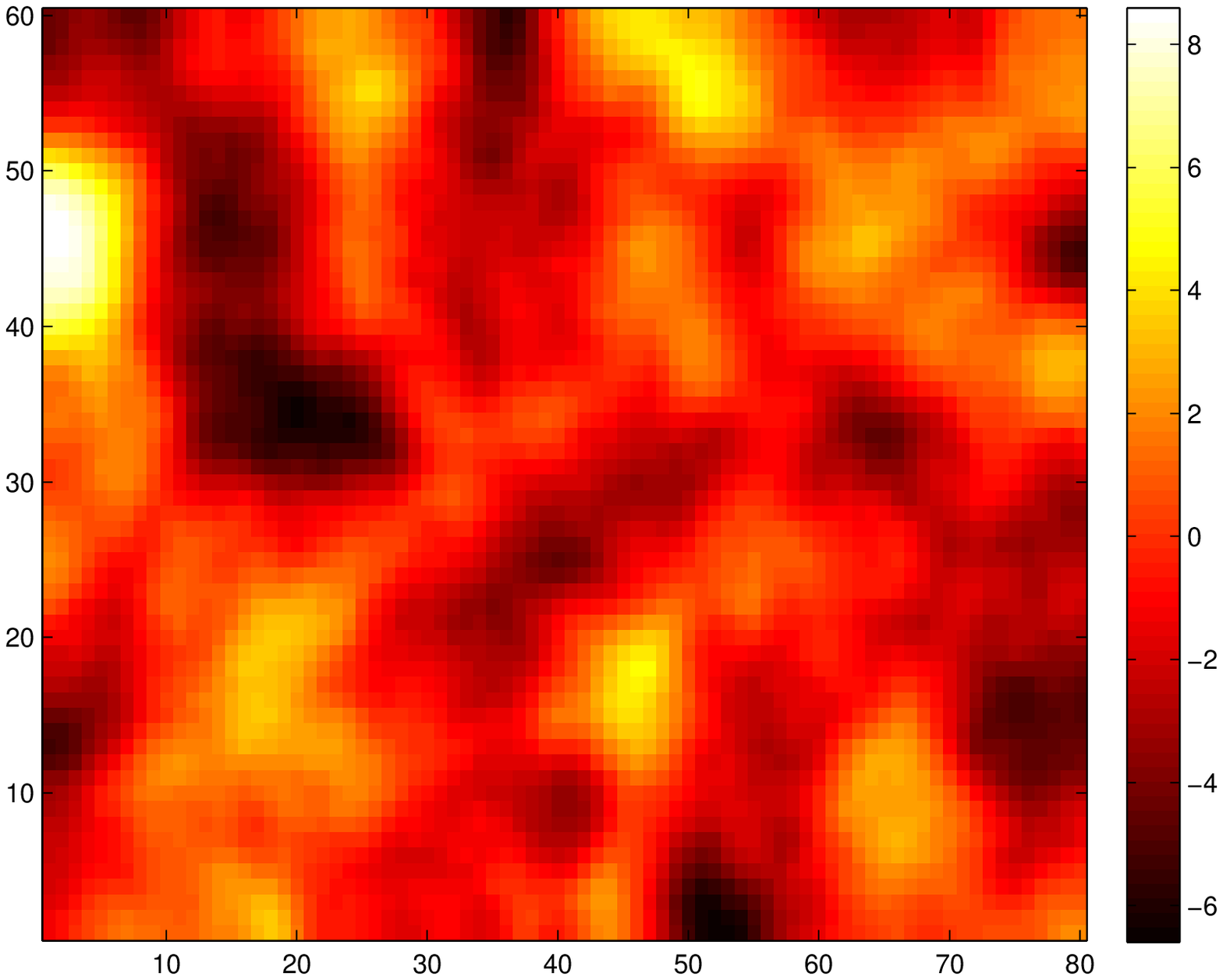} \label{fig: bivariate1_data_mean_2d_1}}
\subfigure[]{\includegraphics[width=0.45\textwidth,height=0.45\textwidth]{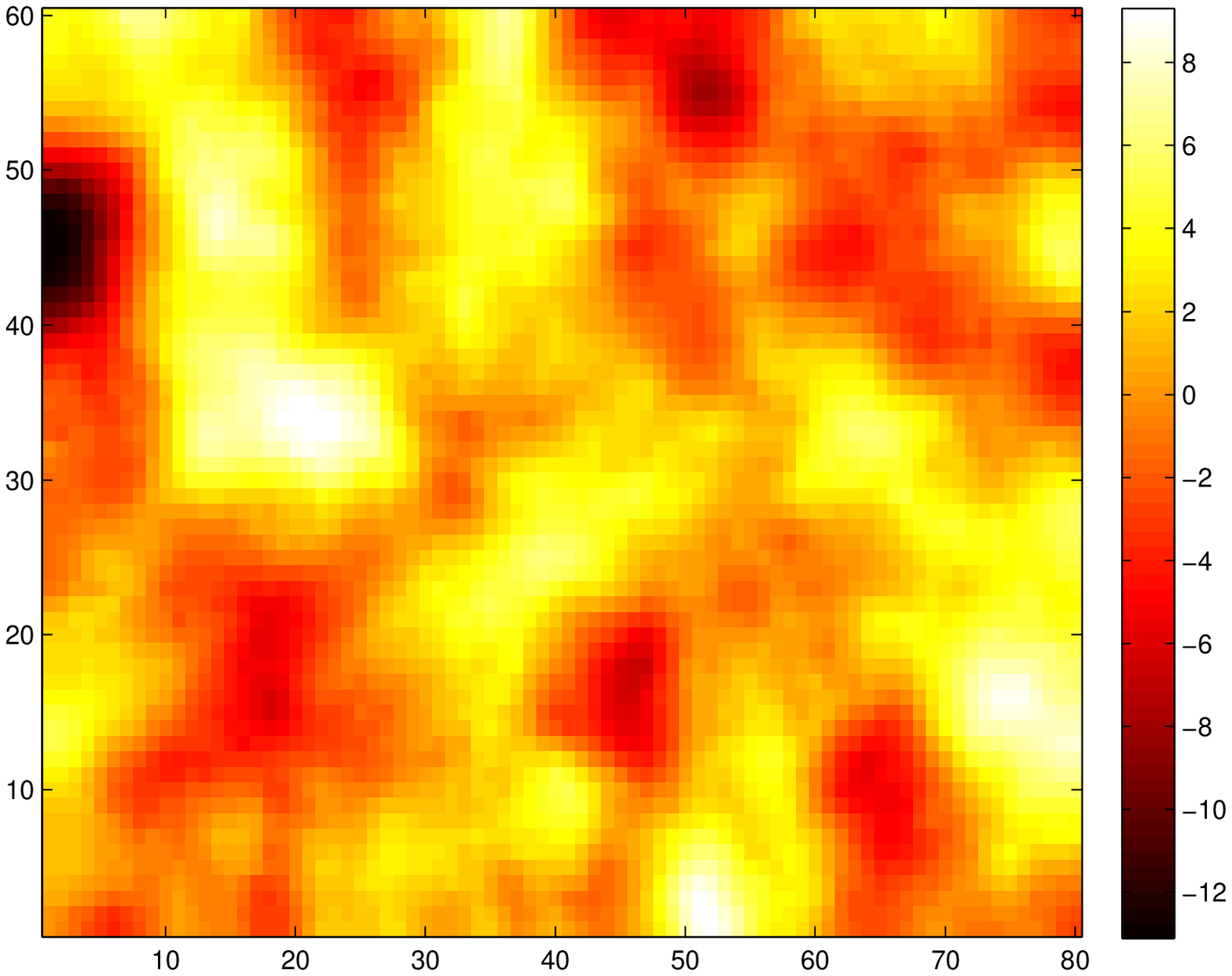} \label{fig: bivariate1_data_mean_2d_2}}
\caption{The true bivariate GRFs (a) - (b) and the estimated conditional mean for the bivariate GRFs (c) - (d) for simulated dataset $1$.}
\label{fig: bivariate1_data_simulated}
\end{figure}

In the first example, the two GRFs are negatively correlated and the realizations are shown in Figure \ref{fig: bivariate1_data_simulated_1} and Figure \ref{fig: bivariate1_data_simulated_2}. The corresponding estimated conditional mean 
for the negatively correlated GRFs are given in Figure \ref{fig: bivariate1_data_mean_2d_1} and Figure \ref{fig: bivariate1_data_mean_2d_2}. 
We can notice that there are no large differences between the estimated conditional mean for the GRFs and the true bivariate GRFs. The estimates for the parameters are given in Table \ref{tab: simulated_result1} with their 
standard derivations. From this table, we notice that the estimates for all the parameters are quite accurate with accuracy to $2$ digits. The true values of the parameters are within $1$ standard deviation from
the estimates.

\begin{figure}
 \centering
\subfigure[]{\includegraphics[width=0.45\textwidth,height=0.45\textwidth]{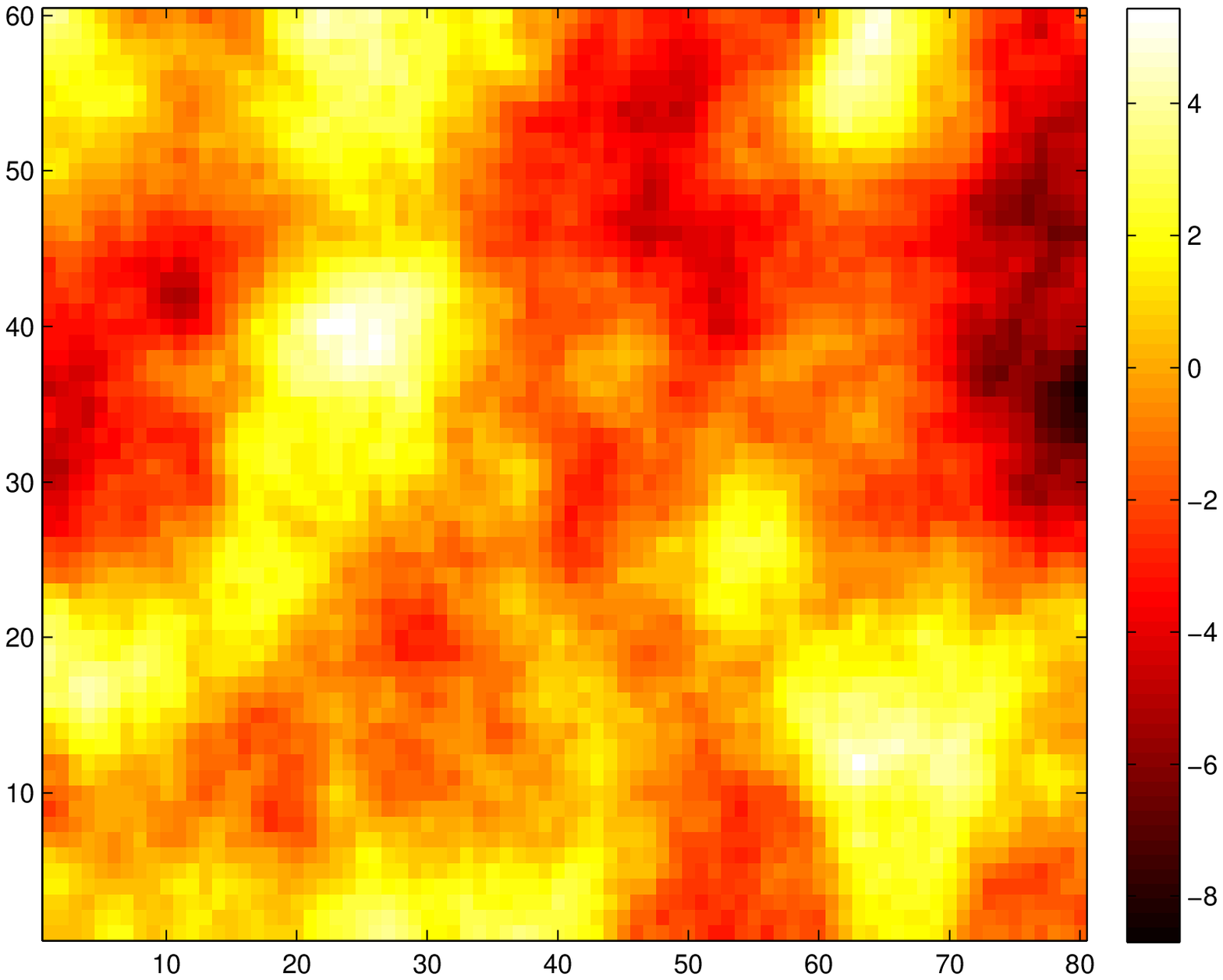} \label{fig: bivariate2_data_simulated_1}}
\subfigure[]{\includegraphics[width=0.45\textwidth,height=0.45\textwidth]{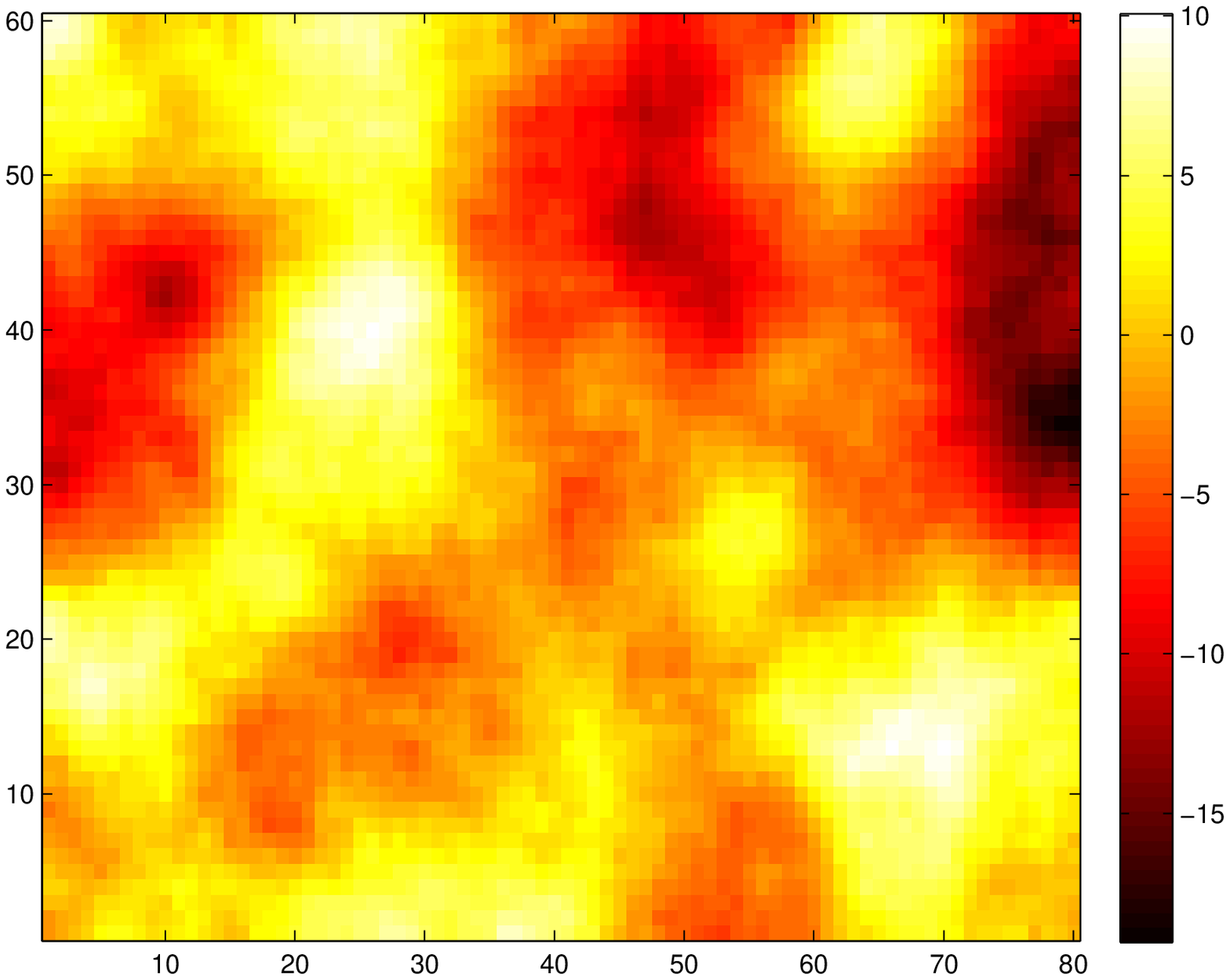} \label{fig: bivariate2_data_simulated_2}} \\
\subfigure[]{\includegraphics[width=0.45\textwidth,height=0.45\textwidth]{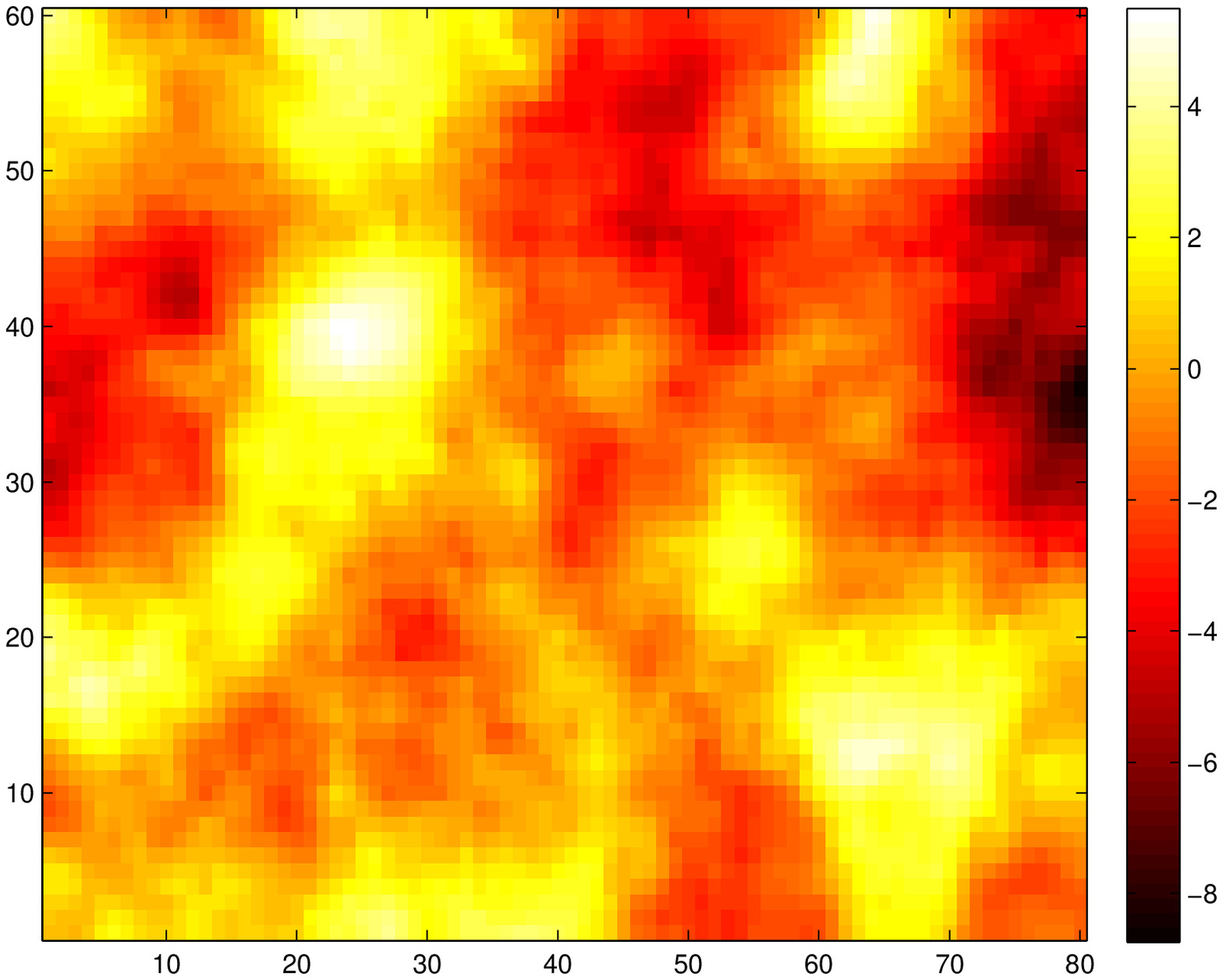} \label{fig: bivariate2_data_mean_2d_1}}
\subfigure[]{\includegraphics[width=0.45\textwidth,height=0.45\textwidth]{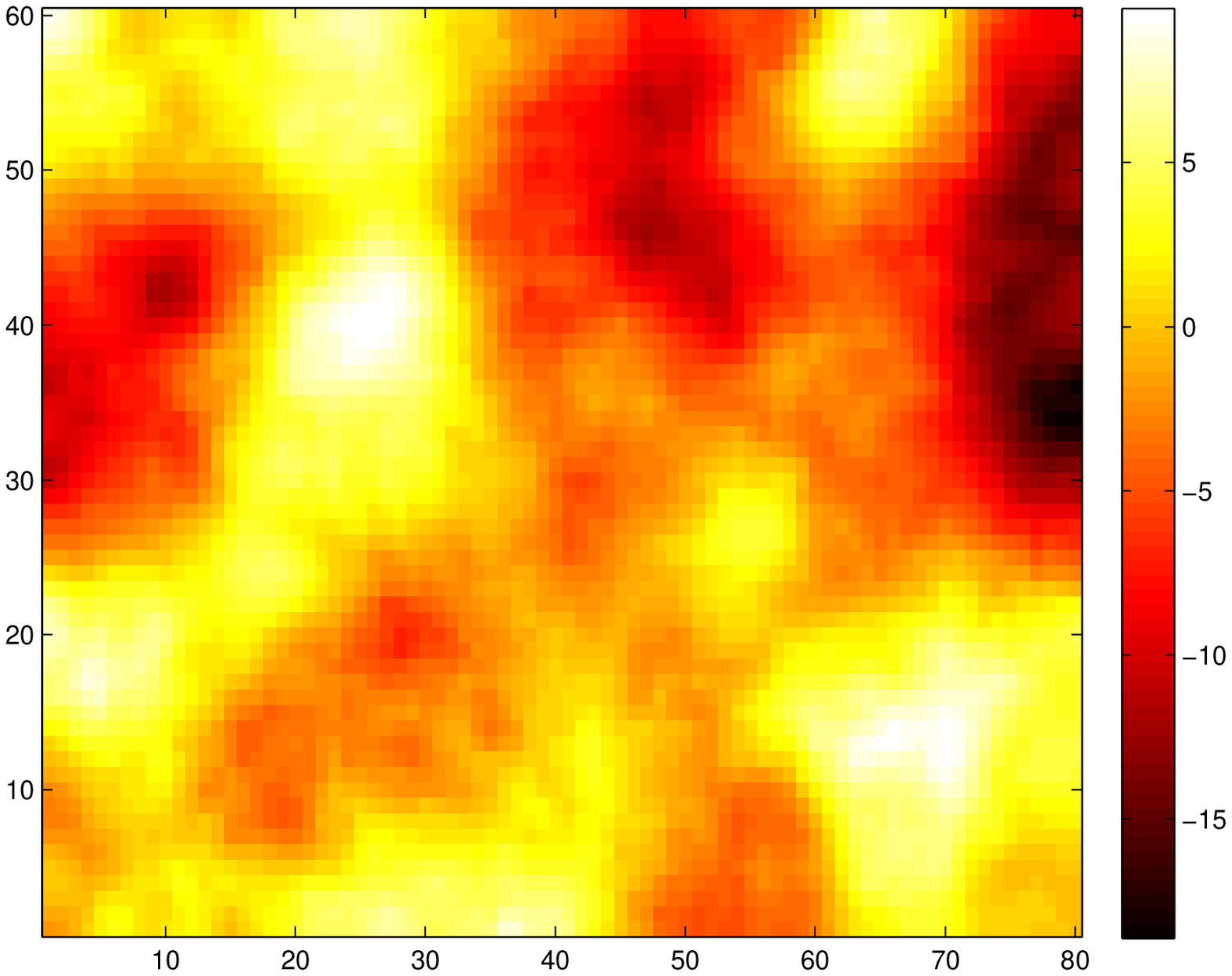} \label{fig: bivariate2_data_mean_2d_2}}
\caption{The true bivariate GRFs (a) - (b) and the estimated conditional mean for the bivariate GRFs (c) - (d) for simulated dataset $2$.}
\label{fig: bivariate2_data_simulated}
\end{figure}

The second example with the simulated data uses two GRFs that are positively correlated, and the realizations are shown in Figure \ref{fig: bivariate2_data_simulated_1} and Figure \ref{fig: bivariate2_data_simulated_2}. 
The corresponding estimated conditional mean for the bivariate GRFs are illustrated in Figure \ref{fig: bivariate2_data_mean_2d_1} and Figure \ref{fig: bivariate2_data_mean_2d_2}. From Figure \ref{fig: bivariate2_data_simulated_1}
to Figure \ref{fig: bivariate2_data_mean_2d_2}, we can again notice that the estimated conditional mean for the positively correlated bivariate GRFs 
are almost the same as the true bivariate GRFs. The estimates are accurate and the true values of the parameters 
are within $1$ standard deviation from the estimates.

\begin{figure}
 \centering
\subfigure[]{\includegraphics[width=0.45\textwidth,height=0.45\textwidth]{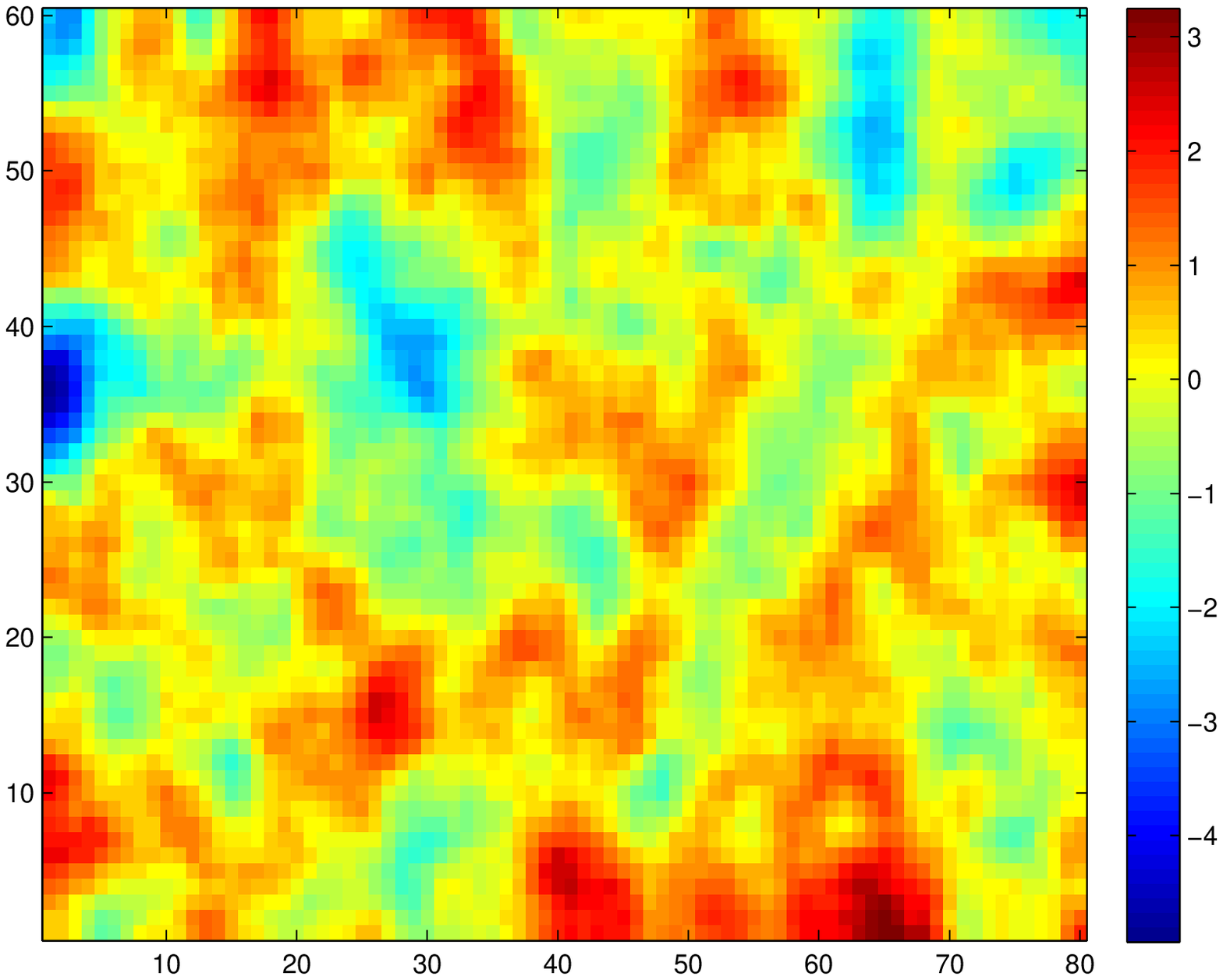} \label{trivariate_mean_2d_1}}
\subfigure[]{\includegraphics[width=0.45\textwidth,height=0.45\textwidth]{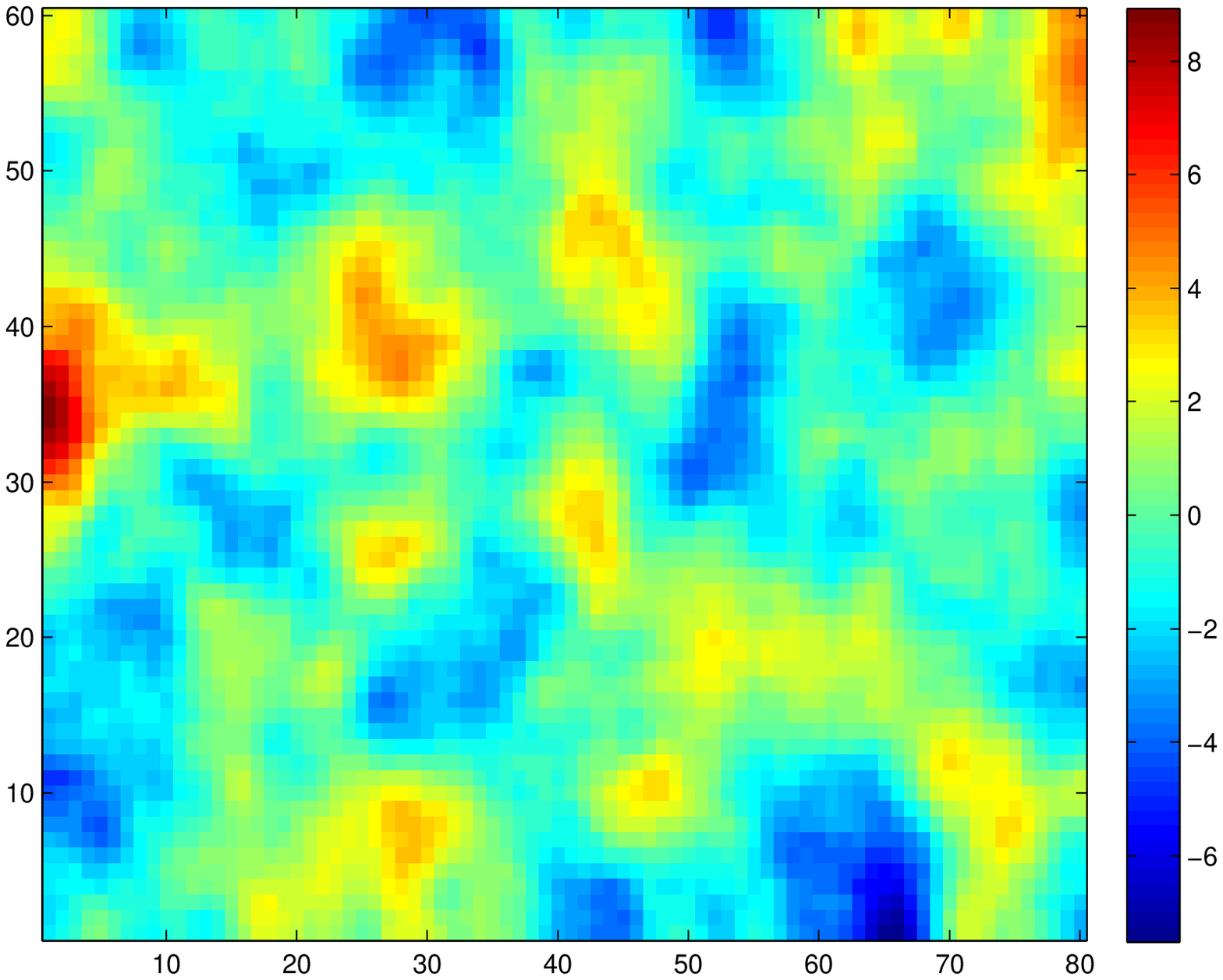} \label{trivariate_mean_2d_2}}
\subfigure[]{\includegraphics[width=0.45\textwidth,height=0.45\textwidth]{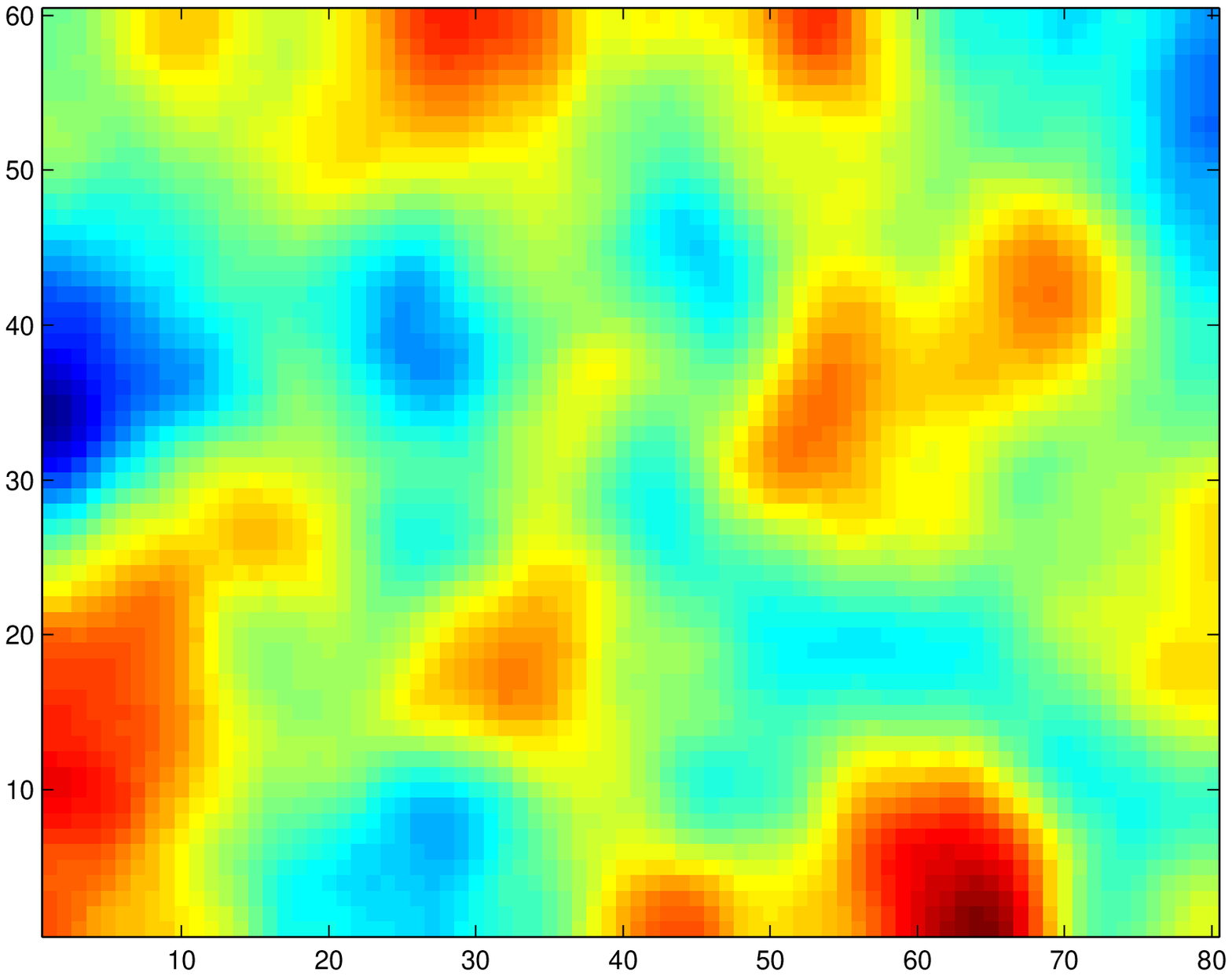} \label{trivariate_mean_2d_3}}
\caption{The estimated conditional mean for the trivariate GRFs.}
\label{fig: trivariate_reconstruction}
\end{figure}

The third example uses a trivariate GRF. We reuse the parameters given in Table \ref{tab: simulated_result3} as true parameters for 
the simulated data. One realization from the trivariate GRF is given in Figure \ref{fig: Tri_Matern} in Section \ref{sec: Sampling}. The nugget effects are assumed to be known as $\tau_1 = \tau_2 = \tau_3 = 0.01$. In this example, we also fix 
$\{ \alpha_{ij}; i, j = 1,2\}$ and $\{ \alpha_{n_i}, i = 1,2,3 \}$. The scaling parameters for the noise processes are fixed with similar settings 
as the bivariate GRFs $\{ \kappa_{n_i} = \kappa_{ii}; i = 1,2,3\}$. Therefore we have $12$ parameters to be estimated. They are $b_{ij}$ and $\{ \kappa_{ij}; i,j = 1,2,3, i \geq j \}$. The estimates for these parameters are given in 
Table \ref{tab: simulated_result3}. We can notice that the estimates are accurate in this example since all the estimated are accurate to $2$ digits, and the true values of the parameters are within $2$ standard deviations from the 
corresponding estimates. The estimated conditional mean for the trivariate GRF are given in Figure \ref{trivariate_mean_2d_1} - Figure \ref{trivariate_mean_2d_3}. Comparing the results given in Figure \ref{fig: trivariate_reconstruction}  
with the corresponding fields given in Figure \ref{fig: Tri_Matern}, we can see that there are no large differences between the true random fields and the estimated conditional mean.
From the examples we can notice that our approach works well not only for bivariate GRFs but also for multivariate GMRFs. 

\begin{table}
\centering
\caption{Parameters for simulating the bivariate GRFs}
 \begin{tabular}{c|c|c||c|c|c}
\hline
\hline
\multicolumn{3}{c||}{ dataset $1$}                             &    \multicolumn{3}{c}{ dataset $2$}       \\
\hline
$\boldsymbol{\alpha}$     &  $\boldsymbol{\kappa}$        &   $\boldsymbol{b}$ &  $\boldsymbol{\alpha}$      &  $\boldsymbol{\kappa}$        &   $\boldsymbol{b}$\\
\hline
$\alpha_{11} = 2 $    &  $\kappa_{11} = 0.3$     & $b_{11} = 1$   &   $\alpha_{11} = 2$     &  $\kappa_{11} = 0.15$     & $b_{11} = 1$ \\
$\alpha_{12} = 0 $    &  $\kappa_{12} = 0$       & $b_{12} = 0$   &   $\alpha_{12} = 0$     &  $\kappa_{12} = 0$        & $b_{12} = 0$ \\
$\alpha_{21} = 2 $    &  $\kappa_{21} = 0.5$     & $b_{21} = 1$   &   $\alpha_{21} = 2$     &  $\kappa_{21} = 0.5$      & $b_{21} = -1$ \\
$\alpha_{22} = 2 $    &  $\kappa_{22} = 0.4$     & $b_{22} = 1$   &   $\alpha_{22} = 2$     &  $\kappa_{22} = 0.3$      & $b_{22} = 1$  \\
$\alpha_{n_1} = 1$    &  $\kappa_{n_1} = 0.3$    &                &   $\alpha_{n_1} = 0$    &  $\kappa_{n_1} = 0.15$    &               \\
$\alpha_{n_1} = 0$    &  $\kappa_{n_2} = 0.4$    &                &   $\alpha_{n_2} = 0$    &  $\kappa_{n_2} = 0.3$     &                \\
\hline
 \end{tabular}
  \label{tab: simulated_parameters}
\end{table}

\begin{table}
\centering
\caption{Inference with simulated dataset 1}
\begin{tabular}{c|c|c|c}
  \hline
  \hline
Parameters     &   True value   &  Estimated      & Standard deviations     \\
\hline
$\kappa_{11}$  &     0.3        &   0.295         &       0.019             \\
$\kappa_{21}$  &     0.5        &   0.471         &       0.044             \\
$\kappa_{22}$  &     0.4        &   0.380         &       0.020             \\ 
$b_{11}$       &      1         &   1.009         &       0.069             \\
$b_{21}$       &      1         &   1.032         &       0.064             \\
$b_{22}$       &      1         &   0.997         &       0.059             \\
\hline
 \end{tabular}
 \label{tab: simulated_result1} 
\end{table}

\begin{table}
\centering
\caption{Inference with simulated dataset 2}
\begin{tabular}{c|c|c|c}
  \hline
 \hline
Parameters     &   True value   &  Estimated      & Standard deviations \\
\hline
$\kappa_{11}$  &     0.15       &   0.139         &       0.124             \\
$\kappa_{21}$  &     0.5        &   0.487         &       0.059             \\
$\kappa_{22}$  &     0.3        &   0.293         &       0.061             \\ 
$b_{11}$       &      1         &   0.991         &       0.017             \\
$b_{21}$       &     -1         &  -1.002         &       0.033             \\
$b_{22}$       &      1         &   1.011         &       0.018             \\
\hline
 \end{tabular}
\label{tab: simulated_result2} 
\end{table}

\begin{table}
\centering
\caption{Inference for the trivariate GRF}
\begin{tabular}{c|c|c|c}
  \hline
  \hline
Parameters     &   True value   &  Estimated      & Standard deviations \\
\hline
$b_{11}$       &      1         &   1.002         &     0.014               \\
$b_{21}$       &     0.8        &   0.807         &     0.024               \\
$b_{22}$       &      1         &   0.984         &     0.013               \\
$b_{31}$       &      1         &   0.958         &     0.031               \\
$b_{32}$       &     0.9        &   0.878         &     0.026               \\
$b_{33}$       &      1         &   0.986         &     0.012                \\
$\kappa_{11}$  &     0.5        &   0.481         &     0.022               \\
$\kappa_{21}$  &     0.6        &   0.582         &     0.044               \\
$\kappa_{22}$  &     0.4        &   0.393         &     0.021               \\ 
$\kappa_{31}$  &     0.5        &   0.564         &     0.050               \\
$\kappa_{32}$  &      1         &   1.008         &     0.034               \\
$\kappa_{33}$  &     0.3        &   0.286         &     0.012               \\
\hline
 \end{tabular}
 \label{tab: simulated_result3} 
\end{table}

\subsection{Inference with real data} \label{sec: real_data}
We illustrate how to use the SPDEs approach for multivariate data in real application in this section. The meteorological dataset used by \citet{gneitingmatern} is analyzed in this paper.  
This meteorological dataset contains one realization consisting of $157$ observations of both temperature and pressure in the north American Pacific Northwest, and 
the temperature and pressure are always observed at the same locations.
It contains the following data: pressure errors (in Pascal) and temperature errors (in Kelvin), measured against longitude and latitude.  The errors are defined as the forecasts minus the observations.
The data are valid on 18th, December of 2003 at 16:00 local time, at a forecast horizon of 48 hours. For information about the data, see, for example, \citet{eckel2005aspects} and \citet{gneitingmatern}.
\citet{gneitingmatern} have chosen this dataset with the aim of doing probabilistic weather field forecasting, 
which is a big research area \citep{gel2004calibrated, berrocal2007combining, sloughter2007probabilistic, berrocal2008probabilistic}. One of the main focuses in this area is to fit 
and sample spatially correlated error fields. This aim fits our motivation well since the SPDEs approach can be applied to construct multivariate random fields in order to
capture the dependence structures not only within the fields but also between the fields.

The main aim of this example is to illustrate how to fit a bivariate random field with the SPDEs approach for pressure and temperature errors data. This bivariate random field
can be used to explain the features of the two random fields as in \citet{gneitingmatern}.
The constructed bivariate spatial random fields are used to represent the error fields for temperature and pressure.
It is known that the temperature and the pressure are negatively correlated \citep{courtier1998ecmwf,ingleby2001statistical}.
As pointed out by \citet{gneitingmatern}, the forecast fields are usually smooth fields. The observation field for the pressure is smooth. However, 
the observation field for temperature is rough. So the model should set up to give the same type of behavior.
Without any confusion, we will call the temperature error field the temperature and the pressure error field the pressure.
In general, we need to choose the order of the fields at the modelling stage. The simple way is fitting the 
data with both the orders and selecting the best one using some predefined scoring rules. For more information about the scoring rules, see for example, \citet{gneiting2005calibrated}.  
In this paper we will set the first field $x_1(\boldsymbol{s})$ as the pressure and the second field $x_2(\boldsymbol{s})$ as the temperature. 

The bivariate random field with our approach is constructed with a system of SPDEs, 
\begin{equation}
\label{eq: bivariate_system_euqations}
 \begin{split}
 & (\kappa_{n_1}^2 - \Delta)^{\alpha_{n_1}/2} f_1(\boldsymbol{s}) =  \mathcal{W}_1(\boldsymbol{s}),   \\
 & (\kappa_{n_2}^2 - \Delta)^{\alpha_{n_2}/2} f_2(\boldsymbol{s}) =  \mathcal{W}_2(\boldsymbol{s}),   \\
 & b_{11}(\kappa_{11}^2 - \Delta)^{\alpha_{11}/2} x_1(\boldsymbol{s}) = f_1(\boldsymbol{s}),   \\
 & b_{22}(\kappa_{22}^2 - \Delta)^{\alpha_{22}/2}x_2(\boldsymbol{s}) +  b_{21}(\kappa_{21}^2 - \Delta)^{\alpha_{21}/2} x_1(\boldsymbol{s}) = f_2(\boldsymbol{s}).
 \end{split}
\end{equation} 
Since the main purpose of this example is to illustrate that we can construct the same (or similar) model as the covariance-based approach presented by \citet{gneitingmatern},  only the models which can result in
similar models as theirs have been chosen from the parameter matching results given in Section \ref{sec: parameter_matching}.
We assume that $\kappa_{n_1} = \kappa_{11}$ and $\kappa_{n_2} = \kappa_{22}$.
This particular setting corresponds to the first random field $x_1(\boldsymbol{s})$ being a Mat\'ern random field and the second random field $x_2(\boldsymbol{s})$ being close to a Mat\'ern random field.
These settings make the model constructed through our SPDEs approach closer to the \citet{gneitingmatern} approach.

\begin{figure}
  \centering
 \subfigure[]{\includegraphics[width=0.7\textwidth,height=0.55\textwidth]{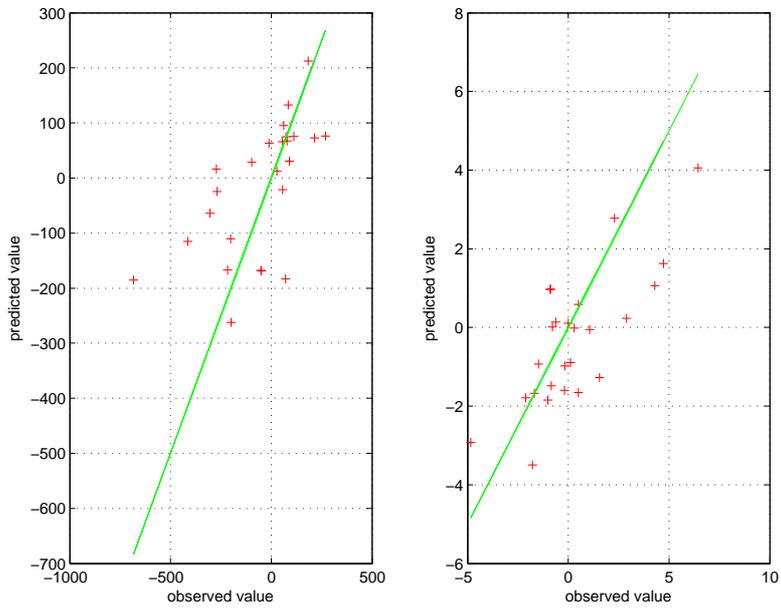}}
  \subfigure[]{\includegraphics[width=0.7\textwidth,height=0.55\textwidth]{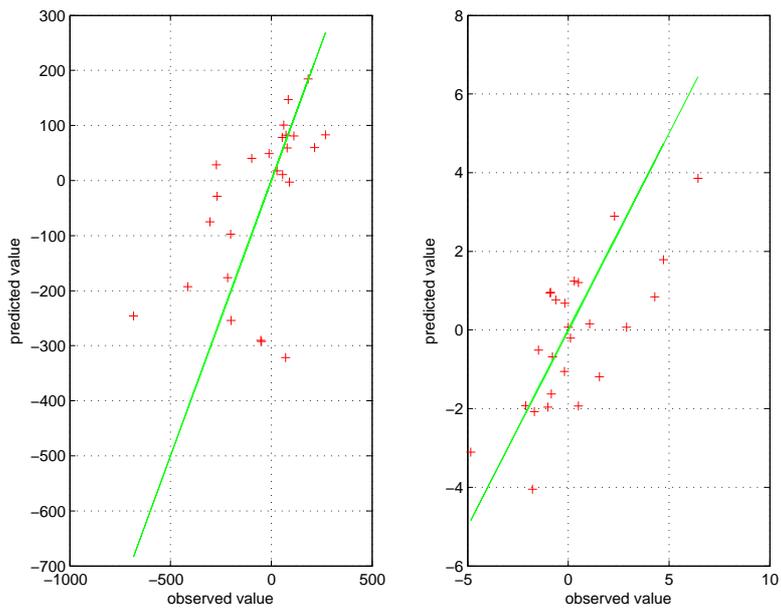}}
 \caption{The predictive performances of SPDEs approach (a) and the covariance-based approach (b) }
   \label{fig: bivariate_predictive_performance}
\end{figure}

\begin{figure}
 \centering
\includegraphics[width=0.8\textwidth,height=0.6\textwidth]{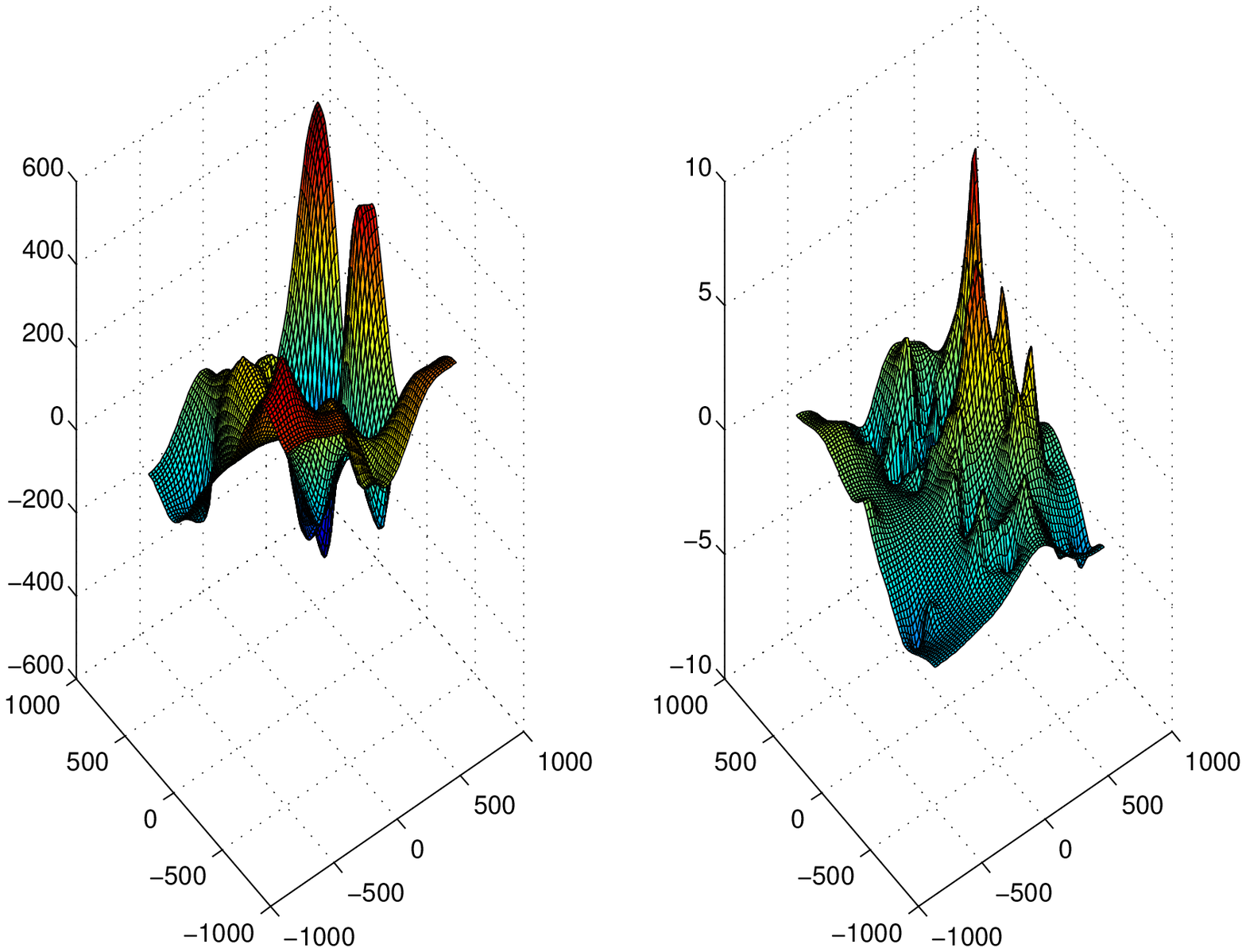}
\caption{Estimated conditional mean for bivariate GRFs for SPDEs approach in $3$D.}
  \label{fig: bivariate_RFs_realdata_3d}
\end{figure}

\begin{figure}
 \centering
 \includegraphics[width=0.9\textwidth,height=0.9\textwidth]{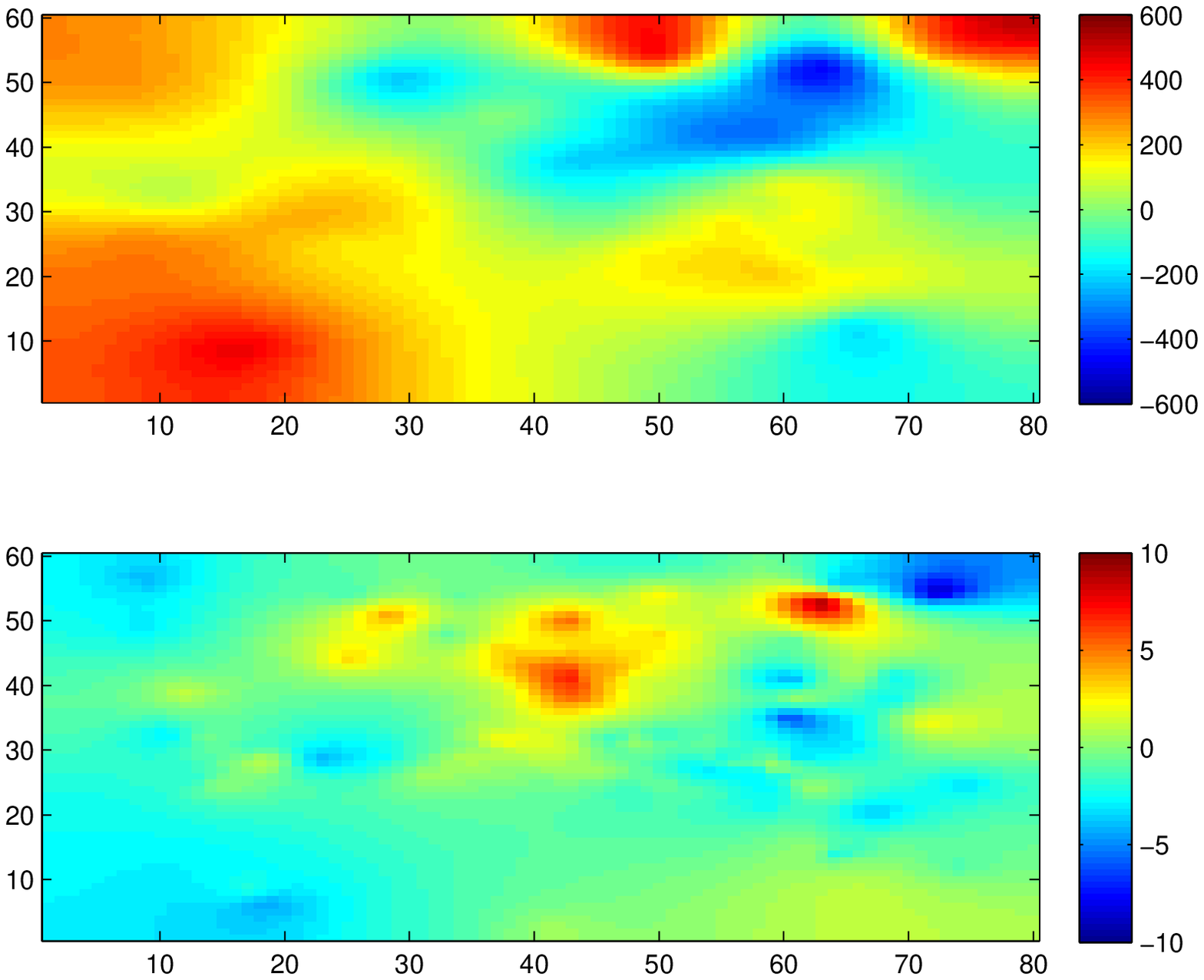}
\caption{Estimated conditional mean for bivariate GRFs for SPDEs approach in  $2$D.}
  \label{fig: bivariate_RFs_realdata_2d}
\end{figure}

\begin{figure}
 \centering
   \includegraphics[width=0.8\textwidth,height=0.6\textwidth]{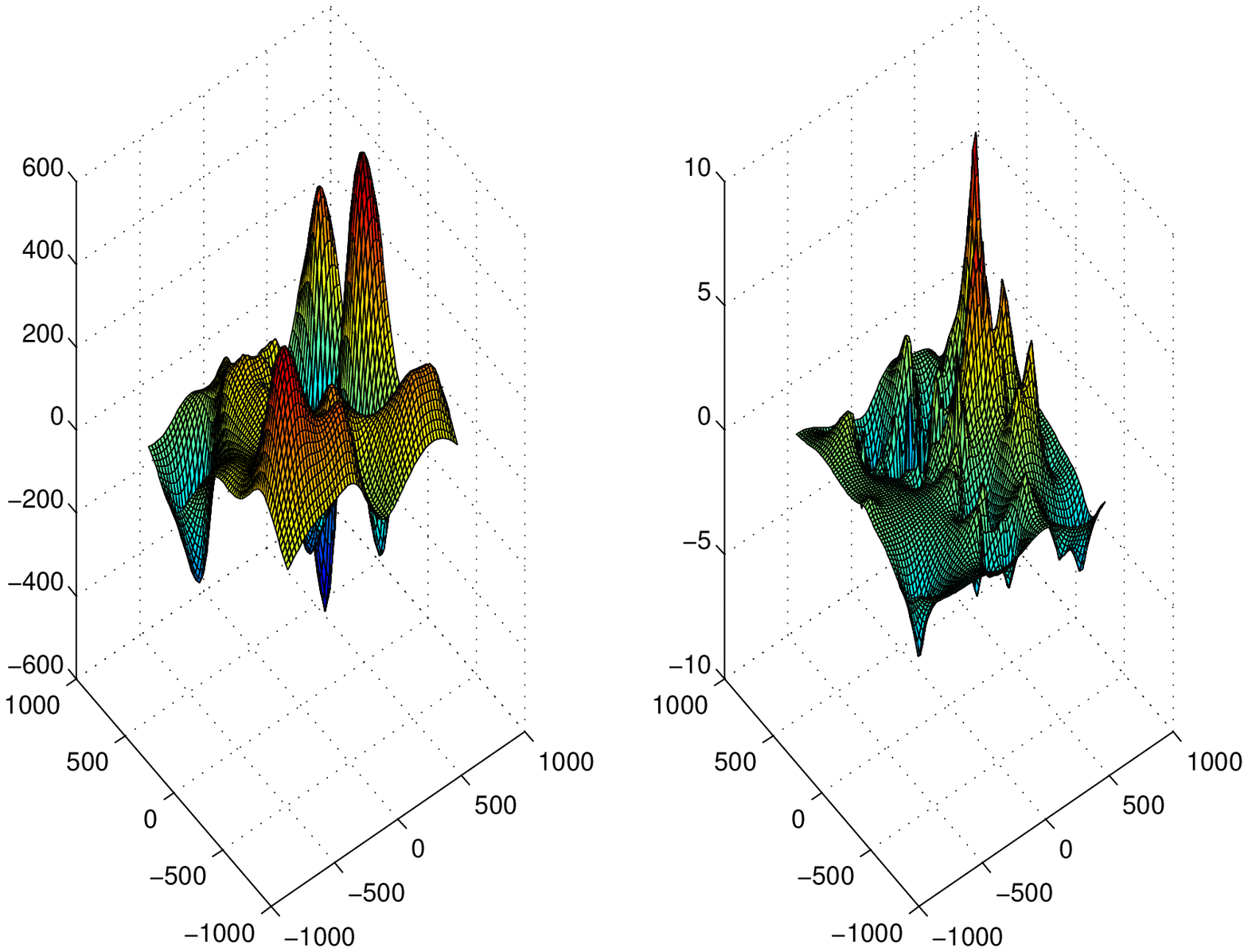}
\caption{Estimated conditional mean for bivariate GRFs for covariance-based model in $3$D.}
 \label{fig: bivariate_RFs_gneiting_3d}
\end{figure}

\begin{figure}
 \centering
   \includegraphics[width=0.9\textwidth,height=0.9\textwidth]{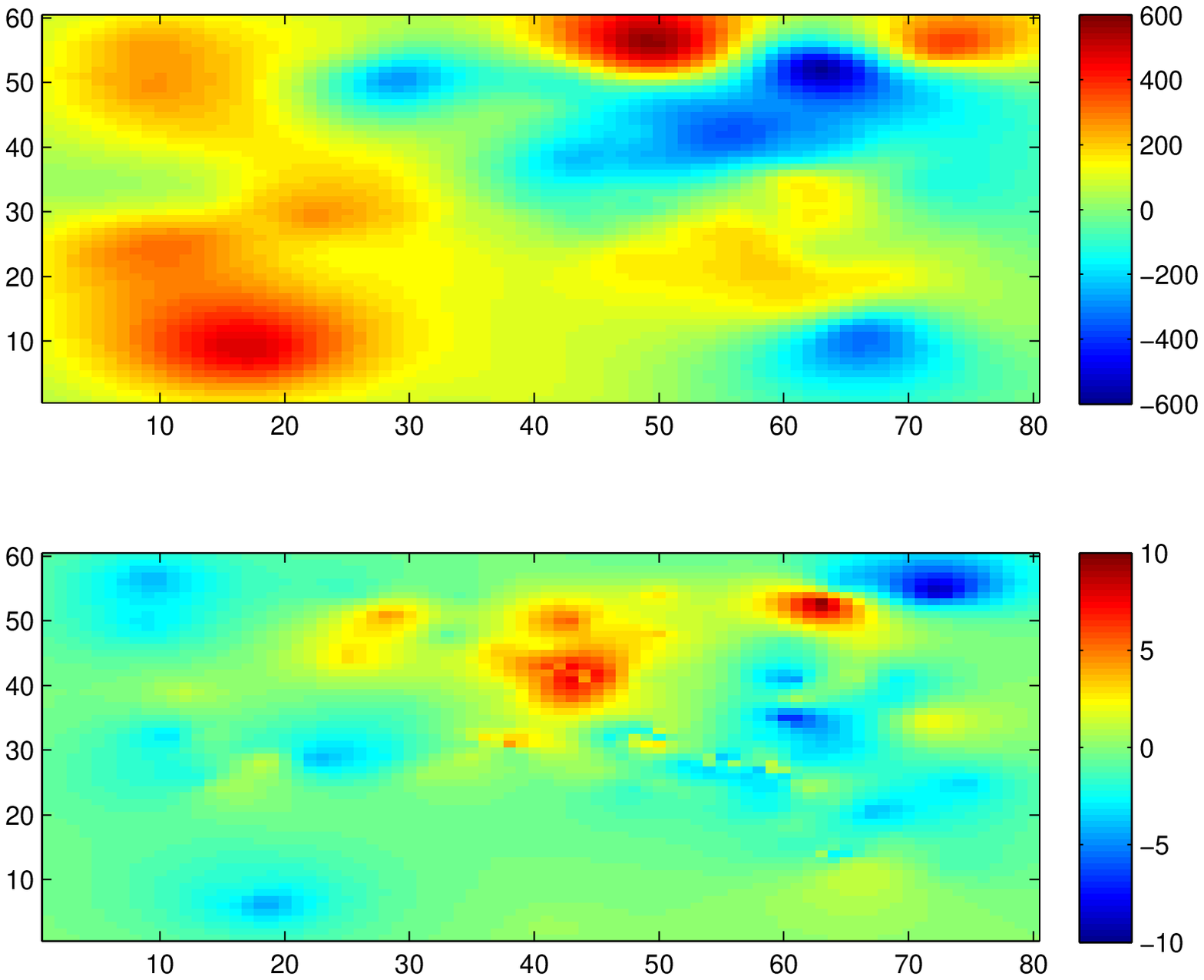}
\caption{Estimated conditional mean for bivariate GRFs for covariance-based model in $2$D.}
 \label{fig: bivariate_RFs_gneiting_2d}
\end{figure}

The results for the estimates with the 
SPDEs approach are given in Table \ref{tab: SPDEs_results_parameters}. From the table we can notice that we can capture the negative dependence structure between temperature and pressure 
since $b_{21} > 0$.  The estimated co-located correlation coefficient $\rho_{pt} = -0.53$ which is quite similar as the result from \citet{gneitingmatern}. The standard deviations for pressure and temperature
are $\sigma_{P} = 202.1$ Pascal and $\sigma_{T} = 2.76$ Celsius degrees. 
In order to compare the predictive perform between our approach and approach proposed by \citet{gneitingmatern}, we randomly leave out $25$ observations from each field
and use only $132$ observations for parameters estimation. The relative error $\mathscr{E}$ has been chosen to compare the predictions and is defined as
\begin{displaymath}
 \mathscr{E}= \frac{\norm{\hat{\boldsymbol{y}} - \boldsymbol{y}}_2}{\norm{\hat{\boldsymbol{y}}}_2}.
\end{displaymath}
where $\norm{\cdot}_2$ denotes the $2$-norm. $\hat{\boldsymbol{y}}$ denotes the vector of predictions for the observations $\boldsymbol{y}$. 
The predictive performances for these two approaches are given  in Figure \ref{fig: bivariate_predictive_performance}. We can notice that the results from our SPDEs approach and the approach in \citet{gneitingmatern} are quite similar.
Table \ref{tab: SPDEs_pred_err} shows the corresponding predictive errors for these two approaches. From this table, we can notice that  
our model gives slightly better results than the covariance-based model but the difference is not statistically significant.
Another merit with our approach which has not been discussed until now is that the SPDEs approach in general is much more computationally efficient since the precision matrix $\boldsymbol{Q}$ is sparse.
The conditional mean in $3$D and $2$D for the bivariate GRF are shown in Figure \ref{fig: bivariate_RFs_realdata_3d} and Figure \ref{fig: bivariate_RFs_realdata_2d} for our approach, respectively.
The corresponding results for the covariance-based approach are shown in Figure \ref{fig: bivariate_RFs_gneiting_3d} and Figure \ref{fig: bivariate_RFs_gneiting_2d}. As we can see from the $3$D and $2$D figures, 
the bivariate GRFs from our approach and from the covariance-based approach give quite similar results.

For the nugget effects (measurement error variances) of pressure and temperature $\boldsymbol{\tau} = \left( {\tau_1}, {\tau_2} \right)^\text{T}$, 
we are going to use an iterative bias correction to estimate them. This is based on formula 
\begin{equation}
{\sigma}_{ij}^2 = \tau_{i}^2 + V_{ij}, \quad i = {1,2}, \hspace{3mm}  j = 1,\dots, n,   
\label{eq: bias_correction}
\end{equation}
where $i$ indicates the pressure with $i = 1$ and the temperature with $i = 2$.  ${\sigma}_{ij}^2 = \mbox{Var}(y_{ij} - \hat{y}_{ij})$ is the variance which contains 
the nugget effects and the kriging variances. $\{ \tau_{i}^2 = \mbox{Var}(y_{ij} -x_{ij}); i = 1,2, j = 1, \dots, n\}$ are the nugget effects for the pressure when $i = 1$
or temperature when $i = 2$.  $\{ V_{ij} = \mbox{Var}(x_{ij} - \hat{y}_{ij});  i = 1,2, j = 1, \dots, n\}$ are the kriging variances and they are from the model bias. $n$ denotes the total number of data points in each field. 
$y_{ij}$ denotes the observed value at data point for each field. $\hat{y}_{ij}$ is the predicated values from a given model and ${x_{ij}}$ denotes the true value which is unknown.
See Appendix for a simple proof of \eqref{eq: bias_correction}.

\begin{figure}
 \centering
\includegraphics[width=0.8\textwidth,height=0.8\textwidth]{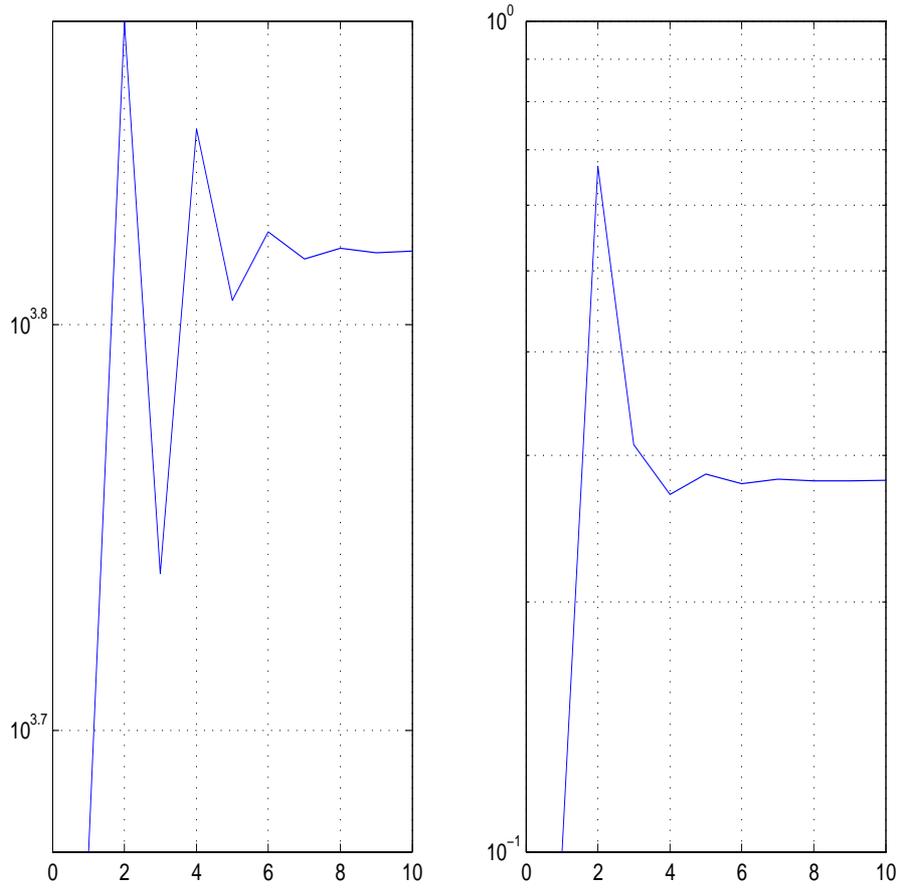}
 \caption{Bias correction iteration. The curves are for $\tau_1^2$ and $\tau_2^2$ respectively}
  \label{fig: bivariate_bias_correction}                                    
\end{figure}

One simple way is to use Equation \eqref{eq: bias_correction_simple} where we just subtract the kriging variance $V_{ij}$ from the empirical variance $\hat{\sigma}_{ij}^2$ to 
get the estimate of nugget effect $\hat{\tau}_{i}^2$ for each field 
\begin{equation}
 \hat{\tau}_{i}^2 = \frac{1}{n} \sum_j  (\hat{\sigma}_{ij}^2 -V_{ij}), 
\label{eq: bias_correction_simple}
\end{equation}
where  $\hat{\sigma}_{ij}^2 = (y_{ij} -\hat{y}_{ij})^2$. But another preferable way which we have used in this paper is with the formula 
\begin{equation}
  \hat{\tau}_{i}^2 = \sum_j \left\{ w_{ij} \left( \hat{\sigma}_{ij}^2 - V_{ij} \right) \right\},
 \label{eq: bias_correction_better}
\end{equation}
where $w_{ij} = \frac{1/(\tau_{i}^2 +V_{ij})}{\sum_j \left( 1/(\tau_{i}^2 +V_{ij}) \right)}$. This can give an unbiased estimator of nugget effects.
See Appendix for the proof.
The initial values are chosen to be similar to the results given in \citet{gneitingmatern}, and the results from the bias correction approach are shown in Table \ref{tab: bias_correction}.
The convergence curves of the nugget effects for the fields are illustrated in Figure \ref{fig: bivariate_bias_correction}.

\begin{table}
 \caption{Estimated parameters for the SPDEs approach}
\centering
 \begin{tabular}{c|c|c|c|c|c}
  \hline
 \hline                            
   $\kappa_{11}$        &    $\kappa_{21}$             &   $\kappa_{22}$                   &   $b_{11}$                &        $b_{21}$            &    $b_{22}$     \\
\hline                                                                                                                            
   $6.353 \times 10^{-3}$ &   $0.413$     &     $2.243 \times 10^{-2}$        &    $0.2165$  &   $1.298 \times 10^{-4}$  &     $5.458$        \\
\hline
 \end{tabular}
\label{tab: SPDEs_results_parameters}
\end{table}

\begin{table}
 \caption{Predictive relative errors for the SPDEs approach and the covariance-based models}
\centering
 \begin{tabular}{c|c|c}
  \hline
   \hline
                          \multicolumn{3}{c}{ relative errors}                       \\
\hline
  Models                   &     random field $1$          &   random field $2$    \\
\hline
   SPDEs approach          &         $0.777$               &      $0.690$           \\
\hline
  Covariance-based model  &         $0.821$               &      $0.716$            \\
\hline
 \end{tabular}
\label{tab: SPDEs_pred_err}
\end{table}

\begin{table}
 \centering
\caption{The nugget effects for pressure and temperature}
 \begin{tabular}{c|c|c}
  \hline
  \hline
              &       initial values          &    estimated values     \\
  \hline
  $\tau_1$    &      68.4                    &         81.1               \\
  \hline
  $\tau_2$    &       0.01                   &         0.53                \\
  \hline
\end{tabular}
 \label{tab: bias_correction}
\end{table}

\section{Discussion} \label{sec: discussion}
Spatial modelling of multivariate data are of demand in many areas, such as weather forecasting \citep{courtier1998ecmwf,reich2007multivariate}, air quality \citep{brown1994multivariate, schmidt2003bayesian}, 
economics \citep{gelfand2004nonstationary, sain2007spatial}. The important issue for modelling spatial multivariate data is that the approach can not only model the marginal covariances  
within each field, but also has the ability to model the cross-covariances between the random fields. In addition, we need to solve the theoretical challenge for the positive definite constraint of the covariance functions
and the computational challenge for large dataset.

The main aim of this work is to illustrated the possibility of constructing \emph{multivariate} GRFs through the SPDEs approach.
We notice that the parameters in the SPDEs approach is interpretable and can link our approach to the covariance-based approach.
By using the approximate weak solution of the corresponding system of SPDEs, we can represent multivariate GRFs by GMRFs. Since the precision matrices of the GMRFs are sparse, 
numerical algorithms for sparse matrices can be applied, and therefore fast sampling and inference are feasible. Our approach inherited the properties from the approach discussed by \citet{lindgren2011explicit}.
There are three main advantages for our newly proposed SPDEs approach. The first advantage is that our new SPDEs approach does not depend on direct construction of positive definite matrix.
The notorious requirement of positive definite covariance matrix is fulfilled automatically.
The second advantage is that we can remove the symmetry property constraint shared by both the covariance-based approach \citet{gneitingmatern}
and the LMC approach \citep{gelfand2004nonstationary, gneitingmatern}. 
The third advantage is that we can construct the multivariate GRFs on manifolds, such as on the sphere $\mathbb{S}^2$. The extension follows the discussion given by \citet{lindgren2011explicit}.
We just need to reinterpret the systems of SPDEs to be defined on the manifold. 

One issue that needs to be pointed out is that we have chosen $\kappa_{n_1} = \kappa_{11} \text{ and } \kappa_{n_2} = \kappa_{22}$ in the model. This restriction used in this paper is to make the model closer to the models
presented by \citet{gneitingmatern} and also make the inference easier. This may not be needed in other applications and $\kappa_{n_1}$ and $\kappa_{n_2}$ can be estimated from the data. However, 
$\kappa_{n_1}$ and $\kappa_{11}$ might be not distinguishable with the triangular systems of SPDEs. So one suggestion is that if we are use the triangular systems of SPDEs, then we fix $\kappa_{n_1}$ 
based on some other information or set $\kappa_{n_1} = \kappa_{11}$ when doing inference.

Another issue that needed to be pointed out is that we have prosecuted the \emph{full} version of the SPDEs approach but have not used in all the examples. This version could give us more flexibility to construct multivariate GRFs. The 
modelling procedures and the inference are the same as for the \emph{triangular} version of the SPDEs which were used and discussed extensively in this paper.

We also want to point out that the proposed approach costs effort during the implementation and pre-processing stages since we need to build the system of SPDEs, discretize the fields and do the approximation to 
obtain a GMRF representation. But we believe, as pointed out by \citet{lindgren2011explicit}, that `` \emph{such costs are unavoidable when efficient computations are required}''.  

Similarly, as pointed out by \citet{lindgren2011explicit}, although the SPDEs approach presented here is not generally applicable for all covariance functions, 
it covers many useful model in spatial statistics. And it is 
possible to extend our approach to construct even richer class of models by following the discussion given by, for instance, \citet{bolin2011spatial} and \citet{fuglstad2011spatial}. 
Our approach extends the applicability of multivariate GRFs in practical applications since we can build and interpret the model using GRFs but do computation with GMRFs.  
It is further possible to include this approach in the integrated nested Laplace approximation (INLA) framework \citep{rue2009approximate}. These extensions are under study.

\appendix
\section*{Appendix}
In order to prove Equation \eqref{eq: bias_correction}, we need to write down the expression explicitly. 
Let $y_{ij}$ denote the observed value for each field, $\hat{y}_{ij}$ the predicated value from a given model, and ${x_{ij}}$ the true value. Then we have
\begin{equation}
\label{eq: appendix_totalvariace}
 \begin{split}
  \mbox{Var}(y_{ij} - \hat{y}_{ij}) & = \mbox{Var}\left( (y_{ij} - x_{ij}) + (x_{ij} - \hat{y}_{ij}) \right) \\
                                    & = \mbox{Var} \left( y_{ij} - x_{ij} \right) +\mbox{Var} \left( x_{ij} - \hat{y}_{ij} \right) \\
                                    & \hspace{5mm}      + 2\mbox{Cov}\left(y_{ij} - x_{ij}, x_{ij} - \hat{y}_{ij} \right) \\
                                    & = \mbox{Var} \left( y_{ij} - x_{ij} \right) +\mbox{Var} \left( x_{ij} - \hat{y}_{ij} \right),
 \end{split}
\end{equation}
since 
\begin{equation}
\label{eq: appendix_crossterm}
 \begin{split}
  \mbox{Cov}\left(y_{ij} - x_{ij}, x_{ij} - \hat{y}_{ij} \right) & = \mathbb{E}\left((y_{ij} - x_{ij})(x_{ij} - \hat{y}_{ij}) \right) \\
                                                                 & = \mathbb{E}\left[\mathbb{E}\left((y_{ij} - x_{ij})(x_{ij} - \hat{y}_{ij}) |\boldsymbol{y}\right)\right] \\
                                                                 & = \mathbb{E}\left[ (y_{ij} - \hat{y}_{ij}) \times 0 \right] \\ 
                                                                 & = 0.
 \end{split}
\end{equation}
So Equation \eqref{eq: bias_correction} is established. 

Now we show that with the weights $w_{ij} = \frac{1/(\tau_{i}^2 +V_{ij})}{\sum_j (\tau_{i}^2 +V_{ij})}$, the estimator in Equation \eqref{eq: bias_correction_better} is an unbiased estimator for the nugget effects.
Using Equation \eqref{eq: appendix_totalvariace}, we know that $(y_{ij} - \hat{y}_{ij}) \sim \mathcal{N}(0, \tau_{i}^2 +V_{ij})$. Then 
\begin{equation}
 \begin{split}
  \mbox{Var}\left((y_{ij} - \hat{y}_{ij})^2\right) & = \mathbb{E}\left((y_{ij} - \hat{y}_{ij})^4\right) - \left(\mathbb{E}\left((y_{ij} - \hat{y}_{ij})^2\right)\right)^2 \\
                                  & = 3(\tau_{i}^2 + V_{ij})^2 - (\tau_{i}^2 + V_{ij})^2 \\
                                  & = 2 (\tau_{i}^2 + V_{ij})^2
 \end{split}
\end{equation}
 and thus the estimator is unbiased since the scaling constant cancels in the weight coefficient. 
                                                                                                                                         
\bibliographystyle{plainnat}
\linespread{0.8}{\small {\bibliography {../Reference/Ref}}}


\end{document}